\documentstyle[12pt,psfig,aaspp4]{article}

\def\beq{\begin{equation}}
\def\eeq{\end{equation}}
\def\bey{\begin{eqnarray}}
\def\eey{\end{eqnarray}}

\def\v200{V_{200}}

\def\araa{{\rm ARA\&A}, }
\def\aj{{\rm AJ}, }  
\def\apj{{\rm ApJ}, }  
\def\apjs{{\rm ApJS}, }  
\def\mn{{\rm MNRAS}, }      
\def\nat{{\rm Nature}, }      
\def\aa{{\rm A\&A}, }     

\def\spose#1{\hbox to 0pt{#1\hss}}
\def\lta{\mathrel{\spose{\lower 3pt\hbox{$\mathchar"218$}}
     \raise 2.0pt\hbox{$\mathchar"13C$}}}
\def\gta{\mathrel{\spose{\lower 3pt\hbox{$\mathchar"218$}}
     \raise 2.0pt\hbox{$\mathchar"13E$}}}

\def\clock{\count0=\time \divide\count0 by 60
     \count1=\count0 \multiply\count1 by -60 \advance\count1 by \time
     \number\count0:\ifnum\count1<10{0\number\count1}\else\number\count1\fi}

\begin{document}
\singlespace
\title{The Dynamical Evolution of Substructure}

\author{Bing Zhang\altaffilmark{1},   Rosemary F.G. Wyse\altaffilmark{1,2}}
\affil{Department of Physics and Astronomy, Johns Hopkins University, 3400 N.Charles Street, 
Baltimore, MD 21218, USA}

\author{Massimo Stiavelli\altaffilmark{1}}
\affil{Space Telescope Science Institute, 3700 San Martin Drive, Baltimore, MD 21218, USA}

\author{Joseph Silk\altaffilmark{1}}
\affil{Department of Astrophysics, University of Oxford, Keble Road, Oxford OX1 3RH, England}

\altaffiltext{1}{E-mail: bing@epicsystems.com (BZ); wyse@pha.jhu.edu (RFGW); 
mstiavel@stsci.edu (MS); silk@astro.ox.ac.uk (JS)} 
\altaffiltext{2}{Also School of Physics \& Astronomy, University of St Andrews, North Haugh, KY16 9SS, Scotland}

\begin{abstract}

The evolution of substructure embedded in non-dissipative dark halos
is studied through N-body simulations of isolated systems, both in and
out of initial equilibrium, complementing cosmological simulations of
the growth of structure.  We determine by both analytic calculations
and direct analysis of the N-body simulations the relative importance
of various dynamical processes acting on the clumps, such as the
removal of material by global tides, clump-clump heating, clump-clump
merging and dynamical friction.  The ratio of the internal clump
velocity dispersion to that of the dark halo is an important
parameter; as this ratio approaches a value of unity, heating by close
encounters between clumps becomes less important while the other
dynamical processes continue to increase in importance.  Our
comparison between merging and disruption processes implies that
spiral galaxies cannot be formed in a proto-system that contains a few
large clumps, but can be formed through the accretion of many small
clumps; elliptical galaxies form in a more clumpy environment than do
spiral galaxies.  Our results support the idea that the central cusp
in the density profiles of dark halos is the consequence of
self-limiting merging of small, dense halos. This implies that the
collapse of a system of clumps/substructure is not sufficient to form
a cD galaxy, with an extended envelope; plausibly subsequent accretion
of large galaxies is required.  The post-collapse system is in general
triaxial, with rounder systems resulting from fewer, but more massive,
clumps.  Persistent streams of material from disrupted clumps can be
found in the outer regions of the final system, and at an overdensity
of around 0.75, can cover 10\% to 30\% of the sky.

\end{abstract}

\keywords {galaxies: formation - galaxies: structure - galaxies: elliptical - 
galaxies: spiral}

\section{Introduction}

In the context of Cold Dark Matter dominated (CDM), hierarchical
clustering cosmology, large galaxies are formed from the merging and
accretion of small, less massive progenitors (clumps). The exact
role played by this substructure in the formation of galaxies is
still unclear, and there remain unresolved several very important and
fundamental issues that are closely related to the role of these clumps. 

The angular momentum content of galaxies is a fundamental parameter.
Cosmological N-body simulations have verified the proposal that tidal
torques between proto-galaxies generate angular momentum (Peebles
1969), and have shown that the dimensionless spin parameter $\lambda
\equiv J | E |^{1/2} G^{-1} M^{-5/2}$ has a well-defined mean value of
$\sim 0.06$, independent of the details of the power spectrum of
density fluctuations (Efstathiou \& Jones 1979; Aarseth \& Fall 1980;
Barnes \& Efstathiou 1987; Warren {\it et al.} 1992; Steinmetz \&
Bartelmann 1995; Cole \& Lacey 1996).  Since the spin parameter
$\lambda$ is much smaller for present day ellipticals -- indeed equal
to the value of $\sim 0.06$ of a typical dark halo -- than is the
value of $\lambda$ for spirals, while the effective radius for spirals
is comparable to that of ellipticals at a given luminosity (Fall
1980), it is important to know how the ellipticals succeed in losing
angular momentum, compared to the expectation of smooth collapse and
spin-up within a dark halo (Fall \& Efstathiou 1980) while still
dissipating the same amount of binding energy as do proto-spirals. On
the other hand, there exists an angular momentum problem in that while
disks are formed in fully self-consistent hierarchical-clustering
simulations, the angular momentum transport inherent in the merging
process produces disks that are too centrally-concentrated and contain
too little angular momentum compared with observed spirals
(e.g. Navarro \& Benz 1991; Evrard, Summers \& Davis 1994; Vedel,
Hellsten \& Sommer-Larsen 1994; Navarro \& Steinmetz 1997). Indeed
standard analytic and semi-analytic models of the formation of spiral
galaxies show that extended disks, as observed, form through detailed
angular momentum conservation, with proto-disk material retaining the
same value of specific angular momentum as the dark halo (e.g. White
\& Rees 1978; Fall \& Efstathiou 1980; Gunn 1982; Jones \& Wyse 1983;
Dalcanton, Spergel \& Summers 1997; Mo, Mao \& White 1998; Zhang \&
Wyse 2000, Silk 2001).  Hence models must retain angular momentum for spirals,
while losing it for ellipticals.  Mergers and the associated
gravitational torques and angular momentum transport clearly play a
role.  However, detailed merger simulations of two disk galaxies have
shown that there are difficulties with the simplest merger picture of
how ellipticals form (Bendo \& Barnes 2000; Cretton, Naab, Rix \&
Burkert 2001), in that equal mass mergers seem to produce too much
kinematic misalignment, while in unequal mass mergers the strong
signatures of the more massive disk survive.

What are the 
effects of substructure 
such as gas clouds, stellar (globular?) clusters, dwarf galaxies or
sub-halos which fall later into the galactic-scale potential well? Do they
significantly change the angular momentum distribution and further
change the final galaxy classification?  An analytical approach will
have difficulties in dealing with this problem, although the secondary
infall model is successful in describing the smooth post-collapse halo
profile (Gunn \& Gott 1972; Gunn 1977; Filmore \& Goldreich 1984;
Hoffman \& Shaham 1985; Bertschinger \& Watts 1988; Zaroubi, Naim \&
Hoffman 1996). However, numerical simulations have the capability of
treating the formation and the non-linear evolution of dark halos in a
cosmological context (Quinn, Salmon \& Zurek 1986; Zurek, Quinn \&
Salmon 1988; Quinn \& Zurek 1988; Warren {\it et al.} 1992; Dubinski
\& Carlberg 1991). These studies show that the angular momentum and
binding energy are indeed redistributed, in an orderly manner, during
the relaxation of the halos (Quinn \& Zurek 1988).  Further, during
the merging process inherent in hierarchical clustering, baryonic
substructure can lose  angular momentum 
to the dark halo 
(Frenk {\it et al.} 1985; Zurek, Quinn \& Salmon 1988; Barnes 1988;
Navarro \& Steinmetz 1997; Navarro, Frenk \& White 1995;).  Zurek,
Quinn \& Salmon (1988) suggest that whether the baryonic clumps are gas
clouds (and hence dissipative) or stellar  could be an important
criterion in determining whether the forming  galaxy will become a spiral or
an elliptical. 

Proposed solutions to the problem with disks, within the context of
CDM cosmologies, have included delaying the onset of disk formation
until after the merging process is essentially complete, by appeal to
suitable `feedback' from the first stars (Weil, Eke \& Efstathiou
1998), but this solution must address the old age for disk stars in
the solar neighborhood (e.g. Edvardsson {\it et al.} 1993; Wyse 1999)
and in the outer parts of the M31 disk (Ferguson  \& Johnson 2001),
in addition to the observation of apparently fully-formed disks at
redshifts above unity (Brinchmann \& Ellis 2000) with 
a Tully-Fisher relation  that is essentially
indistinguishable from that of present-day disks, apart from passive
evolution.   The density and mass of the
substructure are clearly important, and a major motivation for the
present work is to quantify the effects of substructure of a range of
each.  Further, when is substructure disrupted and why?

Another issue we seek to clarify is how much of the angular momentum
change seen in the cosmological simulations is due to the boundary
conditions, namely that the system under study is not closed.  It is
hard to separate the effects of the real angular momentum transfer
from clumps to halo from the angular momentum gain by the accretion of
material or the torques applied by the neighboring environment or the
later infall of dark matter.  Thus it is worthwhile to study, as here,
the evolution of angular momentum in isolated systems.  In the
isolated N-body simulations by Barnes (1988) of the encounter of two
galaxies, it is found that the dark halo can extract some angular
momentum from other components. 

In this paper, we shall extend such
N-body studies of the dynamical evolution to analyse the evolution of
a system of clumps in a larger, smooth, isolated system. 
Our simulations have relevance beyond a forming proto-galaxy.  We may
address the dynamical processes occurring in a virialized system with
substructure, such as a cluster of galaxies or a single galaxy
and its retinue of dwarf companion galaxies.  The relative importance
of the effects of the dissipationless collapse of the overall system, 
tidal disruption of clumps, dynamical friction of clumps,
clump-clump heating and merging can possibly be related to the formation
of cD galaxies in clusters of galaxies, to the formation of central
density cusps in dark halos, to the triaxial nature of dark halos, to the
formation of substructure composed of disrupted debris in the outer
halo regions of galaxies, etc.
Thus we initiated a consistent comparison among these dynamical
processes in the context of galaxy formation.  

In section 2, we present the parameters of our N-body simulations.
Section 3 contains a discussion of our expectations for the dynamics
of the substructure based on analytic calculations and comparisons
with the N-body simulations, while section 4 discusses the angular
momentum content and transport.  The internal kinematics of the
different components are discussed in Section 5, while their
morphology is discussed in Section 6.  Section 7 summarizes our
conclusions.  The more technical aspects are given in the Appendices.


\section{The N-body Models}                   
\subsection{The Technique}
We adopt a numerical N-body code based on the hierarchical tree
algorithm (Barnes \& Hut 1986; Hernquist 1987). Our simulations study
the dynamical evolution of a number of clumps embedded in a smooth
dark halo. Due to the collisionless property of the particles used in
our N-body simulation, it should be emphasized that the clump can
represent either stellar or dark matter depending on our
interpretation of the physical entity.  The initial dark halo profile
is modelled with either the Hernquist profile (Hernquist 1990) or the
Plummer profile (Plummer 1911), the details of which are given in Appendix~A.  The Hernquist profile, with $\rho \sim
r^{-1}$ at the centre and $\rho \sim r^{-4}$ at large radius, was
introduced as being, when projected, close to the observed surface
brightness profiles of elliptical galaxies, and is close to the mass
profile seen in hierarchical clustering simulations (the NFW profile;
Navarro, Frenk \& White 1997), which has a cuspy centre, $\rho \sim
r^{-1}$ (though flatter than the $\sim r^{-1.5}$ behaviour of
high-resolution simulations; Ghigna {\it et al.} 2000), and $\rho \sim
r^{-3}$ at large radius.  The Plummer profile has close to 
a harmonic potential
in the core (only slowly varying density) and $\rho \sim r^{-5}$ at large
radius. In all realisations, each clump is simulated with a Plummer
profile.

We ran simulations of systems in initial equilibrium plus
out-of-equilibrium systems that underwent an initial collapse. 
In order to study the angular momentum behaviour of the clumps and the
dark halo, we require that the system contain enough angular momentum,
consistent with the cosmological initial conditions expected, i.e.
$\lambda \sim 0.1$ (e.g. Efstathiou \& Jones 1979; Barnes \&
Efstathiou 1987). First, we build the dark halo with no streaming
velocity. Then, we add some angular momentum about the $z$ direction,
which is defined to be the axis of the halo. The orbital rotation
velocity added to each particle is the component projected on the
equatorial plane of a constant fraction, $\beta$, of the circular
velocity where the particle is located, i.e., $V_{rot}=\beta V_c(r)
cos \theta $, where $\theta$ is the angle between $\hat{\bf r}$ and
$\hat{\bf z}$. We further stretch the spherical profile to an oblate
profile with ellipticity $\epsilon_0 =0.53$ (Warren {\it et al.}
1992).  Assuming, if the system is virialized, that the flattening is
only caused by the addition of rotation, we adopt the empirical
relation $\beta=\sqrt{\epsilon_0/5}$, which may be seen by noting the
scaling of the rotation parameter $v_0/\sigma \propto
\sqrt{\epsilon_0}$ for isotropic systems flattened by rotation (Binney
\& Tremaine 1987).  With this choice of $\beta$, the initial spin
parameter of the halo in the virialized simulations (with the
virial ratio $2T/W \sim 1$, where $T$ is the total kinetic energy
and $W$ is the total potential energy) is $\lambda \sim 0.16$.
The
collapse simulation models (with the virial ratio $2T/W \sim 0.1$)
have $\lambda \sim 0.07$.  In order to achieve an initial condition
suitable for pre-collapse, i.e. with small $2T/W$, we decrease the
total velocity of the dark halo particles and the velocity of each
clump by a factor of $\sqrt{10}$ for $2T/W \sim 0.1$.

No internal rotation is
added to the clumps, which are spherical.  
The distribution of clumps is chosen to follow
the dark halo profile with  similar treatment of the flattening and
rotation velocity. For the simulation models with the same number 
of clumps, we require the  orbital characteristics 
for the clumps be the same with the exception of models B 
and E (see Table~1).

\subsection{The Parameters and Units}

The parameter values of interest for our simulations are given in
Table~\ref{tab:parameter}.  The total number of particles in our
simulations ranged from 20,000 to 120,000, with smooth dark halo
particles and clump particles having an equal mass.  The total number
of clumps, $n_c$, varied from 5 to 80 and the total clump mass
fraction, $f$, varied from 5\% to 100\%.  These ranges were chosen to
study the effect of clumpiness on the outcome of the simulations. The
core radius of a clump (a Plummer sphere) is set to achieve the chosen
density contrast, $\rho_{cl}/\rho_0$, the ratio of the mean density
within the half mass radius of the clump to that of the whole system
(dark matter plus the clumps).  We vary the density parameter
$\rho_{cl}/\rho_{DM}$ (with $\rho_{cl}/\rho_0 = (1-f)
\rho_{cl}/\rho_{DM}$) over the range of 3 to 64 in our simulations, to
study the effects of global tides in disrupting the clumps.  A value
of three for the density contrast is the minimum expected for a system
to withstand global tides (the Roche criterion).

The total mass of the system is normalized to be unity.  The unit of
distance is the core size of the dark halo ($a=1$, see Appendix~A) and
we further set the gravitation constant $G=1$.

With these units and as detailed in Appendix A, the mean crossing time
for a Hernquist halo is 5.3, while that for a Plummer halo is 2.1, and
the simulations are run for $\sim 25$ crossing times, or approximately
a Hubble time in physical units.

It is convenient  to express some of our parameters,  such as the  
clump mass fraction $f$, the number of clumps  $n_c$, density contrast 
of each clump  
$\rho_{cl}/\rho_0$, 
and crossing time $t_{\frac{1}{2} h c}$,  in terms of 
dimensionless quantities.  
From our definition of the density contrast of the clump, 
the ratio of the half mass radius of a  
clump to that of the whole system can be written as
\beq
r_{\frac{1}{2} c} / r_{\frac{1}{2} h} = f^{1/3} n^{-1/3}_c \left( \frac{\rho_{cl}}{\rho_0} \right)^{-1/3}.
\eeq

The ratio of the one-dimensional internal velocity dispersion of the clump and the halo is 
$ \sigma_c/\sigma_h =\frac{v_{\frac{1}{2} c}}{v_{\frac{1}{2} h}}
=\frac{r_{\frac{1}{2} c}}{r_{\frac{1}{2} h}} \frac{t_{\frac{1}{2}hc}}{t_{\frac{1}{2}cc}}
=\left( \frac{r_{\frac{1}{2} c}}{r_{\frac{1}{2} h}} \right) \left( \frac{\rho_{cl}}{\rho_0}\right)^{1/2}$. This can be 
further written as
\beq
 \sigma_c/\sigma_h = f^{1/3} n^{-1/3}_c \left(\frac{\rho_{cl}}{\rho_0}\right)^{1/6}.
\eeq

\subsection{Numerical Effects}

As demonstrated in Appendix~B,  while there may well be spurious clump
heating effects in our simulations, due to the limited number of
particles used, they do not pose a problem for our analysis. 
  The softening-length in the treecode is set at $r_s=0.5
r_{\frac{1}{2} cl} N^{-1/3}_{cl}$, where $N_{cl}$ is the number of
particles used in each clump.  This relatively large value is chosen
to suppress the unphysical two-body relaxation (White 1978) which
otherwise could occur due to the small number of particles used in
each clump (ranging from 150 to 2400).

\begin{figure}[htbp]
\centerline{\psfig{file=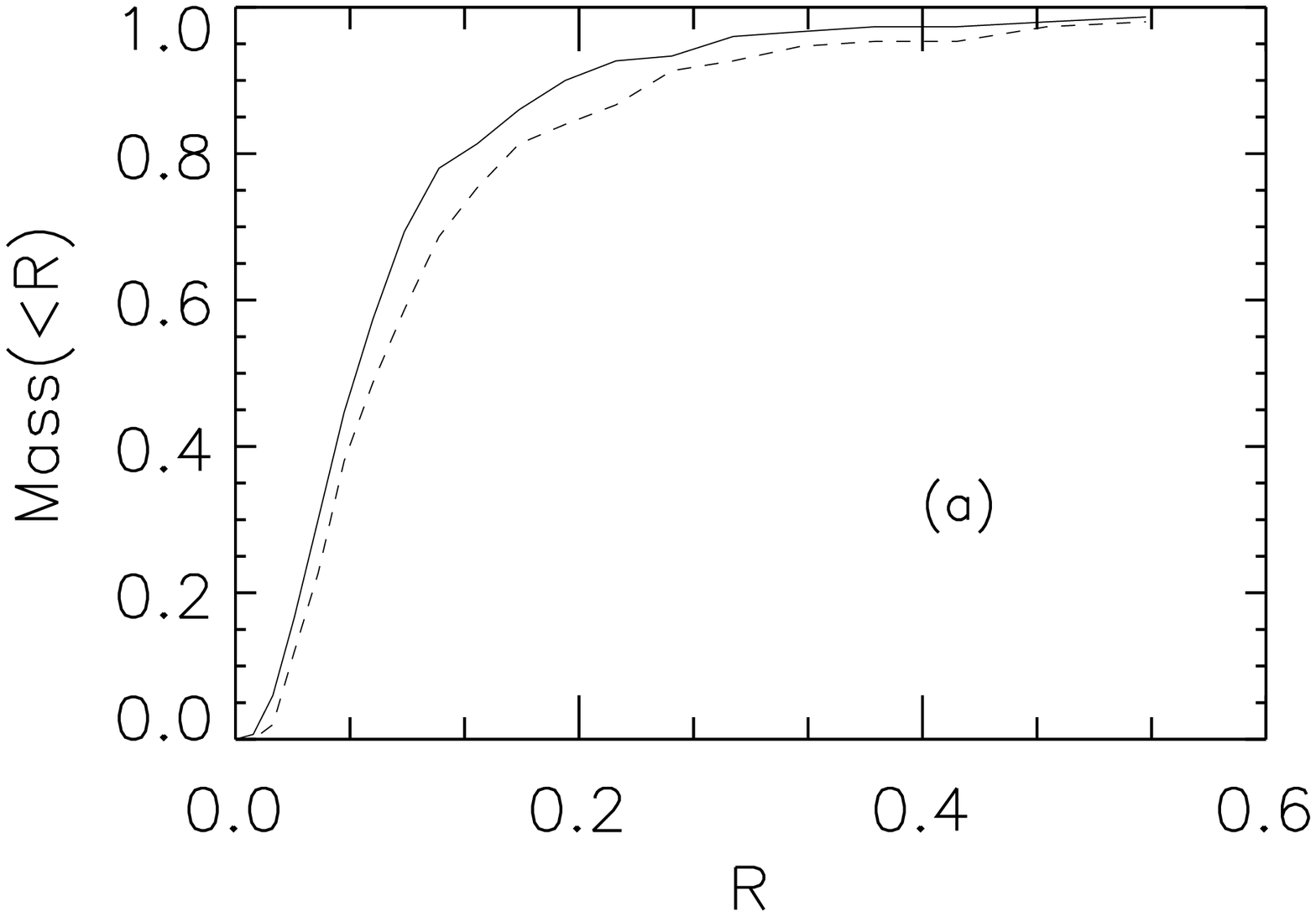,width=0.6\textwidth,angle=0}}
\vspace{1.0cm}
\centerline{\psfig{file=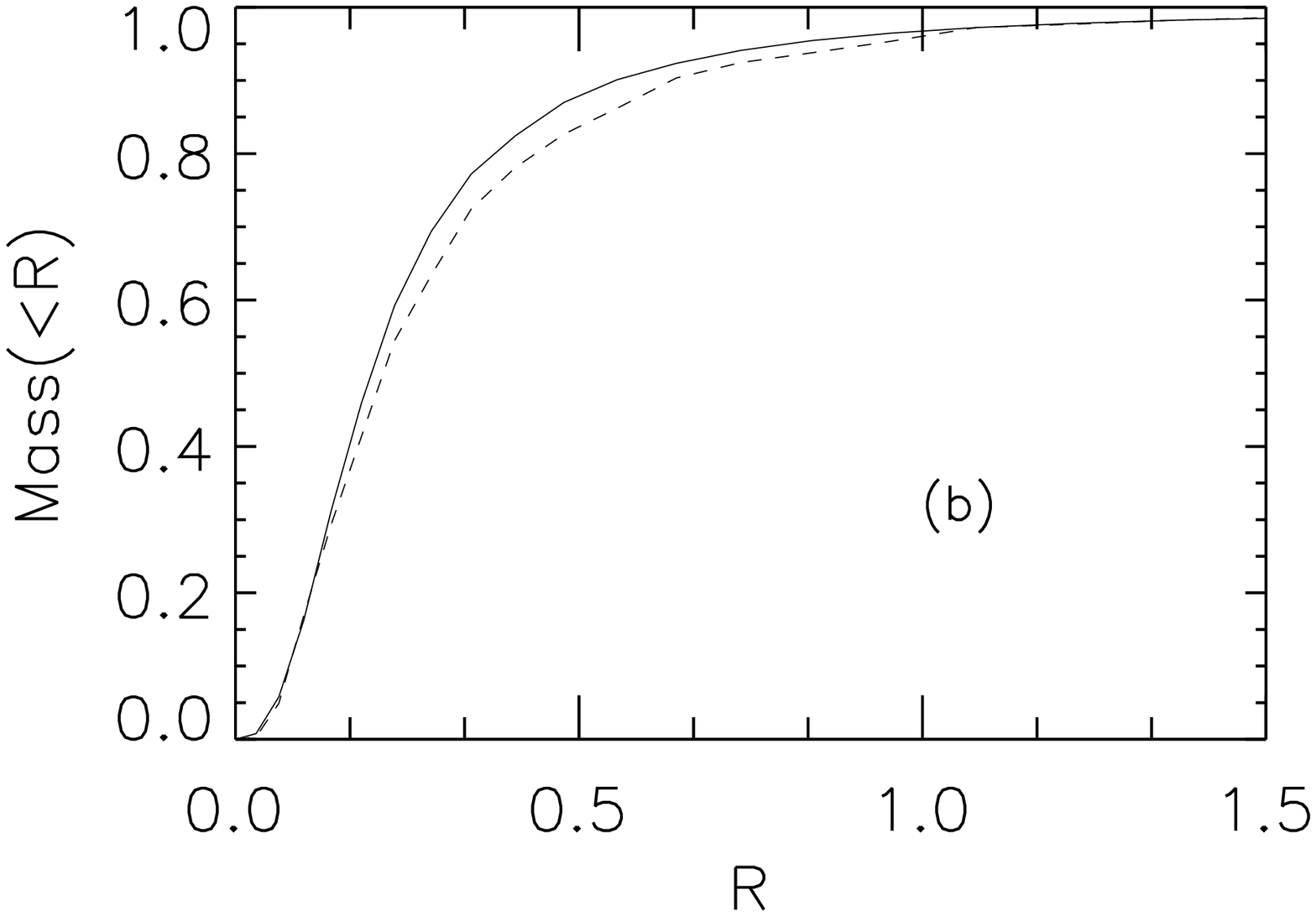,width=0.6\textwidth,angle=0}}
\vspace{1.0cm}
\caption[relax]
{The mass profile, in arbitrary units,  
for an  unperturbed clump with (a) 150 particles, from simulation A  (b) 900 particles, from simulation H,  both 
at the beginning (solid line) and at the end (dashed line) of 
our simulations. There is negligible change with time, showing that relaxation effects are not important.} 
\label{fig:relax}
\end{figure}

Figure~\ref{fig:relax} shows the mass profiles  
for examples of unperturbed clumps 
consisting of 150 particles (from simulation A) and 900 
particles (from simulation H)  
at the beginning (solid line) and the end (dashed line) of 
our simulations. The radius of a given mass fraction does not change much even 
for the clump with 150 particles, demonstrating that indeed  
relaxation effects are not important.

\subsection{The Analysis Method}
In this section, we describe the methods we used in calculating the
quantities of physical interest, such as the angular momentum, the properties of the debris from the disrupted clumps, and the 
ellipticity or triaxiality, density and kinematics of the clumps and the halo.

In calculating the angular momentum, one needs to choose the origin
carefully.  Here we calculate the angular momentum of the different
components of the system in two ways, relative to the centre of mass
of the whole system and relative to the highest density point of the
dark halo. The centre of mass of the whole system can be calculated
directly, while the highest density point of the dark halo can be
searched for in an iterative way: first we choose the centre of mass
of the whole system as the starting point $C_0$, and calculate the
centre of mass $C_1$ of the dark halo particles contained within a
radius $r_1=20$ relative to the point $C_0$.  Then we adopt the point
$C_1$ as the next starting point to repeat the above process until the
centre of mass converges; then we decrease the radius $r_1$ by half
and repeat the first step.  After several iterations of these two
steps, the highest density point is located when the search radius
$r_1$ has decreased greatly.  Throughout this paper we adopt the
highest density point as the origin when calculating other physical
quantities such as the density profile, ellipticity and kinematics.

The debris disrupted from each clump is calculated in an iterative
way: at some timestep, e.g. initially, for each clump we know those
particles still bound to that clump; at the next timestep we check
which of those particles remain bound to the clump, and we assign the
unbound particles to the debris.  We repeat this process at each
timestep until the end of the simulation is reached or the clump is
completely disrupted. It should be noted that our simulation code is
not designed to search for merging events, which would require a more
sophisticated algorithm that includes a check on a merging criterion
at each timestep of the integration.  Thus the `debris' we identify
could include some clump particles that are no longer bound to their
initial clump due to the fact that this initial clump has merged with
another clump, and the particles are bound to the new merger remnant,
rather than being truely unbound.  In other words, 
we do not re-assign particles to a new, larger clump
should one clump be subsumed in another. We simply keep track of
whether the particles remain bound or not to the initial clump. 
However, our calculation
of the debris would be inaccurate only in a very few cases in which
the merging process is very efficient.  

In addition, when identifying debris, we ignore the
possibility that a clump can capture particles that were removed from
another clump, but this is not an important process for our
simulations which do not include a dissipative component.

The intrinsic triaxiality of the final dark halo or debris component
is calculated in a simple way.  We calculate the moment of inertia
formed by the relevant particles within a chosen radius 
(e.g. half mass radius).  From the
eigenvalues of the moment of inertia $I_1 < I_2 < I_3$, 
we can fit the moment of inertia with a triaxial ellipsoid with
the intrinsic axial
ratios $\epsilon_1=b/a$ and $\epsilon_2=c/a$ where $a \ge b \ge c$. 
It can be calculated that
$\epsilon_1=\sqrt{\frac{-PQ-P+Q}{PQ-P-Q}}$ and 
$\epsilon_2=\sqrt{\frac{-PQ+P-Q}{PQ-P-Q}}$ where $P=I_1/I_2$ and $Q=I_1/I_3$.
The intrinsic axial
ratios calculated in this way can be slightly overestimated compared
to their real values, but provide a consistent comparison among our
simulations.  The ellipticity of the projected (surface density)
images of the final dark halo or debris can be calculated in a more
direct way, for a range of viewing angles.  We calculate an isodensity
contour map by assigning a Gaussian density profile to each
particle, with width  proportional to the softening length. 
Then we simply fit the contour with the
ellipse to obtain the axial ratio $b/a$ and the ellipticity $\epsilon
=1-b/a$.

The radial density profile is calculated by dividing the particles
into radial bins, each containing  equal numbers of particles (about 16). 
The velocity dispersion and rotation velocity are
calculated by the conventional approach. 


\section{The Dynamics of Clumps}
  We can study the dynamical evolution of clumps that contain enough
particles.  The study of the dynamics of substructure has applications
on both the galaxy size scale, where star clusters, gas clouds and
dwarf satellite galaxies play the role of clumps (Fall \& Rees 1977),
and the galaxy cluster scale, where galaxies play the role of clumps
(Dressler 1984).  It is also related to the `undermerging' problem
that has emerged recently, in that high-resolution CDM cosmological
simulations predict too many surviving dark matter satellites around
halos of the size of our Galaxy (Klypin {\it et al.} 1999; Moore {\it
et al.} 1999)).

The dynamical processes associated with the evolution of these clumps
we shall investigate are global tidal stripping, dynamical friction,
and merging and close encounters between clumps.  We first derive
expectations, based on analytic arguments, for the roles these
dynamical processes have played and on which parameters they depend,
for the virialized models.  A preliminary comparison between these
theoretical estimates and our simulations is provided at the end of
the section. Snapshots of each model are plotted in Figures~2--7.

\subsection{Analytic Expectations}

We derived analytic expectations for the amplitudes of the various
processes we believe are operating in the simulations, to gain an
understanding of the results. Many processes are working together,
often with the same net result (disruption of the clumps), which
would greatly complicate the interpretation of the N-body simulations in the absence of analytic insight.

\subsubsection{The Effects of Global Tides} 
Global tides due to the spatial variation of the underlying
large-scale dark halo potential provide an important mechanism for the
disruption of substructures as they orbit through the dark halo.
Analytical studies of the tidal effect on a clump show that, for a given
clump orbit, the disruption efficiency depends on the ratio between the
clump density to the mean density of halo within the pericentre of
the clump orbit (e.g. King 1962; Binney \& Tremaine 1987; 
Johnston, Hernquist \& Bolte 1996).
The tidal radius of the clump, during its disruption, is the Roche
radius at which  its gravity is equal to the tidal force exerted by
the dark halo, which roughly scales as
\beq
r_{tide} \sim \left(\frac{M_c}{M_{tot}}\right)^{1/3} R,
\eeq
where $R$ is the pericentre distance.  For our simulations, we expect
that the clumps with  density contrast $\left( \rho_{cl}/\rho_0
\right) =3$ should suffer more efficient tidal disruption
than those with $\left( \rho_{cl}/\rho_0 \right) = 64$. For a
clump in a circular orbit at the half mass radius of the dark halo, the
fraction of mass contained within the tidal radius is $ \sim 97\% $
for $\left( \rho_{cl}/\rho_0 \right) = 64$ and $ \sim 74\% $ for
$\left( \rho_{cl}/\rho_0 \right) = 3$. Since the clumps in our models are each 
represented by a Plummer profile, which has a fairly  constant density core (see Appendix~A), 
while the dark halo is represented typically by a Hernquist profile, which has
a cuspy central density profile, the minimum pericentre distance for
the approximately constant density core to be disrupted is $\sim 0.085$ for $\left(
\rho_{cl}/\rho_0 \right) = 64$ and $\sim 0.72$ for $\left(
\rho_{cl}/\rho_0 \right) = 3$.

\subsubsection{Merging Effects}
The merging cross section between a pair of identical spherical clumps
has been studied by Makino \& Hut (1997) numerically, using a variety
of mass profiles for the clumps.  They find that in a cluster of
clumps with one-dimensional velocity dispersion $\sigma_{h}$ the
number of merging events per unit time per unit volume, ${\cal R}_m$,
is given by \beq {\cal R}_m = \frac{18}{\sqrt{\pi}}\frac{1}{x^3} n^2
r^2_{c vi} \sigma_{c} {\cal R}_0(x) = \frac{18}{\sqrt{\pi}}\frac{1}{x^4} n^2 r^2_{c
vi} \sigma_{h} {\cal R}_0(x), \eeq where $x = \sigma_h/\sigma_c $,
with $\sigma_h$ and $\sigma_c$ being the one-dimensional velocity
dispersion of the system of clumps (the halo) and the internal
velocity dispersion of a clump respectively, $n$ is the number density
of the clumps within the half mass radius of the dark halo, $r_{c vi}$
is the virial radius of the clump, and the dimensionless quantity
${\cal R}_0(x) =\frac{Ax^2}{x^2+B},$ with $A=12 $ and $B=0.4$ for
clumps with a Plummer profile.  Written this way, one can see
explicitly a strong dependence of the merging rate on the parameter
$x$ (the ratio of the internal velocity dispersions of halo and
clump).

The number of merging events per unit time, $R_m$, expected in our
simulations can be estimated by the product of ${\cal R}_m$ and the
volume within the half mass radius of the halo profile (equal to the
half mass radius of the system of clumps). After some algebraic
manipulations using equations (1) and (2) given in section~2.2 and the
scalings in Appendix~A, we obtain \beq R_m =\frac{4}{3}\pi
r^3_{\frac{1}{2} h} {\cal R}_m \simeq 0.47 f^2 {\cal R}_0(x)
\frac{1}{t_{\frac{1}{2}hc}}.  \eeq

Thus the number of merging events per halo crossing time is $0.47 f^2
{\cal R}_0(x)$, and depends fairly strongly on the clump mass
fraction, $f$, but with only a weak residual explicit dependence on
the parameter $x$ (many parameters are interdependent).  In the limit
of a high relative velocity compared to internal velocity dispersion,
$x^2 \gg 0.4$, which is approximately valid for all our models, ${\cal
R}_0(x) \sim 12$.  This limit does not favour mergers, and results in
the number of merging events per halo crossing time being $\sim 0.05$
for $f=0.1$, $\sim 1$ for $f=0.4$ and $\sim 5$ for $f = 1$.

We can also estimate the merging timescale for a clump. 
For a given clump within the half-mass radius of the halo, 
the probability of it merging with 
another clump per unit time is ${\cal R}_m /n$. Thus  the merging time scale is
\beq
t_m=n/{\cal R}_m=\frac{n_c}{f^2 {\cal R}_0(x)} t_{\frac{1}{2}hc}. 
\eeq
It should be emphasized that the internal velocity dispersion ratio of
the halo to the clump, $x$, itself depends on the clump mass fraction
$f$, the clump number $n_c$ and the density contrast
$\rho_{cl}/\rho_0$ by equation (2). We can see here that, given the values of the model parameters, the merging
time scale is only very weakly dependent
on the density contrast, which is instead a crucial factor for the
efficiency of the global tidal stripping. Merging between clumps  
is an important effect for large values of the 
clump mass fraction, $f$, and for small numbers of clumps, ${n_c}$. In our
models with $f \sim 0.1$, the merging effect is very small with $t_m
\gg 50 t_{\frac{1}{2}hc}$, while for larger $f > 0.4$,  merging
can be important, and the minimum merging timescale is $t_m
\sim 1.7 t_{\frac{1}{2} hc}$ (model~I).

The merging rate calculated above is the mean quantity at the initial
half mass radius.  Further, it does not take into account other
dynamical effects, such as global tidal effects and close encounters, that can
greatly decrease the calculated merging rate (Makino \& Hut 1997). The
effects of global tides can decrease the merging cross section by
tending to tear apart a pair of clumps which, if isolated, would have
merged.  Similarly, the local tides from a third clump can also act to
tear apart a merging pair of clumps. Furthermore, the subsequent
nonlinear evolution in the presence of other more important dynamical
processes can make the estimated merging timescale change quickly with
time. We will see the limitation of this analytic estimate when we
compare with the simulations below.

\subsubsection{Clump-Clump Heating}
Disruption of clumps can also be caused by close encounters with other
clumps.  Similarly to the discussion given in section 2.3.2, we can
estimate this clump disruption timescale as follows: For a given
clump, the heating rate is \beq \dot E = \frac{4 \sqrt{\pi} G^2 m^3_c
n \overline{ r^2_c}}{3 g \sigma_h r^2_{\frac{1}{2} c}}, \eeq where $n$
is the mean clump number density within the half mass radius of the
dark halo, and all other parameters are as defined above. As discussed in Appendix B2, we
adopt $g \sim 3$ and $ \overline{r^2_c} \sim 2 r^2_{\frac{1}{2}
c}$. Thus the clump disruption timescale due to clump-clump heating is
given by: 

\beq t_{c-c} =0.03 f^{-4/3} n^{1/3}_c \left(
\rho_{cl}/\rho_0 \right)^{1/3} t_{\frac{1}{2} hc}.  \eeq

Thus  clump-clump heating is important for a large clump mass
fraction, $f$,  small clump number, $n_c$, and low density contrast,
$\rho_{cl}/\rho_0$.  We have $t_{c-c} \sim 11 t_{\frac{1}{2} hc}$ for
model A, and $t_{c-c} \sim 1.5 t_{\frac{1}{2} hc}$ for
model B.

However, we should keep in mind that the above calculation could
overestimate the effects of clump-clump heating. 
The choice of $\overline{r^2_c}$ is uncertain.  Further, the
impulse approximation holds only for $\sigma_c/\sigma_h
\ll 1$ if the impact parameter $b$ is chosen to be $\sim
r_{\frac{1}{2} c}$.  If the encounter time is too long compared with
the clump internal crossing time, particles in the clump may
adiabatically respond to the encounter, hence the encounter could
leave no net effects (Binney \& Tremaine 1987).  This could possibly
explain the apparent inconsistency between our analytic estimates for
the clump-clump merging effect and the clump-clump heating effect
(equations (6) and (8)), in that they both become more important
with large clump mass fraction and small numbers of clumps.  
From equation (2),
we can see that $\sigma_c/\sigma_h $ also tends to be large for a large
clump mass fraction and a small number of clumps (provided that 
$\sigma_c/\sigma_h \ll 1$).  When
$\sigma_c/\sigma_h$ increases above $0.5$, our prediction of the 
clump-clump heating effect in equation (8) becomes invalid, and
the clump-clump heating effect will decrease quickly while the merging
effect will increase continuously. 

The above arguments can be understood in terms of 
three different regimes depending on the value of the  encounter speed, 
basically equivalent to the role of $\sigma_h$.  (1) In the case that the 
encounter speed $V$ is very high, the two clumps will not merge and
the clumps are just heated somewhat during the fast encounter. (2) In
the case that the encounter speed is just slightly reduced,  it
can be envisioned that the chance for merging is slightly increased,
and the heating effect is also slightly enhanced since the duration of
the encounter is slightly longer. (3) In the case that the encounter
velocity is very low, the merging rate  is greatly increased, and the
heating effect can be inhibited, since if the duration of the encounter
is long enough compared to the internal crossing timescale in the
clump, the clump just adiabatically responds during the encounter and
returns to the initial state after the encounter.

\subsubsection{Dynamical Friction}
Dynamical friction can drive clumps to the centre of the dark halo,
where they can be tidally disrupted more easily. The slowing down of
the clumps also can increase their merging cross-section.  At the same
time this process can extract angular momentum from the clumps, and
transfer it to the dark halo. Since Chandrasekhar (1943) introduced
the concept of dynamical friction, namely that an object moving
through an infinite and homogeneous medium made of small mass
particles suffers a drag force, there have been many studies on this
subject by numerical simulations (e.g. White 1978, 1983; Lin \&
Tremaine 1983; Bontekoe \& van Albada 1987; Zaritsky \& White 1988;
Hernquist \& Weinberg 1989; van den Bosch {\it et al.}  1999) and
analytical methods (e.g. Tremaine 1981; Tremaine \& Weinberg 1984;
Weinberg 1989; Maoz 1993; Dom\'{\i}nguez-Tenreiro \& G\'omez-Flechoso
1998; Colpi, Mayer \& Governato 1999; Tsuchiya \& Shimada 2000). An
overview of past work is given by Cora, Muzzio \& Vergne (1997).  In
spite of the difficulties encountered in the study of dynamical
friction, many authors find Chandrasekhar's formula is a remarkably
good approximation (e.g. Velazquez \& White 1999).

Adopting Chandrasekhar's formula the deceleration of a clump
is then 
\beq
\frac{d {\bf v}_M}{dt} = - \frac{{\bf v}_M}{t_{df}},
\eeq
where 
\beq
 t^{-1}_{df}=16 \pi^2 \ln \Lambda G^2 m_p (M_c+m_p) 
        \frac{\int^{v_M}_0 f(v) v^2 dv }{v^3_M}, 
\eeq  
and $m_p$ is the mass of a background halo particle, $M_c$ is the mass of
the clump, $v_M$ is the velocity of the clump, $f(v)$ is the
phase-space number density of the background halo medium, and
$\Lambda$ is the ratio between the maximum and minimum impact parameters 
$b_{max}/b_{min}$.  Using the values at the half mass radius, we can
immediately see that the characteristic dynamical friction time, 
\beq t_{df} \propto \frac{n_c}{f(1-f)} t_{\frac{1}{2}hc} \propto \frac {1}{M_c (1-f)} t_{\frac{1}{2}hc}
\eeq  scales
inversely with the mass of a clump ($ f/n_c$,  with the total mass normalized to unity as here).

In applying these formulae to  our 
simulations with virialized initial conditions, $2T/W \sim 1$,
and for which the dark halo is taken to follow the  Hernquist profile, we have chosen to
use the phase space distribution function given by Hernquist (1990), hence
ignoring the angular momentum dependence of the distribution
function. The phase-space density of the dark halo is then as detailed in Appendix~C. 
For a clump initially on a circular orbit at the half mass radius, the
dynamical friction time calculated as in Appendix~C  is
$$t_{df,circ}=\frac{0.16 n_c}{f(1-f)}t_{\frac{1}{2} hc}.$$  Analyses of the orbital eccentricities of substructure in spherical potentials have found that for isotropic distribution functions   the typical 
orbital  eccentricity is  $\sim 0.6$ (van den Bosch {\it et al.} 1999). With this value  the dynamical
friction time is decreased by a factor of up to 2, and the typical
dynamical friction time for our computed models should then be
\beq
t_{df} \sim 0.08 \frac{n_c}{f(1-f)}t_{\frac{1}{2} hc}.
\eeq

Thus dynamical friction is more important for a large clump mass
fraction (provided $f < 0.5$) and a small number of clumps, giving a
large mass for each clump.  This dependence on $f$ and $n_c$ in a
general sense is consistent with that of clump-clump merging in all
parameter ranges and that of clump-clump heating in some restricted
parameter ranges.  For our models with $n_c =80$, dynamical friction
is not important, with $t_{df} > 20 t_{\frac{1}{2} hc}$. For our
models with $n_c =20$, dynamical friction is not important for
$f=0.1$, with $t_{df} \sim 18 t_{\frac{1}{2} hc}$, but for $f=0.4$ it
becomes important, with $t_{df} \sim 6.7 t_{\frac{1}{2} hc}$.  For our
models with $n_c =5$, dynamical friction is important, with $t_{df}
\sim 8.4 t_{\frac{1}{2} hc}$ for $f=0.05$ and $t_{df} \sim 1.7
t_{\frac{1}{2} hc}$ for $f=0.4$.

It should be emphasized that efficient dynamical friction  leads to
significant angular momentum loss from the clumps, and thus 
can drive clumps to the centre, where  global tidal
effects are stronger.  Thus  the net efficiency of tidal stripping by the 
global potential includes a dependence on 
 the clump mass fraction
and number of clumps  similar to that of dynamical friction.

\subsection{Comparison with Simulations}
It should be noted that the above analysis is mainly applicable to virialized
models. For the pre-collapse models, the disruption of clumps due to
the collapse process itself dominates.  For the Plummer dark halo profile, 
the details of the  
calculation of dynamical friction from section 3.1.4 are not
applicable, but the scaling should be the same as in the Hernquist models. 
Thus, for ease of comparison, the dynamical friction time 
is  normalized to the crossing time scale at the half
mass radius, as are  all timescales.

As noted earlier, all models with the same
number of clumps have the same orbital characteristics for the clumps,
with the exception of models B 
and E. 
Thus it should be the different choices of our three free
parameters -- the density contrast $ \rho_{cl}/\rho_0$, the clump
mass fraction $f$ and the number of clumps $n_c$ -- that are
responsible for differences in the evolution of the different models.

The most important uncertainty in the comparison of the theoretical
predictions with our simulation results comes from two factors that
enter the determination of the theoretical clump-clump heating
timescale, namely the normalization quantity, $\overline{r^2_c}$, and
the use of the impulse approximation even for relatively large values
of $\sigma_c/\sigma_h$.  However, there are also uncertainties in the
analysis of our simulations. Our code is not designed to follow the
merging events that occur throughout the simulations, and as mentioned
above, the `debris' includes both particles genuinely removed from
clumps and orbiting freely in the global potential and the 
remnants of merging clumps. Thus in our analysis, both heating and 
merging  produce debris. As a rough estimate of
the mean disruption timescale of a clump in the simulations, 
we simply use the time required for the 
disruption  of  half of the clump mass.

Table~\ref{tab:timescale} lists, for the virialized models only, all
the timescales, in units of $t_{\frac{1}{2}hc}$, calculated from our
analytic expressions, together with the disruption time of half of the
mass in clumps measured from our simulations. The reader should note
that the relative importance of the processes has not be derived from
the simulations explicitly. Fig.~\ref{timescales} shows the dependence
of various processes on the mass of the clump, and also on the mass
fraction in clumps, $f$, the latter denoted by the symbol size.

The discussion of each virialized model is given in turn below: 

(1) For model A, since $\sigma_c/\sigma_h =0.21$, the
impulse approximation is valid in our calculation of the clump-clump
heating timescale $t_{c-c}$. Indeed, the  dominant dynamical process is
clump-clump heating and the analytic prediction is quite consistent with the 
simulation result.  

(2) For model B,   
again clump-clump heating is  the  dominant dynamical process. 
However our analytic estimate of  $t_{c-c}$ is  a factor of 3 below 
the simulation result. This can be understood since 
$\sigma_c/\sigma_h
=0.31$ and thus we are in the regime of marginal applicability of the impulse approximation, and we may have  underestimated 
$t_{c-c}$.

(3) For model C, we can see $\sigma_c/\sigma_h =0.46$ and thus
our underestimate of $t_{c-c}$ could be significant, possibly by as much as the 
factor of 30 needed for  consistency with the simulation result. 
It is also
possible that $t_{merge}$ is underestimated by the factor of 2 discrepancy 
with the simulation result. The dominant processes appear to be 
merging and clump-clump heating.  

(4) For model G,
$\sigma_c/\sigma_h =0.34$, and  again we find our apparent underestimate of
$t_{c-c}$ is about  a factor of 3.  The dominant process 
is apparently  dynamical friction or possibly clump-clump heating. Note that
dynamical friction acts to  drive clumps to the centre, where they will be  
tidally
disrupted more easily.  

(5) For model H, $\sigma_c/\sigma_h =0.5$, and 
again we have an apparent  underestimate of $t_{c-c}$ by  about a
factor of 20, and an  underestimate of $t_{merge}$ by  a factor of 2.
The dominant process appears to be dynamical friction.  

(6) For model
I, $\sigma_c/\sigma_h =0.74$, sufficiently close to unity that  
our underestimate of
$t_{c-c}$ could be very significant. We can simply assume that there is no
clump-clump heating for this case (adiabatic encounters). 
The dominant physical process is then merging of the clumps. 
As  can be seen from Figure~\ref{fig:vic20hern}, in  this model 
all the clumps finally merge into two large systems. Since in this case the
`debris' is mainly in these two merging remnants, we do not 
calculate the disruption
time in the simulation. 

 (7) For model M, $\sigma_c/\sigma_h =0.34$, and 
we again assume our underestimate of $t_{c-c}$ is about a factor of 3.  As
noted above, our analytic estimates of the dynamical friction
timescale are not applicable for Plummer profiles, but as a rough
estimate we can adopt the value from the equivalent Hernquist
profile model G.  The dominant processes are 
clump-clump heating and possibly dynamical friction.

 (8) For model
N, as above,  we adopt the
dynamical friction timescale from the corresponding Hernquist profile model 
H.
The velocity dispersion ratio is $\sigma_c/\sigma_h =0.5$, and again the anaytic expressions have apparently 
underestimed  $t_{c-c}$ by  about a factor of 20, and 
 $t_{merge}$ by about  a factor of 2.   The dominant
process is dynamical friction.  

(9) For model
O, again the ratio of velocity dispersion is sufficiently close to unity ($\sigma_c/\sigma_h =0.74$) that we assume there is
no clump-clump heating.  The only possible dominant physical 
process is merging of clumps,  and the analytic estimate of 
$t_{merge}$ is apparently  underestimated  by a factor of
4.  

(10) For model S, $\sigma_c/\sigma_h =0.20$.  The
dominant disruption processes are global tidal effects 
and clump-clump heating. 

 (11) For
model T, $\sigma_c/\sigma_h =0.30$.  Again we assume our
underestimate of $t_{c-c}$ is about a factor of 3.  The dominant
disruption processes are global tidal effects and clump-clump heating.  

(12) For model W, 
$\sigma_c/\sigma_h =0.43$, and the analytic 
underestimate of $t_{c-c}$ is at least  a factor of 10.  The
only dominant process is dynamical friction.  

(13) For model X, 
$\sigma_c/\sigma_h =0.79$, again sufficiently high that we 
ignore the
clump-clump heating effect.  Again we assume our underestimate of
$t_{merge}$ is about a factor of 2.  The possible dominant processes
are dynamical friction and merging.

From the above, we can see that the modification to 
our analytic expression for the clump-clump heating timescale $t_{c-c}$ 
should be  performed in a consistent way:
for $\sigma_c/\sigma_h < 0.20$, no modification is needed; 
for $\sigma_c/\sigma_h \sim 0.3$, we multiply by a factor of 3; 
for $\sigma_c/\sigma_h \sim 0.4$, we multiply by a factor of 10; 
for $\sigma_c/\sigma_h \sim 0.5$, we multiply by a a factor of 30;
for $\sigma_c/\sigma_h > 0.75$, we ignore the clump-clump heating effect (i.e.~make the timescale infinite).

From the above we can see that the merging timescale $t_{merge}$ should be multiplied by a factor of 2 for
all values of the velocity-dispersion ratio. 
From the comparison of the analytic theoretical predictions with our simulation
results, we can see that with some reasonable and consistent
modifications to our theoretical estimate  of the timescale for
clump-clump heating and merging, we can reach 
rough  consistency  with the estimates  measured from our simulations.

This allows us to develop an understanding of the 
dependence of the various processes of global tidal stripping, clump-clump heating, clump-clump
merging and dynamical friction on the number of clumps, mass fraction
of clumps and density contrast of a clump in the halo.  In  general,
a  decrease in the number of clumps and an  increase in  the clump mass
fraction can increase the velocity dispersion ratio of clump to halo,
and can further increase directly or indirectly the efficiency of all
four processes, if the ratio is much smaller than the order of unity. For larger values of this velocity dispersion ratio, 
the efficiency of the clump-clump
heating process drops quickly, while the merging process becomes more
important, and the dynamical friction becomes more important as long
as clump mass fraction is less than 50\%.  A  decrease in the density
contrast between clump and  halo can enhance the disruptive effects of 
global tides,  and of
clump-clump heating  (if the velocity dispersion ratio of clump to halo
is much less than unity).

\section{The Angular Momentum}

All the simulations have good conservation of total angular momentum
and of total energy. The total angular momentum is conserved relative
to the initial origin and to the centre of the total mass. When significant 
substructure is involved, as here, the calculation of angular momentum
can be complicated by two factors. The first is that the smooth dark
matter component and the clump component do not always share the same
centre of mass.  The second is that the centre of mass of the smooth
dark matter component does not always correspond to its highest
density point -- which is usually taken to be the centre of the
relevant system, the galaxy or the cluster of galaxies -- due to the
fact that the dark halo density profile to be simulated does not
decrease with radius fast enough, e.g. for Hernquist models, $\rho
\sim r^{-4}$, and the simulation can only allow a limited number of
particles to be distributed to infinity.  The first factor is possibly
non-trivial, since though using many clumps can make their centre of
mass more closely coincide with that of the smooth component, there do
exist physical environments, such as clusters of galaxies, in which
indeed a small number of clumps are embedded in a smooth component.
The second factor however is artificial; the few particles in the
outermost regions have a distribution that is more extended and
asymmetric, and thus have more weight in the determination of the
centre of mass. This problem could possibly be avoided by making the
density profile decrease more rapidly with radius, e.g., applying a
cutoff radius. Indeed, our simulations show that the coincidence
between the centre of mass of the dark halo and  the highest density
point is better for the Plummer profile than for the Hernquist
profile.

Table~\ref{tab:ang1} lists the different measures of
the angular momentum contents of our computed models, with the exception of the
models with a clump mass fraction $f=1$. The quantity 
$\delta_{CL}$, in Column~7, is defined as 
\beq \delta_{CL} = \frac{\sum_{k=1}^{n_c} |J^i_{zk} -
J^f_{zk}|}{\sum_{k=1}^{n_c} |J^i_{zk}|}, \eeq where $J^i_{zk}$ and
$J^f_{zk}$ are the initial and final angular momentum, calculated
relative to the highest density point of the dark matter, of the $k$th
clump. 
 It should be noted
that any direct comparison between the Plummer models and the
Hernquist models should use the halo crossing time as the time unit;
the final times here are  fixed  in these units  for
each of the Plummer models and the Hernquist models.

We can see from the entries in this Table that shifting the origin
from the centre of mass of the total system, to the highest density
point of the dark matter, can produce apparent non-conservation of the
total angular momentum; this arises since the highest density point of the dark
matter defines a non-inertial frame. 

A real physical change in angular momentum with respect to the centre
of mass of the total system occurs for each of the clumps and  the dark
matter (compare the changes with the estimated errors). 
This ensures
that the angular momentum exchange between the clump component and DM
component is real (Barnes 1988). However, from the observational point
of view, or in cosmological simulations, the angular momentum is
rather calculated relative to the highest density point of the DM. Thus
the question we want to ask is whether the angular momentum of the
clumps changes significantly when measured in this way.  
This remains non-trivial since, for example, if a clump is extremely far from
the centre, and thus it contains most of the angular momentum of the
clumpy component, no matter how efficient is the angular momentum
transfer from the remaining clumps to the dark matter, the angular
momentum change of the total clump system will be very small. Thus we
must address not only the total angular momentum change of the clumps
but also the angular momentum change of each individual clump.

The parameter $\delta_{CL}$ is useful in quantifying the mean angular
momentum change of an individual clump.  Figures~\ref{fig:clumpejvi}
and \ref{fig:clumpejco} show the binding energy and angular
momentum for each individual clump.  It should be noted that in our
simulations in which the smooth (dark matter) component has a Plummer
profile, the initial direction of the angular momentum vector of the
clumpy component is in the opposite direction to that of the DM
component. This is simply a result of the random assignment of the
initial velocity, leading to negative values of the angular momentum
for (several of) the clumps at large radius.  From the plots of models
M and P (Figure~\ref{fig:clumpejvi}
and Figure~\ref{fig:clumpejco} respectively) we can see this
fluctuation leads to a lower initial specific angular momentum than in
the Hernquist models. Thus the total angular momentum of the clumps
can be changed drastically compared to its initial value, even if the
change of angular momentum of an individual clump is slight. Thus we
conclude that $\delta_{CL}$ is a more robust parameter to measure
changes in angular momentum than is the percentage change of the total
angular momentum of the clumps, especially for the Plummer models.

From the entries in Table~\ref{tab:ang1} we can see that
$\delta_{CL}$ for all the collapse models is much larger than most of
the virialized models.  By checking individual clumps, as illustrated
in Figures~\ref{fig:clumpejvi} and \ref{fig:clumpejco}, we find
that clumps close to the centre have significant angular momentum loss, 
while the clumps at large distance can  have significant angular
momentum change --  gains and losses -- in collapse cases.

For the virialized models A, S, B,
G, W and T, $\delta_{CL}$ is small,
while for the remaining virialized models $\delta_{CL}$ is large. We list the dynamical
friction time and the time for half of the mass in clumps to be
disrupted, in units of $t_{\frac{1}{2}hc}$, for these models, in
increasing order of the value of $\delta_{CL}$, in
Table~\ref{tab:smallangchange}. We can see that there are two factors
that determine the angular momentum change $\delta_{CL}$: the
dynamical friction time and the disruption time. Generally, long 
dynamical friction times and short  disruption times produce  small
values of $\delta_{CL}$. As discussed in section 3.1.4, dynamical
friction acts to drive  clumps to the centre, with loss of angular
momentum of the clumps. On the other hand, efficient disruption of a
clump, especially due to processes unrelated to dynamical friction
such as clump-clump heating or global tidal effects, helps to decrease the
mass of the clump, which slows down the dynamical friction process,
and thus can decrease the angular momentum loss. In the extreme case,
a clump will no longer suffer dynamical friction after being disrupted
completely. Thus the angular momentum change is most significant for
those massive clumps moving slowly, close to the centre, that are
dense enough to withstand disruption by tides.

For the collapse models, the large value of $\delta_{CL}$
is caused by not only dynamical friction, but also by dynamical mixing
during the post-collapse relaxation, which can redistribute both the
binding energy and the angular momentum (Quinn \& Zurek 1988).  We 
plot the Lindblad diagram for
the smooth DM component for models K and D in
Figure~\ref{fig:lindblad}. The most
bound $90\%$ of the DM particles are divided into nine equal size
bins, sorted by the binding energy of each particle at the end of the 
simulation.  We plot the vector angular momentum $J_z=|\Sigma j_z|$, 
and the scalar angular momentum $J_s=\Sigma |j|$, for each bin. We can see
that for all the collapse models, regardless of the clump mass
fraction or number of clumps, they show the same behaviour, in that
inner, more-bound material loses angular momentum $J_z$, while outer,
less-bound material gains angular momentum $J_z$. The scalar angular
momentum $J_s$ increases for each bin. This result is consistent with
previous results for the redistribution of angular momentum in
protogalaxies (Quinn \& Zurek 1988).  The clumps close to the centre
are disrupted quickly during the collapse, but still  suffer significant 
angular momentum loss, both while still bound in clumps and after 
disruption. This material becomes more bound to the halo. 
For those clumps that survive
the collapse,  subsequent disruption and dynamical friction effects
occur, as discussed above for the case of an initially virialized system.
In addition, the final dark matter distribution is triaxial (see
section 6.2 below) and exerts a torque that helps transfer angular
momentum between clumps and the dark matter, and between the inner and outer
regions (resulting in the angular momentum loss from the inner debris).

In the context of galaxy formation, the quantity of interest is the
total angular momentum change, over a Hubble time, of the baryonic
component (here given by the clumps).  However as can be seen from
Table~\ref{tab:ang1}, the total angular momentum $J_z$ of all the clumps does
not change as much as would be expected from the parameter
$\delta_{CL}$, except for model Y and all the Plummer models. For the
Plummer models, as we discussed above, the large change in the total
angular momentum of the clumpy component is due to the particular
random choice of the initial (low) specific angular momentum of this
component. As an upper limit, we may adopt the parameter $\delta_{CL}$
as the estimate of the angular momentum change for the clump system.
Our simulations show that the total angular momentum of the clumpy
component does not change significantly in a Hubble time for a small
clump mass fraction and a large number of clumps, but does change
significantly (quantified by $\delta_{CL}$ in Table~3 being greater
than $\sim 30\%$) in models with few and massive clumps; compare model
G with model H, or model S with model T, or model B with model X.
Such a high amplitude of angular momentum loss from proto-disk
material cannot allow a spiral to form that matches observation.  Thus
we conclude that spiral galaxies {\it can\/} be formed through the
accretion of many, but small, clumps, that conserve angular momentum
in the process, but cannot be formed in an environment that contains
only a few, but large, clumps (cf. Silk \& Wyse 1993).

\section{The Kinematics}

In all of the collapse models the kinematic properties for the dark
halo component are similar, regardless of the different parameter
values. From Figure~\ref{fig:dmvre10coc20d3f04}, we can see that after
collapse the radial velocity dispersion $\sigma_r$ increases at all
radii, but the increase is larger in the central regions than at
large radius. The other components of the velocity dispersion tensor,
$\sigma_\phi$ and $\sigma_\theta$, increase less than does $\sigma_r$,
with $\sigma_\phi$ slightly larger than $\sigma_\theta$, and they do
not increase significantly at large radius. Thus the final ratios
$\sigma_\theta/\sigma_r$ and $\sigma_\phi/\sigma_r$ decrease with
increasing radius.  However, at the centre, the velocity ellipsoid
approaches isotropy, i.e. $\sigma_\theta \sim \sigma_\phi \sim
\sigma_r$.  The rotation velocity, $V_{rot}$, increases, while
$V_{rot}/\sigma_r$ decreases.  $V_{rot}/\sigma_\phi$ decreases during the collapse in the 
central regions but is fairly constant in time  at large radius.

The Lindblad diagram for the dark halo component is shown in
Figure~\ref{fig:lindblad}, and shows that for the most bound energy
bin the scalar angular momentum $J_s=\Sigma |j_z|$ increases, despite
the fact that the amplitude of the vector angular momentum
$J_z=|\Sigma j_z|$ decreases.  The different behaviours of the scalar
and the vector angular momenta for the most bound energy bin can be
explained by noting that individual particles can gain angular
momentum in addition to kinetic energy, but can be deflected
isotropically in direction during the post-collapse relaxation
process, thus the vector angular momentum of all the particles in that
energy bin decreases.  As for the virialized models, there are no
significant changes for the dark halo component.
 
The kinematic properties of the clump debris for each model are
somewhat different, but regardless of the parameter values, the
velocity ellipsoid properties are similar to those of the dark halo
component, especially in the central regions.
Figure~\ref{fig:dedmvre10coc20d3f04} shows the same quantities for the
debris as does Figure~\ref{fig:dmvre10coc20d3f04} for its dark
halo. In the outer regions, the debris is largely still composed of
many streams formed from the disrupted clumps.  To illustrate,
Figure~\ref{fig:phase} plots $v_x$ and $v_y$ at the final time for the
clump particles within a box of diameter of $\sim 3,$ located at the
coordinate (4,4,0) (beyond the half-mass radius) for model U. The
different symbols indicate membership of different initial individual
clumps; in this example the box contains several streams, which are
composed of eight different disrupted clumps.
 
\clearpage
\section{The Morphology}
\subsection{The Density Profile}          
The density profiles of the disrupted clumps for all the collapse
simulations follow the profiles of the dark matter, the surface
density profile of which can be best fit by the $R^{1/4}$ law.  This
result is consistent with the early studies of van Albada (1982),
which demonstrated that no matter how clumpy the pre-collapse initial
condition is, the final profile is consistent with the de Vaucouleurs
$R^{1/4}$ law, provided the collapse factor is large enough (the
virial ratio $2T/W \sim 0.1$).  However, in the virialized
simulations, the density profile of the debris can be different from
that of the dark matter.  Figures~\ref{fig:rhodensity} and 
\ref{fig:sigdensity} show the density profile and surface density
profile for the debris (normalized to 
their values at the debris half-mass radius).  The solid lines
correspond to the collapse cases, while the dashed lines correspond to
the virialized cases. The density profile for model H shows a peak at
some distance from the centre, due to the presence of a merger remnant
there.  We can see that, once normalized, the profiles of the debris in
all the models are very similar. At the central regions, the density
profile of the virialized models is slightly shallower than that of
the collapse models.  At the normalized distance $r=0.1 $, i.e. at
one-tenth of the half-mass radius, $\rho \sim r^{-2}$ for collapse
models while $\rho \sim r^{-1.7}$ for virialized models.  At the very
centre, due to the different resolutions for our different models, we
can only infer that the central density profile is cuspy.

The models with the Plummer profile for the dark halo behave somewhat
differently.  Figure~\ref{fig:plumdendedm} shows the final density
profiles of the debris and of the dark matter for models M and
N. These demonstrate that the central density of the debris can be
larger than that of the dark halo component, and the central density
profile of the debris is cuspy and does not have a (approximately) 
constant density
core, in contrast to the dark halo profile. The initial and final
density profiles of each component for models M and N are shown in
Figure~\ref{fig:plumdenall}.  Since the distribution of the clumpy
components initially follows the same Plummer profile as does the dark
matter, the initial density profile including all components is
approximately a Plummer profile. The final density profile, including
all components, shows a cuspy structure at the central region, for
example at $r \sim 0.1$, model M has central profile $\rho \sim
r^{-1.1}$ and model N has central profile $\rho \sim r^{-1.7}$.  A
larger clump mass fraction (model N) leads to a steeper central
density profile and a higher normalization, compared to the results
with a lower clump mass fraction (model M).  This is because in the
former case dense and massive clumps can move to the centre through 
dynamical friction without being significantly disrupted, and thus can
contribute to the mass and density profile at the centre.  This result
has the application, in CDM cosmology, that the merging into a lower
density parent halo of a number of higher density subhalos can
produce a final cuspy and dense halo.  Furthermore, how cuspy and
dense the final halo is depends mainly on the number of subhalos in
the parent halo and the density contrast of a subhalo to the parent halo,
both of which are related to the density fluctuation 
power spectrum index, $n$, for $P(k)
\sim k^n$. This provides an explanation for  the central cuspy density
profiles of halos found in cosmological numerical simulations, as  the
consequence of the self-limiting merging/accretion of small, dense halos, as
suggested by Syer and White (1998).

It is well-known that the surface density profile of a cD galaxy in
its outer regions is systematically above the $R^{1/4}$ law that fits
the more central parts.  Recent N-body simulations of groups of 50
galaxies studied by Garijo, Athanassoula \& Garc\'{\i}a-G\'omez (1997)
show that a giant central galaxy can be formed if there is a initial
seed galaxy at the centre. Further, these authors find that their
virialized models produce a giant central galaxy with such a cD
envelope.  Previous studies of the formation of cD galaxies, using
non-cosmological N-body simulations of isolated clusters of galaxies,
(Richstone \& Malumuth 1983; Bode {\it et al.}  1994) concentrated on
identifying the relative roles of stripping, merging and collapse.
Can a cD galaxy be formed by the galaxies that were disrupted at the
cluster centre or by galaxies that merged with the central galaxy?
Does the formation of a cD galaxy happen before, or after, the
collapse of the cluster?  Dubinski (1998) explored the formation of
brightest cluster galaxies through the merging of several massive
galaxies, in the context of hierarchical clustering cosmologies, but
still failed to create the characteristic extended envelope of cD
galaxies. One might expect that a larger clump mass fraction and a
small number of clumps can lead to efficient merging or disruption of
galaxies, which are plausibly important processes to form cD galaxies.
Though our simulations do only explore a limited parameter space, they
include a simulation with only five clumps with a mass fraction of
40\%, and also one with only twenty 
clumps with a mass fraction of unity.  Neither formed a cD. 
All our giant central galaxies can be fit by an $R^{1/4}$ law
very well, over a range of more than 4 orders of magnitude in surface
density. We speculate that collapse and small-scale structure alone
cannot produce cD galaxies.  It is plausible that the continuous later
infall of large galaxies could  provide a mechanism to form cD
galaxies.

\subsection{Triaxiality}
CDM-dominated cosmological N-body simulations show the protohalos are
usually triaxial (Warren {\it et al.} 1992; Dubinski \& Carlberg 1991)
with the tendency to be prolate, and with the mean ratio of minor to
major axes, $c/a$, approximately 0.5 at small radius.  The study of
Tremblay \& Merritt (1995; Merritt \& Tremblay 1996) of the intrinsic shapes of elliptical
galaxies found that bright ellipticals ($M_B < -20$) are
systematically rounder than faint ellipticals, with the ratio of minor
to major axes, $c/a$, peaked at 0.75 for bright ellipticals, while for
faint ellipticals $c/a$ is peaked at 0.65. Small ellipticals also tend
to be flattened by rotation (Davies {\it et al.}~1983; Rix, Carollo \&
Freeman 1999) rather than by anisotropic stellar velocity dispersions.
Studies of the collapse of a proto-galaxy with some initial
triaxiality have explored the possibility that angular momentum can be
transferred from the inner halo to the outer parts (Subramanian 1988;
Curir, Diaferio \& de Felice 1993; Curir \& Diaferio 1994). Our
simulations study the evolution of a system of clumps embedded in a
smooth dark halo with an initial oblate profile, and thus we do not
address the question of the role of any initial triaxiality, but
rather investigate what parameters determine the triaxiality of the
final system.

We first investigate the level of flattening of various components
that can be due to rotation, by comparing the ellipticity with the
ratio of rotation to velocity dispersion.  The ellipticity,
$\epsilon$, is calculated from the projected isodensity contour of the
simulation particles from sixteen different viewing angles.  The
central line-of-sight velocity dispersion, $\sigma_0$, and rotational
streaming velocity, $v_0$, are also calculated for these viewing
angles, and the ratio $v_0/\sigma_0$ formed.  In
Figure~\ref{fig:contourvrvde} we plot the location of each viewing
angle on the ($v_0/\sigma_0$,$\epsilon$) plane for models
J, G, U and K.  Since the
morphology of the debris generally follows that of dark halo, here we
just show the dark halo measured in this way. From
Figure~\ref{fig:contourvrvde} we can see for the collapse cases the
value $v_0/\sigma_0$ is decreased significantly and the system becomes
anisotropic. For model U, with low density contrast between the clumps
and the halo, the ellipticity increases, while for model K, with
large clump mass fraction, the ellipticity decreases.  The virialized
cases are actually initially marginally  out of equilbrium, and generally
evolve such that the ellipticity and $v_0/\sigma_0$ are
decreased, to make the system rounder and more isotropic (remember
these simulations are constructed to be initially close to isotropic
and flattened by rotation).

The three-dimensional shapes can be calculated from the moments of
inertia, giving the triaxiality for both dark halo and debris.  As we
discussed in section 2.4, this technique for calculating the axial
ratio systematically underestimates the flattening. For example, the
initial flattening of the dark halo in all simulations is $c/a \sim
0.53$, while the calculation from the moment of inertia gives $c/a
\sim 0.63$. However, since we are interested more in {\it trends\/} in
the dependence of final triaxiality on the initial conditions, such a
systematic overestimate is not a problem.  Table~\ref{tab:triaxial}
summarizes the values of $b/a$ and $c/a$ in the simulations, with as
usual $a > b > c$ where a, b and c are calculated from the eigenvalues
of the moment of inertia tensor within the half mass radius as
described in section 2.4.

 From Table~\ref{tab:triaxial}, and Fig.~\ref{shapes} 
we can see that:

(a) all the collapse models show some level of triaxiality for both
the dark halo component and the debris component.  Cases with a large
clump mass fraction ($f =0.4$) are slightly rounder than the cases
with small clump mass fraction ($f=0.1$). For the completely clumpy
cases ($f=1$) the final shape of the debris is very spherical, with
the exception of the collapse model with a initial Plummer profile
(model R). The triaxiality for the low density contrast cases, such as
models U and V, is more significant than that of high density contrast
cases, such as models J and K.  The triaxiality for models with a
small number of clumps, such as models Y and Z, is less significant
than that for those with a large number of clumps, such as models J
and K, or D and E. Also we should note that, for our choice of initial
configuration, the clump mass fraction is weakly correlated with the
value of the virial ratio, $2T/W$, for the collapse models. Actually
the real collapse factor for all the collapse models is almost a
constant. For the pure clump models with $f=1$, calculating a
modified virial ratio (denoted $2T'/W'$) by excluding the clump
internal kinetic energy and potential energy, we have
$2T'/W'=0.13$. So, for our models, it is the level of clumpiness,
rather than the collapse factor, that determines the triaxiality of
the final systems.  Large values of the clump density contrast, large
clump mass fractions, and small numbers of (massive) clumps all indicate that
clumpiness is important; a higher level of `clumpiness' makes the
final system rounder. Clearly, the collapse factor will also affect
the triaxiality of the final system.

(b) for the virialized cases, the final shape of the dark matter
component is oblate with $c/a > 0.8$, slightly rounder than its
initial configuration; the shape of the debris component is close to
oblate, with $b/a >0.9$ and $c/a >0.7$.  

 As we can see from the virialized cases, which is an extreme case of
approximately zero collapse factor, the final system tends to be
oblate or round. This result implies that the generation of
triaxiality requires thin box orbits, or radial orbits, that get very
close to the centre. In this case the collapse is mostly radial and
creates a lot of triaxiality; more clumpiness disrupts these radial
orbits.  This suggests that giant ellipticals, which tend to be
rounder than small ellipticals, may be formed in a more clumpy
environment than are small ellipticals, but this inference is
complicated by the fact that the small ellipticals tend to be oblate,
with significant dissipation being implicated (e.g. Wyse \& Jones
1984; Rix {\it et al.}~1999).
                        
\subsection{The Substructure in the Outer Halo}
The existence of a population of intracluster stars in the core region
of the Virgo cluster, detected through planetary nebulae (M\'endez
{\it et al.} 1997) and red giant stars (Ferguson, Tanvir \& von Hippel
1998) indicates that the stripping of stars from cluster galaxies,
through the multiple disruption processes that  we discussed above, is a
reality during the formation of clusters of galaxies.  In addition,
M87, the giant elliptical in the Virgo cluster, exhibits a broad,
diffuse and extended `fan' in surface brightness at a projected
distance of around 100~kpc from its centre.  The accretion of a small
stellar satellite with $\sim 0.1\%$ of the mass of M87, with small
impact parameter, can allow the formation of a surface brightness
feature like this (Weil, Bland-Hawthorn \& Malin 1997).  Even in our
Galaxy, large overdensities of A-stars (Yanny {\it et al.}
2000) and RR Lyrae stars (Ivezic et al.~2000) are observed in the
halo at distance of up to 60 kpc from the Galactic centre, through the
analysis of Sloan Digital Sky Survey imaging data (Yanny {\it et al.}
2000). This structure  may be tidal debris from the Sagittarius dwarf
spheroidal galaxy (Ibata {\it  et al.} 2001).  The observation of
substructure in the outer halo can be useful in constraining the
formation history of galaxies.  From Figures 21 and \ref{fig:substructure} we
can see that in the outer halo region of our simulations there are
indeed prominent streams  from the disrupted clumps, seen
in surface density. These unrelaxed streams can account for the
over-density above the $R^{1/4}$ law seen at the very outer regions in
Figure~\ref{fig:sigdensity}.

A detailed study of the substructure is a mathematical challenge
needing a serious treatment, but as a first step we adopt a 
simple approach to explore its presence and evolution.  To
quantify the lumpiness of substructure as it might be observed, the
concept of the filling factor of disrupted clumps - the fraction of sky
containing one or more streams - was introduced by Tremaine (1993) and
used by many authors subsequently in studying the evolution of tidally 
disrupted stellar systems  (Johnston 1998; Helmi \& White 1999). The
shortcoming of the filling factor as a measure of streaming is that it
can become significantly larger than unity as streams wrap around the
sky.

Here we introduce a different measure of inhomogenities in the debris,
restricting our analysis to the spatial part of phase space (the
complications in velocity space are illustrated in Figure~\ref{fig:phase}).  We
define the quantity $\eta$, defined as the fraction of the debris
which is in regions with overdensity above some critical value, or
$\delta n/\overline{n} > \delta_n$, restricting the calculation to
regions beyond a critical radius $r_{cr}$ projected onto the celestial
sphere.  At a given time $t$, the critical radius $r_{cr}$ is chosen
to be the radius where $t_c(r_c)=c_tt=2\pi r/v_c(r)$, where $v_c(r)$
is the circular velocity at radius $r$, and $c_t$ is a free parameter
of the order of unity.  This critical radius can be regarded as the
division between the inner, smooth debris where mixing is complete,
and the outer regions where mixing is incomplete.  Though the
assumption is over-simplified, a visual check of
Figure~\ref{fig:cparameter} justifies a choice of $c_t \sim 2$. For
collapse cases, we have approximately $r_{cr}=0.7,2.3,4.3$
respectively for $t=14, 56, 126$, while for virialized cases
$r_{cr}=0.6,2.1,4.1$,  for times $t=14, 56, 126 $.  

It should be noted that we do not have any `baryons' that were not
initially in clumps, so that all the unbound clump particles are
`debris'; in the interpretation of the evolution of substructure in 
a galaxy like the Milky Way, the entire field halo would be debris,
and we are quantifying the structure in that debris.

We simply
divide the sky into a number of cells of equal solid angle. The
average number of debris particles in each cell is $\overline{n}$. If
the cell has more than $(1+\delta_n) \overline{n}$ debris particles, then 
it is counted as a region of overdensity.  If $\overline{n} \sim 16$,
the overdensity of $\delta_n =0.75$ can be used to indicate the
existence of streaming structure in that cell, since the density
fluctuation is then greater than 3$\sigma$, with 
$\sigma = \sqrt{\overline{n}}$. Thus an accurate calculation of
$\eta$ requires that both the number of cells and the number of debris
particles be large enough. However, this condition is not satisfied in 
many of our simulations. Thus here we can only give a rough estimate
of $\eta$ for the simulations with large clump mass fraction, using $\delta_n
=0.75$ and $c_t=2$.  

Table~\ref{tab:eta} summarizes the values of
$\eta$ at different times for our simulations, with the conclusion
that for our models $\eta$ is more or less a constant, with a value
varying only from about $10\%$ to $20\%$.  By checking the number of
distinct overdense regions on the celestial sphere used in calculating
$\eta$, and also by a visual check of the snapshots of the models
for 5, 20, and 80 clumps, we can see that, not surprisingly, 
the number of streams
correlates with the number of clumps. The disruption of clumps
is more efficient for collapse cases, resulting in  streams that are
more radially aligned compared to those formed in our virialized models.  
The observations of streams in phase space in the outer
halos of galaxies are thus potentially 
important in constraining the formation history
of galaxies, especially in determining the abundance of accreted substructure 
of different sizes and densities, through comparison with simulations. 

\section{Summary}

In this paper, we use N-body simulations to study the evolution of
clumps, or substructure, embedded in non-dissipative, isolated dark halos. We
ran 26 different combinations of 4 independent free parameters, namely
the clump mass fraction, the number of clumps, the density contrast
of clump to halo and the initial virial ratio.  We also investigated
two different density profiles for the smooth dark halo component, the
cuspy Hernquist profile and the (approximately) constant-density core Plummer
profile.  For the initial virialized systems, we derived analytic
estimates for the importances of various dynamical processes that
occured in our simulations, such as global tidal stripping,
clump-clump heating, clump-clump merging and dynamical friction,
including explicit dependence on the 4 independent free parameters. 
These analytic estimates were used to gain physical insight, and a  
comparison between our numerical results and our analytical estimate
showed  general agreement. 

With an increase in the mass fraction of clumps and a decrease in the
number of clumps, the velocity dispersion ratio of a clump to the halo
increases, and all the dynamical processes become more efficient in a
general sense, provided this velocity dispersion ratio is still much
less than one. As this velocity dispersion ratio increases close to
unity, the heating effect by close encounters between clumps becomes
less important, while other dynamical processes continue to become
more important.

We find that due to the angular momentum transport seen, spiral
galaxies cannot be formed in a system with a large clump mass fraction
and a small number of clumps.  Our simulations show that the angular
momentum of the clumpy component does not change significantly in a
Hubble time for a small mass fraction in the clumps and a large number
of clumps. This result confirms the standard formation scenario for
disk galaxies, in that the merging history of spirals is restricted to
the accretion of many, small systems, still conserving their specific
angular momentum.  This is similar in some aspects to the model of
Vitvitska et al.~(2001), who find that one may also end up with
ellipticals with rapidly spinning halos, especially if only a small
number of massive clumps are involved in the final merging events.

The final density profile, starting with clumps embedded in a smooth
Plummer component, shows a rather steep inner part, depending on the
clump mass fraction.  Thus our experiments support the explanation for
the origin of the central cusps in the density profiles of halos in
cosmological numerical simulation as the consequence of self-limiting
merging of small dense halos, as suggested by Syer and White (1998).
The surface density profile of the debris formed from the disrupted
clumps can be best fit by the de Vaucouleurs law for all our
simulations.  Thus our simulations do not produce cD galaxies, a
deficiency which suggests that the formation of cD galaxies reauires
conditions outside the parameter ranges covered by our
simulations. Plausibly the formation of the outer envelope
characteristic of cDs requires some later infall of large galaxies
into the central regions of the cluster, possibly enhanced by
dynamical friction.
 
The iso-density contours of both the dark halo and debris are
consistently triaxial after the collapse, even though the initial
configurations are oblate. The triaxiality of the final system is
found to depend on the clumpiness of the initial system.  A high level
of clumpiness, such as having only a few, but massive, clumps, helps
to produce a round system.  This further suggests that giant
ellipticals, which tend to be rounder than small ellipticals, may be
formed in a more clumpy environment than small ellipticals.

We introduced a new measure of `streams' in the final system, and
found that persistent signatures of disrupted clumps in the outer halo
can, at an overdensity of $>0.75$, cover 10\% to 20\% of the sky.  The
properties of these streams depend on the initial conditions in an
intuitive way, in that the number of streams correlates with the
number of initial clumps, and their orientation is more radial in
collapse simulations.

Some extensions of our work that would further our understanding of
the role of substructure in the formation of galaxies and of cluster
of galaxies include N-body simulations with (1) a mass spectrum of
clumps rather than identical clumps; (2) less restricted coverage of
parameter space, in particular to larger collapse factors and (3)
explicit incorporation of a merging criterion into the calculation at
each time step.

\acknowledgements

We acknowledge support from NSF grant AST-9804706 (RFGW) and the PPARC
visitor programme. RFGW thanks all in the Astronomy group at St
Andrews for providing a pleasant and stimulating environment.

\clearpage 

\appendix
\section{Parameters and Units of the Simulations}

 The density profile, mass profile, corresponding potential,   
and the total potential
 energy of a Hernquist sphere are:
\begin{eqnarray}
\rho (r) &=& \frac{M}{2\pi} \frac{a}{r(a+r)^3} ,\\
M(r) &=& \frac{M r^2}{(r+a)^2},\\
\Psi (r) &=& - \frac{G M}{a+r},\\
\Phi     &=& - \frac{G M^2}{6a},
 \end{eqnarray} 
where $a$ is the core radius. 

The density profile, mass profile, corresponding  potential  and the 
total potential energy of a Plummer sphere are: 
\begin{eqnarray}
\rho (r) &=& \frac{3M}{4\pi a^3} \left( 1+\frac{r^2}{a^2}\right)^{-5/2} ,\\
M(r) &=& \frac{M r^3}{(r^2+a^2)^{3/2}},\\
\Psi (r) &=& - \frac{G M}{\sqrt{r^2+a^2}},\\
\Phi     &=& -\frac{3\pi}{32} \frac{G M^2}{a},
 \end{eqnarray} 
 where $a$ is the core radius.

The
half mass radius for the Hernquist profile is $r_{\frac{1}{2}} = \left(
1+\sqrt{2} \right) a, $ while for the Plummer profile, $r_{\frac{1}{2}}
\approx 1.3 a $.  The mean crossing time of the dark halo, given by 
the ratio of the half mass radius, $r_{\frac{1}{2} h}$, to the
circular velocity at the half mass radius, $v_{\frac{1}{2} h}$, is,
for the Hernquist dark halo, $t_{\frac{1}{2} h c} =r_{\frac{1}{2}
h}/v_{\frac{1}{2} h} = \sqrt{2} \left( 1+\sqrt{2} \right)^{1.5}=5.3$,
while for the Plummer dark halo $t_{\frac{1}{2} h c}= 2.1$.  The
Hernquist models are allowed to run for about $t_{sim} \sim 126$,
i.e. approximately 23 times the mean halo crossing time while for
Plummer models, the simulation is stopped after $t_{sim} \sim 56$,
i.e.  approximately 27 times the mean halo crossing time. For a
typical dark halo of a galaxy, the mean crossing time is about 0.5
Gyr, for $v_{\frac{1}{2} h} \sim 200 kms^{-1}$ and $r_{\frac{1}{2} h}
\sim 100kpc$. Thus a typical simulation duration of 25 mean crossing
times is approximately a Hubble time.  The timescale for a dark halo
typical of a cluster of galaxies is approximately the same.

There are some simple and convenient relations among the mean
quantities of a given mass profile. For many reasonable spherical mass
profiles, including both the Plummer and Hernquist spheres, there is an empirical relation between the half mass radius
$r_{\frac{1}{2}}$ and the virial radius $r_{vi}$, $r_{\frac{1}{2}}
\approx 0.8 r_{vi}$, where the virial radius is defined by the
potential energy $|\Phi| \equiv \frac{GM^2}{2 r_{vi}}$ (Spitzer 1987).
In the case of an isotropic velocity dispersion tensor, we can use the
mean one-dimensional velocity dispersion $\sigma$, defined by $
|\frac{\Phi}{2}| \equiv \frac{3}{2} M \sigma^2$, instead of the
circular velocity at the half mass radius radius $v_{\frac{1}{2}}$.
These are related by 
$v_{\frac{1}{2}}
=\sqrt{\frac{6}{1.6}} \sigma \approx 2 \sigma $.  

\section{Numerical Effects}

We first  discuss the effects of two-body relaxation,
for which the two bodies can be two particles both from a single
clump, or two particles with one from a clump and the other from the
dark halo, or two particles both from the dark halo. 
Then we 
discuss the heating effect of a clump by dark halo particles.
 
\subsection{Two-body Relaxation}
First we just consider the case that the clump is isolated, i.e. the
two bodies under consideration  are both from a
single clump. We know the evaporation timescale is usually 100 times
the relaxation timescale (Binney \& Tremaine 1987) for a given
Maxwellian system; this sets the timescale over which a clump could 
suffer evaporative mass loss  if the two-body
relaxation time is too short compared to the simulation time. The
two-body relaxation time in a single clump is (Sellwood 1987) 
$$t_{relax} \sim \frac{0.14 N_{cl}}{\ln \left( \frac{r_c}{3r_s}
\right)} t_{\frac{1}{2} c c,}$$ where the $N_{cl}$ is the number of
particles in a single clump, $r_s$ is the softening-length and
$t_{\frac{1}{2} c c}$ is the internal crossing time within a clump.
The softening-length in the treecode is set at $r_s=0.5 r_{\frac{1}{2}
cl} N^{-1/3}_{cl}$; this relatively large value is chosen to suppress
the unphysical two-body relaxation (White 1978), which otherwise could
occur due to the small number of particles used in each clump (ranging
from 150 to 2400).

In
terms of the halo crossing time, this two-body relaxation time can be written
as $$t_{relax} \sim \frac{0.14 N_{cl}}{\ln \left(
\frac{r_c}{3r_s}\right)} \frac{t_{\frac{1}{2} h
c}}{\sqrt{\rho_{cl}/\rho_0}}.$$ 
Thus the two-body relaxation time in a
single clump is shorter, in terms of the halo crossing time, for larger values of the 
density contrast $\rho_{cl}/\rho_0$. This poses a limitation for us in 
exploring the parameters for clumps of  large density contrast, in
addition to the limitation from the CPU time used, which is determined
by the total number of particles.  For our computed models with
$N_{cl}=300$ and $\rho_{cl}/\rho_0=64$, we have $t_{relax} \sim 3.5
t_{\frac{1}{2} h c}$.  In the worst case, with $N_{cl}=150$ and $
\rho_{cl}/\rho_0=64$, we have $t_{relax} \sim 2 t_{\frac{1}{2} h
c}$. Since our simulation time is about $25 t_{\frac{1}{2} h c}$,
generally our clumps should suffer somewhat from two-body relaxation
effects, which can lead to some internal manifestation of collisional
behaviour such as mass segregation (to which
we are insensitive since we have adopted equal mass particles). 
  However, we do not expect there to
be significant mass loss due to evaporation effects, not only because
the estimated evaporation time scale is very long compared to our
simulation time, but also due to the fact that adoption of a 
softening-length can inhibit evaporation.  Furthermore, we are not
interested, in this study, in the evolution of the internal structure of
the clump, and thus our chosen parameter values are acceptable.  

Next, we
turn to consideration of two-body relaxation in the dark halo,
i.e. the two bodies participating in the two-body relaxation are both
from the dark halo. Similarly to the expressions above, 
we have $$t_{relax,h} \sim \frac{0.14
N_{h}}{\ln \left( \frac{r_h}{3 r_s} \right)} t_{\frac{1}{2} h c}$$ 
where $N_h$ is the number of particles in the dark halo. 
Typically $t_{relax,h} \sim 700 t_{\frac{1}{2} h c}$. Thus the two-body relaxation
effect in the halo can be ignored.

In addition to the possible evaporation of clump particles due to two-body
heating between two particles within the same clump, there is another
heating process we should consider, caused by two-body heating between a halo particle and a 
clump particle (Carlberg 1994; van Kampen 1995).  The disruption time
due to this heating process is $$t_{hp-cp} \approx \frac{N_{cl}}{4\ln
N_{h}} \frac{r_{\frac{1}{2}h}}{r_{\frac{1}{2}c}} t_{\frac{1}{2} h c}. $$ 
For our simulations,  this disruption time is much longer than
the total integration time; for a typical model, $t_{hp-cp} \sim 130
t_{\frac{1}{2} h c}$ and  we can safely ignore this heating effect too.

\subsection{Particle-Clump Heating Effect}

Numerical limitations resulting from a small number of particles can
lead not only to unwelcome collisional effects as discussed above, but
also can lead to spurious heating effects within clumps.  While in our
simulations we have both halo particles and clump particles of the
same mass, we shall consider the more general case of unequal particles, and indeed of a heavier
halo particle.  As a (heavy) dark halo particle passes by or through a
clump, it can tidally heat the clump, and the clump can eventually be
disrupted (Carlberg 1994).  One can roughly estimate the magnitude of
this effect by using the impulse approximation, provided the duration
of the encounter is short compared to the internal crossing time of
the clump (Binney \& Tremaine 1987).  The binding energy of the clump
is given by $E_b =|\Phi/2|=\frac{Gm^2_c}{4 r_{c vi}}$. For a given
clump, the heating rate is \beq \dot E = \frac{4 \sqrt{\pi} G^2 m^2_p
n_p m_c \overline{ r^2_c}}{3 \sigma_h b^2_{min}}, \eeq where $m_p$ is
the mass of one dark halo particle, $n_p$ is the mean number density
of the dark halo particles within the half mass radius of the dark
halo, $m_c$ is the mass of the clump, $ \overline{r^2_c}$ is the
mean-square radius of the clump, $\sigma_h$ is the one-dimensional
velocity dispersion of the dark halo and $b_{min}$ is the minimum
impact parameter. The size of the single dark halo particle can be set
to the value of the softening length $r_s$. Since $r_s \ll
r_{\frac{1}{2}c}$, we can choose $b_{min} \sim r_{\frac{1}{2}c}$ to
include head-on encounters, thus modifying the above equation by
multiplying a factor of $g \sim 3$ (Binney \& Tremaine 1987).  The
appropriate value of $\overline{ r^2_c}$ is somewhat uncertain, since,
for example, for the Plummer profile $\overline{r^2_c}$ is divergent.
The heating rate however remains finite. This result can be understood
in the following way: the heating by such a close encounter causes the
particles in the clump to gain a velocity increment proportional to
their projected distance from the centre of the clump, and thus while
the inner particles are essentially unaffected, the outermost
particles, though few in number, can gain most of the kinetic energy,
resulting in their removal from the clump. Thus in the calculation of
$\overline{r^2_c}$, we should ignore those particles lying too far
from the clump centre. Here we adopt $\overline{r^2_c}$ to be $2
r^2_{\frac{1}{2} c}$.  After some algebraic manipulations using the
relations in section~2.2 and Appendix~A, we have that the 
clump disruption timescale,
due to halo particle-clump heating, is \beq t_{p-c} = 0.03 N_{tot}
f^{2/3} n^{-2/3}_c \left( \rho_{cl}/\rho_0 \right)^{1/3}
t_{\frac{1}{2} hc}, \eeq where $N_{tot}$ is the total number of
particles used in the simulation. For the worst case (model S) with
$f=0.1$, $\rho_{cl}/\rho_0 =2.7$, $n_c=20$ and $N_{tot} \sim 60000$,
we have $t_{p-c} \sim 73 t_{\frac{1}{2} hc} $. So again this
particle-clump heating effect can be ignored in our simulations. As
discussed in the text (section~3.1.3), the physical clump-clump
heating effect is much more important.
 
\section{Dynamical Friction}

We estimate the expected amplitude of dynamical friction under the
conditions we simulated by adopting, for Hernquist profile halos, the
phase space distribution function he derived, which neglects angular
momentum.  This is 
\beq f(E)=\frac{(1-f)M_{tot}}{8\sqrt{2} \pi^3 a^3
v^3_g m_p} K_0(q), \eeq where \beq K_0(q)= \frac{1}{\left( 1-q^2
\right)^{5/2}} \times \left[3 \sin^{-1}q + q
\left(1-q^2\right)^{1/2}\left(1-2q^2\right)\left(8q^4-8q^2-3\right)\right],
\eeq 
where $a$ is the core radius of the halo, $v_g=\sqrt{GM_{tot}/a}$,
$q=\sqrt{-E}/v_g$, and $E=\frac{1}{2} v^2 + \Psi (r)$ is the specific
energy of a halo particle.  Define $x \equiv r/a$. Then we have \beq
v^2=\frac{2}{x+1} v^2_g- 2v^2_g q^2.  \eeq Since $f(E)$ is
non-negative for all $v< v_e$, where $v_e=\sqrt{-2\Psi(r)}$ is the
escape velocity, and if the clump is moving with velocity $v_M$ less
than the escape velocity at radius r, we can simply substitute for
$f(E)$ in equation (10). With $M_c \gg m_p$ (symbols defined in section 3.1.4), we have \beq \int^{v_M}_0
f(v) v^2 dv =\frac{(1-f)M_{tot}}{4 \pi^3 a^3} f_0(x, v_M) \eeq with
\beq f_0(x, v_M) =\int^{q_{v=0}}_{q_{v=v_M}} \left( \frac{1}{1+x} -
q^2 \right)^{1/2} K_0(q) q dq \eeq where
$q_{v=0}=\left(1+x\right)^{-1/2}$ and
$q_{v=v_M}=\sqrt{\frac{1}{1+x}-\frac{v^2_M}{2 v^2_g}}$.  Thus the
characteristic dynamical friction timescale becomes \beq t^{-1}_{df}
=\frac{4 \ln \Lambda f (1-f)}{\pi n_c} \frac{v^3_g}{v^3_M}
\frac{v_g}{a} f_0(x,v_M) \eeq

To obtain the exact dynamical friction time from the above equations,
for given initial conditions, one needs to integrate over time for the
time-dependent orbit (e.g. Cora, Muzzio \& Vergne 1997).  While it is
generally assumed that orbits are quickly circularized by dynamical
friction, and thus that the Chandrasekhar formula for circular orbits
may be applied, this is not always a good assumpotion (van den Bosch
{\it et al.}  1999). However, there are some empirical relations
derived from numerical studies (see a short summary in Colpi {\it et al.}~1999)
that relate the dynamical friction time calculated numerically for a
particle on a given initial orbit, i.e. $E$ and $J$, to the prediction
using the circular orbit approximation, giving $t_{df}= t_c
\epsilon^{\alpha}$, where $\epsilon= J(E)/J_c(E)$, and $J_c(E)$ is the
angular momentum of a particle on a circular orbit with the same
binding energy.  The parameter $\epsilon$ is simply dependent on the
orbit eccentricity only, while $\alpha$ can range from 0.4 to 0.8.
For a typical orbital eccentricity of $0.6$, the circular orbit
approximation provides an overestimate by a factor of around 2.
We then assume that a clump is initially moving on 
a circular orbit with $v_M = v_c= v_g\sqrt{x}/(1+x)$ (Tsuchiya \&
Shimada 2000), and modify the estimates of the dynamical friction 
timescales by this factor of two. 

Then from equation (9) in the main text, the specific angular momentum loss rate
should be equal to the torque caused by the dynamical friction at
radius $r$, and we have
\beq
\frac{d\left(r v_c\right)}{dt}=-\frac{v_c}{t_{df}} r.
\eeq
It can be further written as 
\beq
\frac{dr}{dt} =- \frac{2r}{t_{df}} \frac{1}{\frac{\partial \ln M}{\partial \ln r} +1}.
\eeq
With some algebraic manipulation, the dynamical friction time can be expressed as
\beq
t_{df}=\frac{\pi n_c}{4 \ln \Lambda f (1-f)} \frac{a}{v_g} B(x_0)
\eeq
where 
\beq B(x_0)= \int^{x_0}_0 \frac{(3+x) x^{1/2}}{2(1+x)^4}
\frac{1}{f_0(x)} dx, 
\eeq
 and $x_0$ is the initial radius. The appropriate
value of $\Lambda $ can be calculated as (White 1976)
\beq
\ln \Lambda = \frac{1}{M^2_c} \int^{b_{max}}_0 D^3 dD 
\left[ \int^{\infty}_D \frac{M_c(r)dr}{r^2 \left(r^2-D^2\right)^{1/2}} \right]^2,
\eeq
where $M_c(r)$ is the mass profile of a clump, here given by the
Plummer law.  The maximum impact parameter is chosen to be the half
mass radius of the dark halo, $b_{max} = r_{\frac{1}{2} h}$. Thus 
\beq
\ln \Lambda =\frac{1}{2} \ln \left( \frac{r^2_{\frac{1}{2} h}}{a^2_c} + 1 \right) + 
\frac{1}{2} \left( \frac{1}{\frac{r^2_{\frac{1}{2} h}}{a^2_c}+1} - 1 \right)
\eeq
where $a_c$ is the core radius of the clump  Plummer profile. 
For our computed models, typically $\ln \Lambda \sim 3 $.

The above expressions are used to derive the expressions in section~3.1.4 of the text.

\clearpage

\begin{table}[tbh]
      \caption[ ]{The initial parameter values of the computed models}
       \begin{flushleft} 
      \begin{tabular}{lrccccrrr}
            \tableline\tableline\noalign{\smallskip}
             Model & $2T/W$ &  $n_c$ & $\rho_{cl}/\rho_0$ & $f$ & $\sigma_c/\sigma_h$ & Dark halo 
             & $N_{halo}$ & $N_{cl}$ \\        
            \noalign{\smallskip}
            \tableline\noalign{\smallskip}
            A & 1.07 & 80 & 58 & 0.1 & 0.21 & Hernquist & 108000 & 150 \\
            B & 0.91 & 80 & 38 & 0.4&0.31 &Hernquist & 36000 & 300 \\            
            C & 1.35 & 80 & 64 & 1 &0.46  & Hernquist & 0     & 300 \\            
            D & 0.11  & 80 & 58 & 0.1&0.21& Hernquist & 108000 & 150 \\
            E & 0.13 & 80 & 38 & 0.4& 0.31& Hernquist & 36000 & 300\\
            F & 0.31 & 80 & 64 & 1  &0.46& Hernquist & 0     & 300 \\            
            G & 1.05 & 20 & 58 & 0.1&0.34& Hernquist & 54000 & 300 \\
            H & 0.95 & 20 & 38 & 0.4&0.50& Hernquist & 27000 & 900 \\            
            I & 1.20 & 20 & 64 & 1  &0.74& Hernquist & 0 & 1000 \\             
            J & 0.11 & 20 & 58 & 0.1&0.34& Hernquist & 54000 & 300 \\
            K & 0.17 & 20 & 38 & 0.4&0.50& Hernquist & 27000 & 900 \\            
            L  & 0.48 & 20 & 64 & 1  &0.74& Hernquist & 0 & 1000 \\             
            M & 1.05 & 20 & 58 & 0.1&0.34& Plummer & 54000& 300 \\
            N & 0.89 & 20 & 38 & 0.4&0.50& Plummer & 27000 & 900 \\            
            O  & 1.17 & 20 & 64 & 1  &0.74& Plummer & 0 & 1000 \\             
            P & 0.11 & 20 & 58 & 0.1&0.34& Plummer & 54000 & 300 \\
            Q & 0.40 & 20 & 38 & 0.4&0.50& Plummer & 27000 & 900 \\            
            R  & 0.48 & 20 & 64 & 1  &0.74& Plummer & 0 & 1000 \\ 
            S & 1.06 & 20 & 2.7 & 0.1& 0.20&Hernquist & 54000 &300 \\
            T & 0.95 & 20 & 1.8 & 0.4& 0.30&Hernquist & 27000& 900 \\                        
            U & 0.11 & 20 & 2.7 & 0.1&0.20& Hernquist & 54000 & 300\\
            V & 0.13 & 20 & 1.8 & 0.4&0.30& Hernquist & 27000 & 900\\                        
            W & 1.08 & 5 & 61 & 0.05&0.43& Hernquist &114000 & 1200 \\
            X & 0.8 & 5 & 38 & 0.4&0.79& Hernquist & 9000 & 1200 \\                                  
            Y & 0.13 & 5 & 58 & 0.1&0.53& Hernquist &36000 & 800 \\
            Z & 0.27 & 5 & 38 & 0.4&0.79& Hernquist &18000 & 2400 \\                        
            \noalign{\smallskip}
            \tableline
            \tableline
         \end{tabular}
      \end{flushleft}
\tablecomments{
In column 2, $T$ is the total kinetic energy and $W$ is the
total potential energy;  the virial ratio $2T/W$ is a quantity that
approximately describes how far the system deviates from virial equilibrium, being unity for a system in equilibrium.  
Column 3 gives the number of clumps used in the
model. Column 4 is the ratio of the mean density within the half mass
radius of a single clump to that of the total mass. Column 5 shows the
mass fraction in clumps. Column 6 shows the ratio of the one-dimensional
internal velocity dispersion of a clump to that of the halo.  Column 7 
identifies which density profile is adopted for the distribution of both the 
dark
halo particles and the system of clumps.  Columns 8 and 9 give the number of
particles used for the dark halo and for each clump, respectively.}
\label{tab:parameter}
   \end{table}
\normalsize
\clearpage


   \begin{table}
      \caption[ ]{The timescales of different dynamical processes of the computed models}
\footnotesize
       \begin{flushleft} 
      \begin{tabular}{lcccccc}
            \tableline\tableline\noalign{\smallskip}
             Model   &$t_{merge}/t_{\frac{1}{2}hc}$  &$t_{c-c}/t_{\frac{1}{2}hc}$ 
&$t_{df}/t_{\frac{1}{2}hc}$ &$t_{\frac{1}{2}dis}/t_{\frac{1}{2}hc}$ 
& $\sigma_c/\sigma_h$ & inferred dominant processes \\        
            \noalign{\smallskip}
            \tableline\noalign{\smallskip}
            A        & 667  & 10.8 & 71     & 12.8 & 0.21 & c-c\\
            B        & 16.7 & 1.5  & 26.7   & 5.1  & 0.31 & c-c\\            
            C         & 6.7  & 0.52 & $\infty$ & 18.3 & 0.46 & merge, c-c\\            
            G        & 167  & 6.8  & 17.8   & 15.1 & 0.34 & c-c, df\\
            H        & 10.4 & 0.93 & 6.7    & 6.6  & 0.50 &df\\   
            I         & 1.7  & 0.33 & $\infty$ & N/A & 0.74& merge \\             
            M   & 167  & 6.8  & N/A    &26.7 & 0.34  &c-c,df(?) \\
            N   & 10.4 & 0.93 & N/A    &9.0 & 0.50   &df(?) \\            
            O    & 1.7  & 0.33 & $\infty$ &9.0 &0.74  &merge \\             
            S         & 167  & 2.4  & 17.8   & 2.5  &0.20  &tides, c-c  \\
            T         & 10.4 & 0.33 & 6.7    & 2.5  &0.30  &tides, c-c \\                      
            W        & 167  & 11.0 & 8.4    & 11.3 &0.43  &df \\
            X         & 2.6  & 0.6  & 1.7    & 4.5  &0.79  &df, merge\\
            
            \noalign{\smallskip}
            \tableline
            \tableline
         \end{tabular}
      \end{flushleft}
\tablecomments{All the timescales are given in units of the crossing 
time of the halo. The quantities in Columns 2--4 are derived from the analytic 
expressions in the text. 
Column~2 gives the merging timescale,
calculated as described in section 3.1.2; Column~3 gives the clump-clump
disruption timescale, calculated as described in  section 3.1.3; 
Column~4 gives the dynamical friction timescale, calculated as described in 
section 3.1.4.   Column~5 is derived from the simulations and gives the time for the mass of material from 
disrupted clumps to equal  half of the 
total mass in clumps;   Column~6 shows the ratio
of the one-dimensional internal velocity dispersion of a clump to that of 
the halo, while  Column~7 lists the possible dominant dynamical process
inferred from the combination of theoretical predictions and
simulation results, where we use the symbols, `c-c', `merge', `tides' and
`df', to represent the processes of  clump-clump heating, merging,  global
tides and dynamical friction, respectively.
}

\label{tab:timescale}
   \end{table}
\clearpage


   \begin{table}
      \caption[ ]{The angular momentum content of the computed models}
\scriptsize
       \begin{flushleft} 
      \begin{tabular}{lrrrrccrrrc}
            \tableline\tableline\noalign{\smallskip}
        Model & state &$J_{\rho TOT}$ & $J_{\rho CL}$ & $J_{\rho DM}$ 
             & $\frac{J_{\rho CL}}{J_{\rho TOT}}$ & $\delta_{CL}$
             & $J_{cm TOT}$ & $J_{cm CL}$ & $J_{cm DM}$ 
                & $\frac{J_{cm CL}}{J_{cm TOT}}$ \\   

            \noalign{\smallskip}
            \tableline\noalign{\smallskip}
 A 
& i      &  -0.6083 &-0.04536 & -0.5630 & $7.5\%$ &$6\%$ &-0.6007 &-0.03910 &-0.5616 &$6.5\%$ \\
& f      & -0.6244  &-0.04769 & -0.5767 &         &   &-0.6020 &-0.04306 &-0.5589 & \\
&(f-i)/i &$2.6\%$   & $5.1\%$ & $2.4\%$ &         &   &$0.2\%$ & $10\%$  &$0.5\%$ & \\
 B 
& i      &-0.6200   &-0.3169 &-0.3030   &$51\%$ &$13\%$  &-0.6254 &-0.3121  &-0.3132 &$50\%$ \\ 
& f      &-0.6203   &-0.3215 &-0.2988   &       &  &-0.6250 &-0.3083  &-0.3167 & \\        
&(f-i)/i &$0.06\%$  &$1.5\%$ &$-1.4\%$  &       &  &$-0.06\%$&$-1.2\%$ &$1.1\%$ & \\
 D &i  
&-0.1904 &-0.01480 &-0.1756 &$7.7\%$ &$49\%$ &-0.1923 &-0.01511 &-0.1771 &$7.9\%$ \\
& f &-0.1991 &-0.01689 &-0.1822 &     &    &-0.1930 &-0.01869 &-0.1743 &\\
&(f-i)/i &$4.6\%$ &$14\%$   &$3.8\%$ & &       &$0.4\%$ &$24\%$   &$-1.6\%$&\\            
 E 
& i &-0.1951  &-0.09930 &-0.09583 &$51\%$&$47\%$ &-0.1963  &-0.09864 &-0.09762 &$50\%$\\
& f &-0.1948  &-0.08726 &-0.1076  &      & &-0.1980  &-0.08917 &-0.1088  &\\
&(f-i)/i &$-0.16\%$&$-12\%$  &$12\%$   & &      &$0.9\%$  &$-9.6\%$ &$11\%$    &\\  
 G 
& i &-0.6092 &-0.04577 &-0.5635  &$7.5\%$&$16\%$ &-0.6172 &-0.05260 &-0.5646 &$8.5\%$\\ 
& f &-0.6202 &-0.04794 &-0.5722  &       & &-0.6170 &-0.04282 &-0.5742 &\\
&(f-i)/i &$1.8\%$&$4.7\%$ &$1.6\%$ &     &      &$-0.04\%$ &$-19\%$ &$1.7\%$ &\\   
 H 
&i  &-0.4822 &-0.1763  &-0.3059 &$37\%$&$47\%$ &-0.4818   &-0.1686 &-0.3133 &$35\%$\\
&f  &-0.3726 &-0.1135  &-0.2590 &      & &-0.4803   &-0.1990 &-0.2813 & \\ 
&(f-i)/i &$-22\%$&$-36\%$ &$-15\%$&    &   &$-0.3\%$ &$18\%$ &$-10\%$ & \\            
J 
&i  &-0.1943  &-0.01615 &-0.1782  &$8.2\%$&$79\%$ &-0.1968 &-0.01810 &-0.1787  &$9.1\%$\\
&f  &-0.1970  &-0.01725 &-0.1798  &       & &-0.1969 &-0.01557 &-0.1814  &\\
&(f-i)/i &$1.4\%$ &$6.9\%$ &$0.9\%$ &     &   & $0.08\%$  &$-13\%$ &$1.5\%$ &\\
K 
& i &-0.1544 &-0.05766 &-0.09673 &$37\%$&$52\%$ &-0.1530   &-0.05386 &-0.09900 &$35\%$\\
& f &-0.1721 &-0.06048 &-0.1116  &      & &-0.1503   &-0.04253 &-0.1078  &\\
&(f-i)/i &$11\%$ &$4.9\%$ &$15\%$  &    &   &$-1.7\%$ &$-21\%$ &$8.8\%$ &\\                    
M 
&i &-0.2708 &0.002488 &-0.2732  &$-0.9\%$&$26\%$ &-0.2702 &0.003379 &-0.2736 &$-1\%$\\
&f &-0.2700 &0.004062 &-0.2740  &        & &-0.2702 &0.003606 &-0.2738 &\\
&(f-i)/i &$-0.3\%$ &$63\%$   &$0.3\%$ &  &      &$0.02\%$  &$7\%$ &$0.06\%$  &\\
N 
&i &-0.1367 &0.01071 &-0.1474 &$-7.8\%$&$34\%$ &-0.1293 &0.02120 &-0.1505 &$-8\%$ \\
&f &-0.1311 &0.01484 &-0.1459 &         & &-0.1292 &0.01630 &-0.1455 &\\
&(f-i)/i  &$-4.1\%$   &$39\%$ &$-1\%$  & &         &$-0.07\%$ &$-23\%$ &$-3.3\%$ &\\               
P 
&i &-0.08585 & 0.000555 &-0.08641 &$-0.6\%$&$85\%$ & -0.08566 & 0.0008716 &-0.08653  &$-1\%$\\
&f &-0.08550 &-0.003074 & -0.08242&         & & -0.08560 &-0.003251  &-0.08235  &\\
&(f-i)/i &$-0.4\%$  &$-654\%$ &$-4.6\%$ &   &        &$-0.07\%$  & $-473\%$ &$-4.8\%$ &\\
            
 \noalign{\smallskip}
            \tableline
         \end{tabular}
      \end{flushleft}
\label{tab:ang1}
   \end{table}

\normalsize
\clearpage

%
   \begin{table}
\tablenum{3}
      \caption[ ]{(Continued) The angular momentum content of the computed models}
\scriptsize
       \begin{flushleft} 
      \begin{tabular}{lrrrrccrrrc}
            \tableline\tableline\noalign{\smallskip}
             Model & state &$J_{\rho TOT}$ & $J_{\rho CL}$ & $J_{\rho DM}$ 
             & $\frac{J_{\rho CL}}{J_{\rho TOT}}$ & $\delta_{CL}$
             & $J_{cm TOT}$ & $J_{cm CL}$ & $J_{cm DM}$ 
                & $\frac{J_{cm CL}}{J_{cm TOT}}$ \\        
            \noalign{\smallskip}
            \tableline\noalign{\smallskip}
Q 
& i &-0.04355&0.003064 &-0.04661&$-7\%$&$75\%$ &-0.04112&0.006532  &-0.04765 &$-16\%$\\
& f &-0.04193&-0.006918&-0.03501&      & &-0.04143&-0.006248 &-0.03518 &\\
&(f-i)/i &$-4\%$  &$-326\%$ &$-25\%$&  &       &$0.8\%$ &$-196\%$ &$-26\%$  &\\  
S 
&i  &-0.6082  &-0.04469 &-0.5635  &$7.3\%$&$11\%$ &-0.6162 &-0.05164 &-0.5646 &$8.3\%$\\
& f &-0.6189  &-0.04763 &-0.5712  &       & &-0.6159 &-0.04354 &-0.5723 &\\
&(f-i)/i  &$1.8\%$ &$6.6\%$ &$1.4\%$ &    &    &$-0.05\%$      &$-16\%$ &$1.4\%$ &\\
T 
& i &-0.4812  &-0.1754  &-0.3059   &$36\%$&$31\%$ &-0.4817   &-0.1684 &-0.3133 &$35\%$\\
& f &-0.4435  &-0.1548  &-0.2897   &      & &-0.4809   &-0.1666 &-0.3143 &\\
&(f-i)/i &$-7.8\%$&$-12\%$ &$-5.6\%$ &    &   &$-0.2\%$  &$1\%$   &$0.3\%$ &\\
U 
&i  &-0.1933 &-0.01512 &-0.1768  &$7.8\%$&$73\%$ &-0.1958 &-0.01720 &-0.1786 &$8.8\%$\\
& f &-0.1953 &-0.01605 &-0.1792  &       & &-0.1953 &-0.01606 &-0.1792 &\\
&(f-i)/i &$1\%$ &$6.1\%$  &$6.1\%$ &     &      &$-0.3\%$&$-6.6\%$ &$0.3\%$ &\\ 
V 
& i &-0.1533 &-0.05657 &-0.09673 &$37\%$&$65\%$ &-0.1526 &-0.05356 &-0.09907 &$35\%$\\
& f &-0.1730 &-0.05907 &-0.1140  &      & &-0.1519 &-0.05269 &-0.09919 &\\
&(f-i)/i &$13\%$  &$4.4\%$  &$18\%$   & &      &$-0.5\%$&$-1.6\%$ &$0.1\%$ &\\    
            
W
& i  &-0.6251  &-0.01217 &-0.6130  &$2\%$&$21\%$ &-0.6169  &-0.001880 &-0.6151 &$0.3\%$\\
& f  &-0.6117  &-0.01168 &-0.6000  &     & &-0.6174  &-0.01037  &-0.6070 &\\
&(f-i)/i &$-2.1\%$ &$-4\%$   &$-2.1\%$ & &     &$0.08\%$ &$451\%$   &$-1.3\%$ &\\ 
X 
& i &-0.4011  &-0.09584 &-0.3052 &$24\%$&$55\%$ &-0.3678   &-0.08759 &-0.2802 & $24\%$ \\
& f &-0.3306  &-0.1236  &-0.2070 &      & &-0.3675   &-0.1068  &-0.2607 &\\
&(f-i)/i &$-18\%$ &$29\%$   &$-32\%$ &  &     &$-0.08\%$ &$22\%$   &$-7\%$ &\\
Y 
& i &-0.1829   &-0.007196 &-0.1757 &$4\%$&$83\%$ &-0.1786  &-0.004523  &-0.1741 &$2.5\%$\\
& f &-0.1780   &-0.001705 &-0.1763 &     & &-0.1762  &0.0009680 &-0.1771 &\\
&(f-i)/i &$-2.7\%$ &$-76\%$   &$0.3\%$ & &     &$-1.3\%$ &$-121\%$   &$1.8\%$ &\\
Z 
& i  &-0.1264 &-0.03001 &-0.09640 &$24\%$ &$57\%$ &-0.1404 &-0.04144 &-0.09896 &$30\%$\\
& f  &-0.1668 &-0.03264 &-0.1342  &       & &-0.1425 &-0.01930 &-0.1232 &\\
&(f-i)/i &$32\%$  &$8.7\%$  &$39\%$   &   &   &$1.5\%$ &$-53\%$ &$24\%$ &\\                       
            \noalign{\smallskip}
            \tableline\tableline
         \end{tabular}
      \end{flushleft}
\tablecomments{Column~1 identifies the model;
Column~2 indicates whether the values that follow in subsequent columns refer to the initial state, to the final state or to the relative
change.  Columns~3, 4 and 5 give the angular momentum, $J_z$ (measured 
relative to
the highest density point of the dark matter) of the total mass, of the clump
mass and of the dark matter, respectively, while  
column~6 gives the initial angular momentum
fraction contained in the clumpy mass. 
Column~7 gives the value of $\delta_{CL}$, as defined in the text. 
Columns~8, 9, 10 and 11 list similar quantities to those in
columns~3, 4, 5 and 6, but now with the angular momentum calculated
relative to the centre of mass of the total mass. 
}
\label{tab:ang2}
   \end{table}

\normalsize
\clearpage


   \begin{table}
      \caption[ ]{The computed models with little angular momentum change}
\footnotesize
       \begin{flushleft} 
      \begin{tabular}{lrcc}
            \tableline\tableline\noalign{\smallskip}
             Model   &$\delta_{CL}$  &$t_{df}/t_{\frac{1}{2}hc}$ 
&$t_{\frac{1}{2}dis}/t_{\frac{1}{2}hc}$\\        
            \noalign{\smallskip}
            \tableline\noalign{\smallskip}
            A        & 6\% & 71     & 12.8 \\
            S         & 11\%  & 17.8   & 2.5  \\                                   
            B        & 13\%  & 26.7   & 5.1  \\            
            G        & 16\%  & 17.8   & 15.1 \\
            W        & 21\%  & 8.4    & 11.3 \\
            T         & 31\%  & 6.7    & 2.5  \\                                  
            \noalign{\smallskip}
            \tableline\tableline
         \end{tabular}
      \end{flushleft}
\tablecomments{For convenience, we have tabulated particular 
quantities from Tables~2 and 3 together here, to highlight their values for the models with little angular momentum change.}
\label{tab:smallangchange}
   \end{table}
\clearpage


   \begin{table}
      \caption[ ]{The final shapes of the dark halo and of the debris}
\footnotesize
       \begin{flushleft} 
      \begin{tabular}{lcccc}
            \tableline\tableline\noalign{\smallskip}
             Model & $(b/a)_{DM}$ &  $(c/a)_{DM}$ & $(b/a)_{DE}$ & $(c/a)_{DE}$ \\
            \noalign{\smallskip}
            \tableline\tableline\noalign{\smallskip}
            A & 0.99 & 0.81 & 0.95 & 0.71\\
            B & 0.99 & 0.86 & 0.96 & 0.83\\
            C &N/A & N/A &0.98 & 0.88 \\            
            D & 0.66  & 0.56 & 0.63 & 0.54\\
            E & 0.72 & 0.61 & 0.68 & 0.62\\
            F &N/A &N/A & 0.97 &0.80 \\

            G & 0.99 & 0.84 & 0.97 & 0.86 \\
            H & 0.97 & 0.88 & 0.94 & 0.80 \\
            I & N/A &N/A  & N/A & N/A \\
            J & 0.68 & 0.57 & 0.67 & 0.53 \\
            K & 0.88 & 0.70 & 0.83 & 0.62 \\
            L  & N/A  & N/A & 0.96  & 0.92 \\

            M & 0.99 & 0.85 & 0.95  & 0.70 \\
            N & 0.99 & 0.88 & 0.98 & 0.81\\
            O  &N/A  &N/A  &0.95  &0.90 \\
            P &0.69  &0.59  & 0.61 & 0.52\\
            Q &0.82  &0.61  & 0.70 & 0.50\\
            R  &N/A  &N/A  &0.73  &0.70 \\

            S & 0.98 & 0.82 & 0.88 &0.80 \\
            T & 0.98 & 0.82 & 0.91 &0.80 \\
            U & 0.63 & 0.51 &0.61  &0.54 \\
            V & 0.75 & 0.55 &0.76  &0.60 \\

            W & 0.99 & 0.83 & 0.95 &0.55 \\
            X & 0.98 & 0.90 & 0.92 & 0.80 \\
            Y & 0.76 & 0.61 &0.67 &0.59 \\
            Z & 0.88 & 0.84 &0.86 &0.80 \\
            
            \noalign{\smallskip}
            \tableline\tableline
         \end{tabular}
      \end{flushleft}

\tablecomments{Columns~2 and 3 give the axial ratios $b/a$
and $c/a$ of the dark matter, while columns~4 and 5 give the axial
ratios $b/a$ and $c/a$ of the debris. }

\label{tab:triaxial}
   \end{table}
\normalsize
\clearpage


   \begin{table}
      \caption[ ]{The $\eta$ values for our simulations}
\footnotesize
       \begin{flushleft} 
      \begin{tabular}{lccc}
            \tableline\tableline\noalign{\smallskip}
             Model & $\eta (t=126)$ &  $\eta (t=56)$ & $\eta (t=14)$ \\
            \noalign{\smallskip}
            \tableline\noalign{\smallskip}

Z &0.08 & 0.11 & 0.12\\

S  & 0.1 & 0.155 & 0.165\\

H & 0.11 & 0.13 &0.12\\

T &0.115 & 0.12 & 0.135\\

K &0.125 & 0.155  & 0.14\\

V & 0.13  & 0.13   &0.18\\

E &0.135 &  0.175 & 0.16\\

Y & 0.135 & 0.145 & 0.13\\

B & 0.14 & 0.12 & 0.125\\

U &0.145 &0.17 &0.145\\

W & 0.15 & 0.17 & 0.16 \\

G & 0.155 & 0.125 & 0.13\\

J &0.165 &0.16 &0.12\\

D &0.165 &0.14 &0.175\\

X & 0.17 & 0.235 & 0.105\\

A & 0.17 & 0.15 & 0.18\\
            
            \noalign{\smallskip}
            \tableline\tableline
         \end{tabular}
      \end{flushleft}
\tablecomments{The different columns are for values estimated at different times.}
\label{tab:eta}
   \end{table}

\normalsize
\clearpage

\clearpage

\begin{figure}
\caption[The snapshots of the models G, H, I, S and T] {The snapshots
of the models G, H, I, S and T (from top panel to bottom panel) at
times $t =0, 14, 56, 126$, with only those particles belonging to the
clumpy component plotted (the unit of time is the initial halo crossing time).  Each image is the projection onto the XY
plane of the distribution of those particles within a box of size 22
units (the unit is the initial core radius $a$ of the halo mass distribution) on a side.   }
\label{fig:vic20hern}
\end{figure}

\begin{figure}
\caption[The snapshots of the models A, B, C, W and X]
{The snapshots of the models A, B, C, W and 
X (from top panel to bottom panel) at times $t =0, 14, 56, 126$ (the unit of time is the initial halo crossing time). 
Each image  is the projection onto the XY plane of the particles in the 
clumpy component 
within a box of size 22 units (the unit is the initial core radius $a$ of the halo mass distribution) on a side.
}
\label{fig:vic80c5hern}
\end{figure}

\begin{figure}

\caption[The snapshots of the models M, N and O]
{The snapshots of the models M, N and O
(from top panel to bottom panel) at times $t =0, 14, 56, 126$ (the unit of time is the initial halo crossing time). 
Each image  is the projection onto the XY plane of the particles in the 
clumpy component 
within a box of size 10 units  (the unit is the initial core radius $a$ of the halo mass distribution) on a side.
}
\label{fig:vic20plum}
\end{figure}

\begin{figure}
\caption[The snapshots of the models J, K, L, U and V]
{The snapshots of the models J, K, L, U and V 
(from top panel to bottom panel) at times $t =0, 14, 56, 126$ (the unit of time is the initial halo crossing time). 
Each image  is the projection onto the XY plane of the particles in the 
clumpy component 
within a box of size 22 units (the unit is the initial core radius $a$ of the halo mass distribution) on a side.
}
\label{fig:coc20hern}
\end{figure}

\begin{figure}
\caption[The snapshots of the models D, E, F, Y and Z]
{The snapshots of the models D, E, F, Y and 
Z 
(from top panel to bottom panel) at times $t =0, 14, 56, 126$ (the unit of time is the initial halo crossing time). 
Each image  is the projection onto the XY plane of the particles in the 
clumpy component 
within a box of size 22 units (the unit is the initial core radius $a$ of the halo mass distribution) on a side.
}
\label{fig:coc80c5hern}
\end{figure}

\begin{figure}

\caption[The snapshots of the models P, Q and R]
{The snapshots of the models P, Q and R
(from top panel to bottom panel) at times $t =0, 14, 56, 126$ (the unit of time is the initial halo crossing time). 
Each image  is the projection onto the XY plane of the particles in the 
clumpy component 
within a box of size 10 units (the unit is the initial core radius $a$ of the halo mass distribution) on a side.
}
\label{fig:coc20plum}
\end{figure}

\begin{figure}[htbp]
\centerline{\psfig{file=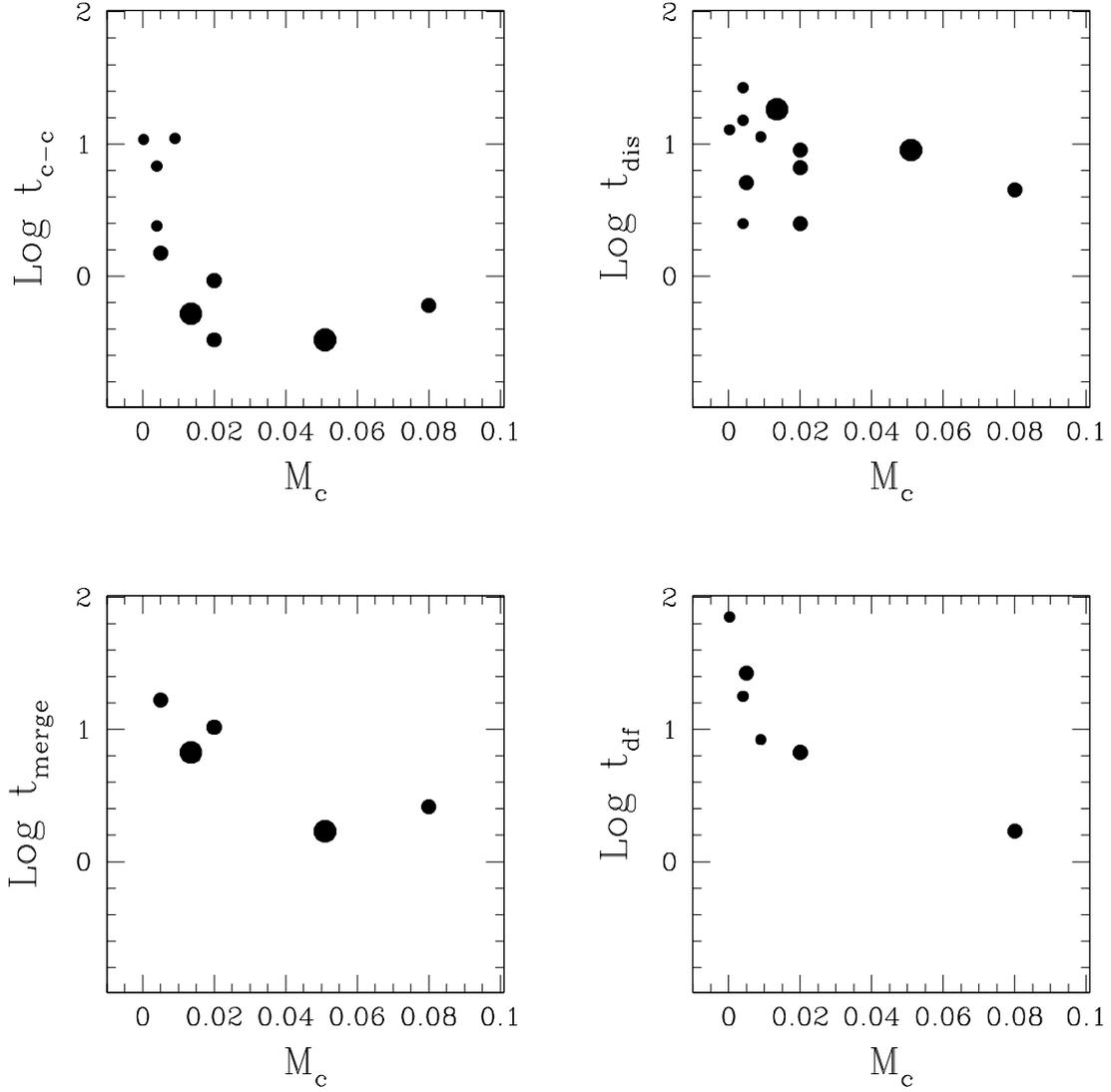,width=\textwidth,angle=0}} 
\caption{The timescales (again in units of halo crossing times) of
various processes plotted against the mass of a clump, for the
virialized models only.  The symbol size is larger for larger clump
mass fraction $f$. }
\label{timescales}
\end{figure}
\clearpage 
\begin{figure}[htbp]
\centerline{\psfig{file=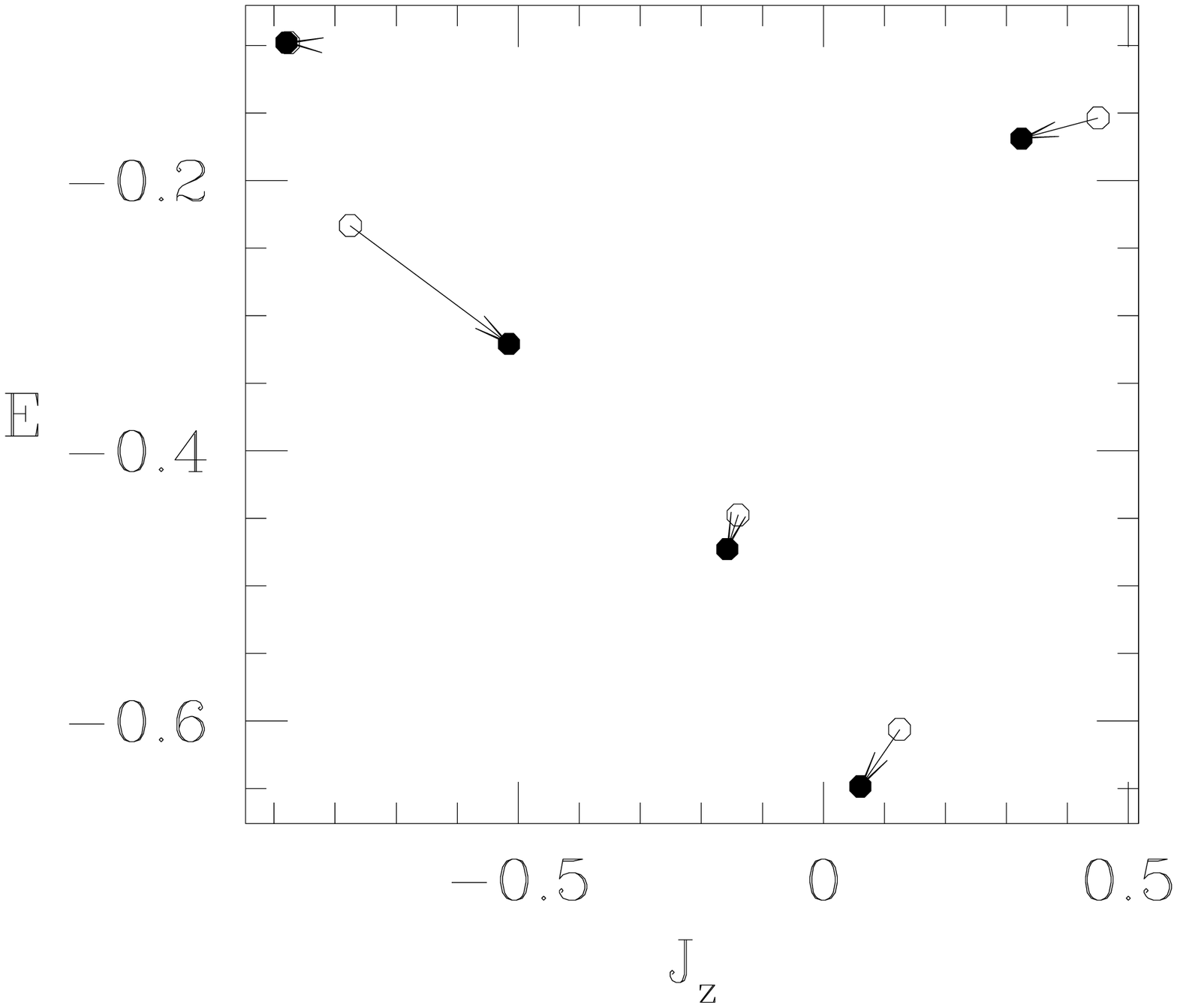,width=0.45\textwidth,angle=0} 
\psfig{file=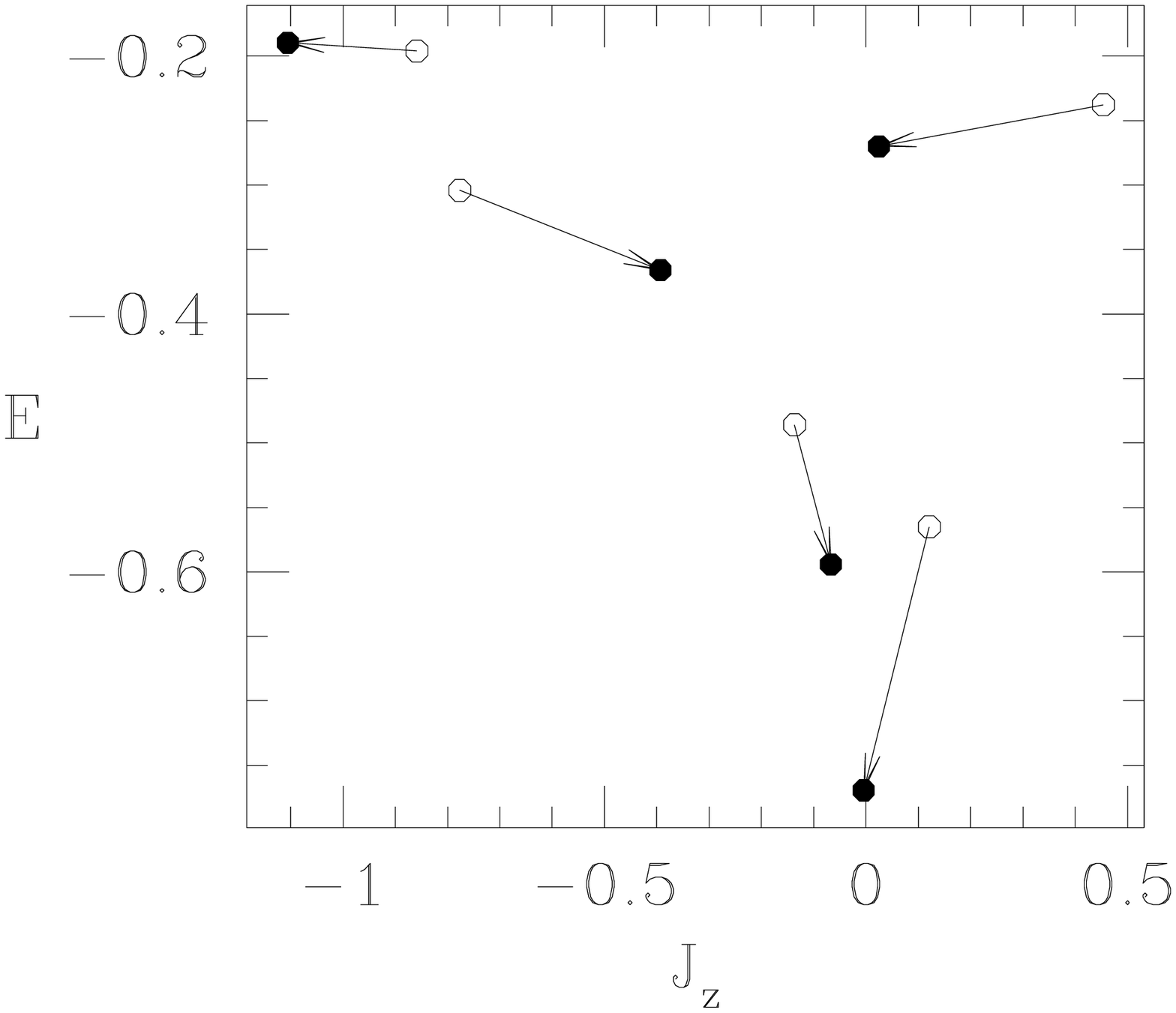,width=0.45\textwidth,angle=0}}
\vspace{0.5cm}
\centerline{\psfig{file=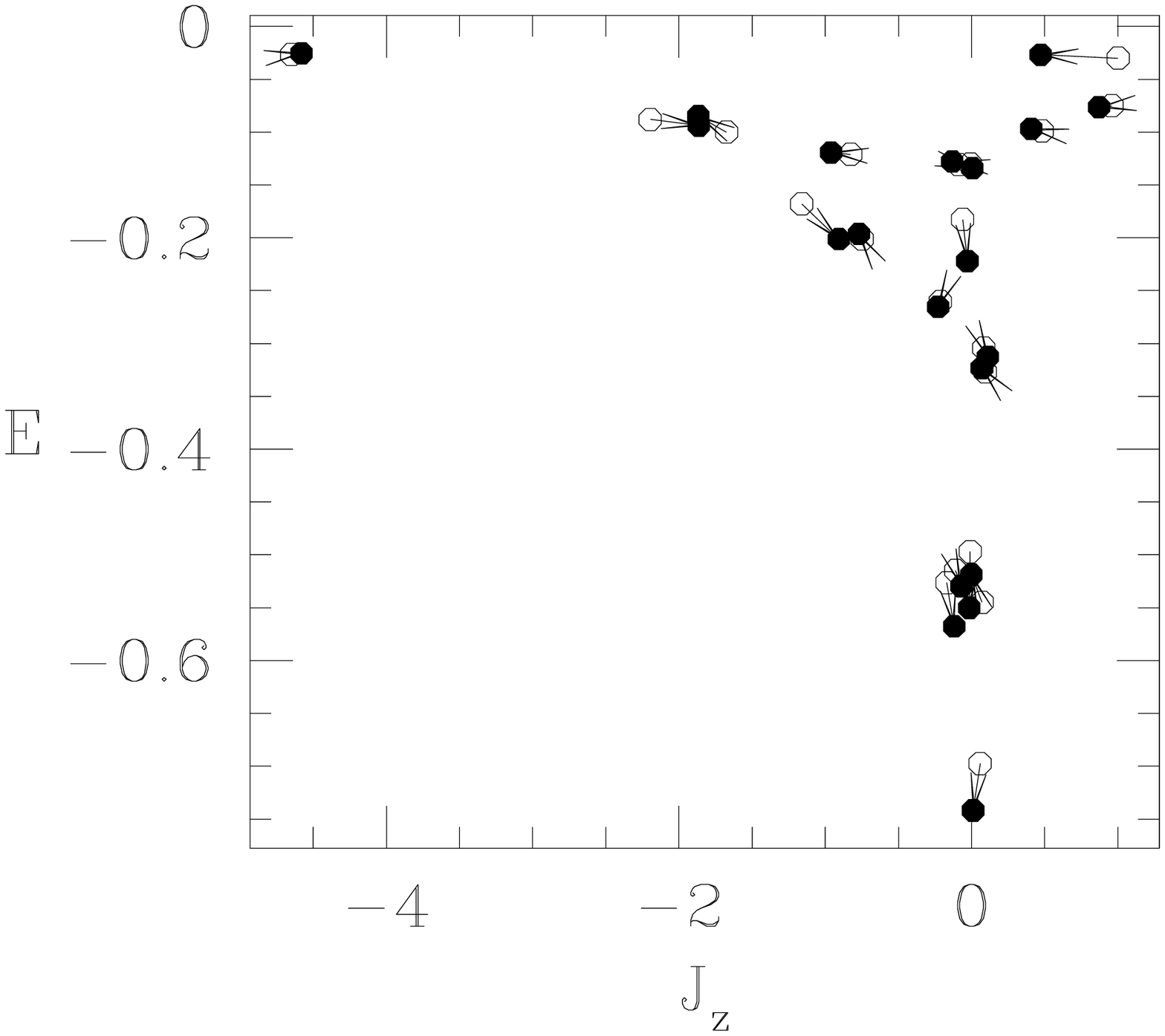,width=0.45\textwidth,angle=0}
\psfig{file=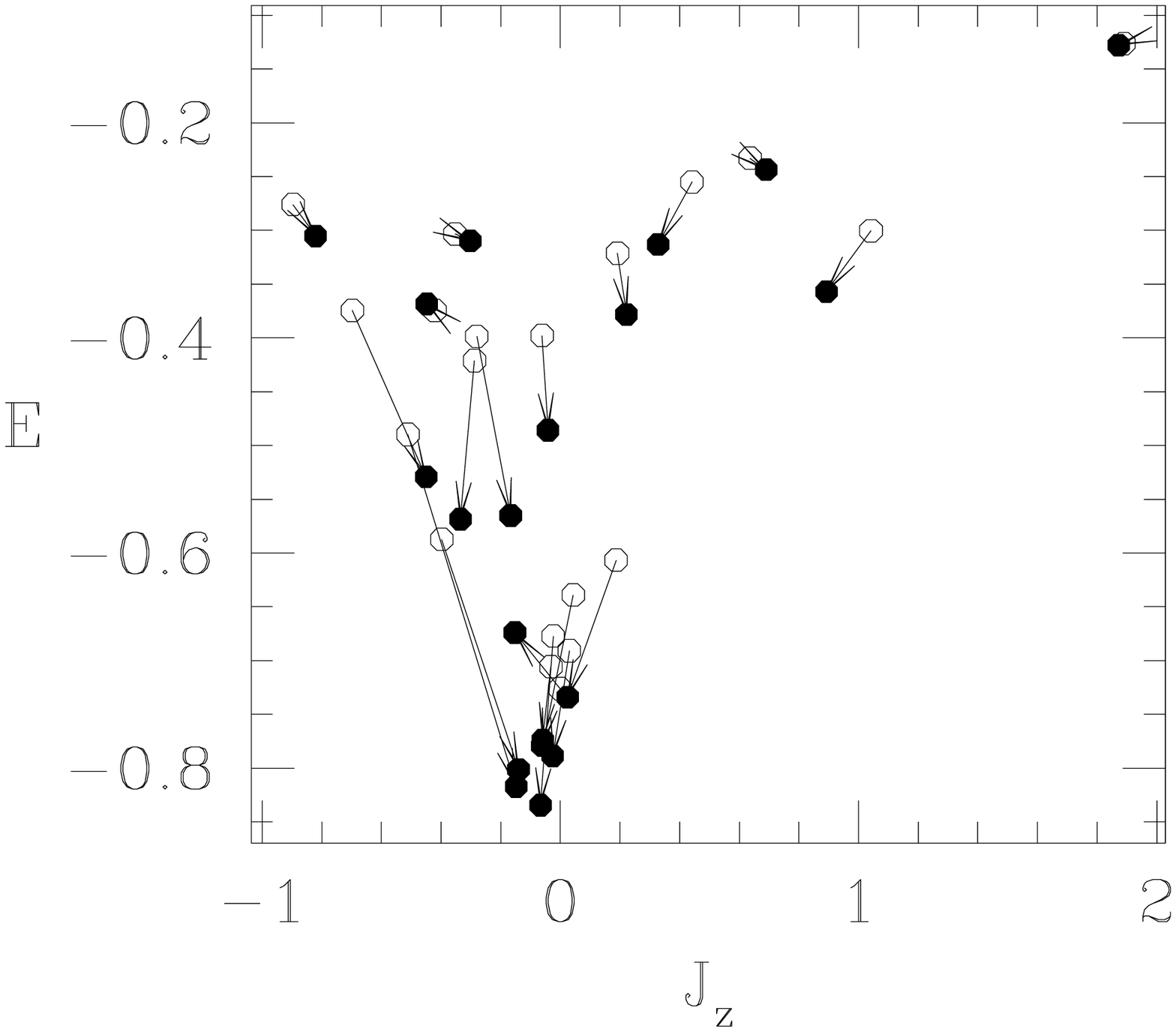,width=0.45\textwidth,angle=0}}
\vspace{0.5cm}
\caption[Clump E-J diagram for virialized cases]
{The binding energy and angular momentum for each  clump for the virialized cases.
The upper left panel is for model W; the 
upper right panel is for model X; the 
lower left panel is for model G, while the 
lower right panel is for model M. The open circles represent initial values and the 
filled circles represent final values.}
\label{fig:clumpejvi}
\end{figure}
\clearpage 

\begin{figure}[htbp]
\centerline{\psfig{file=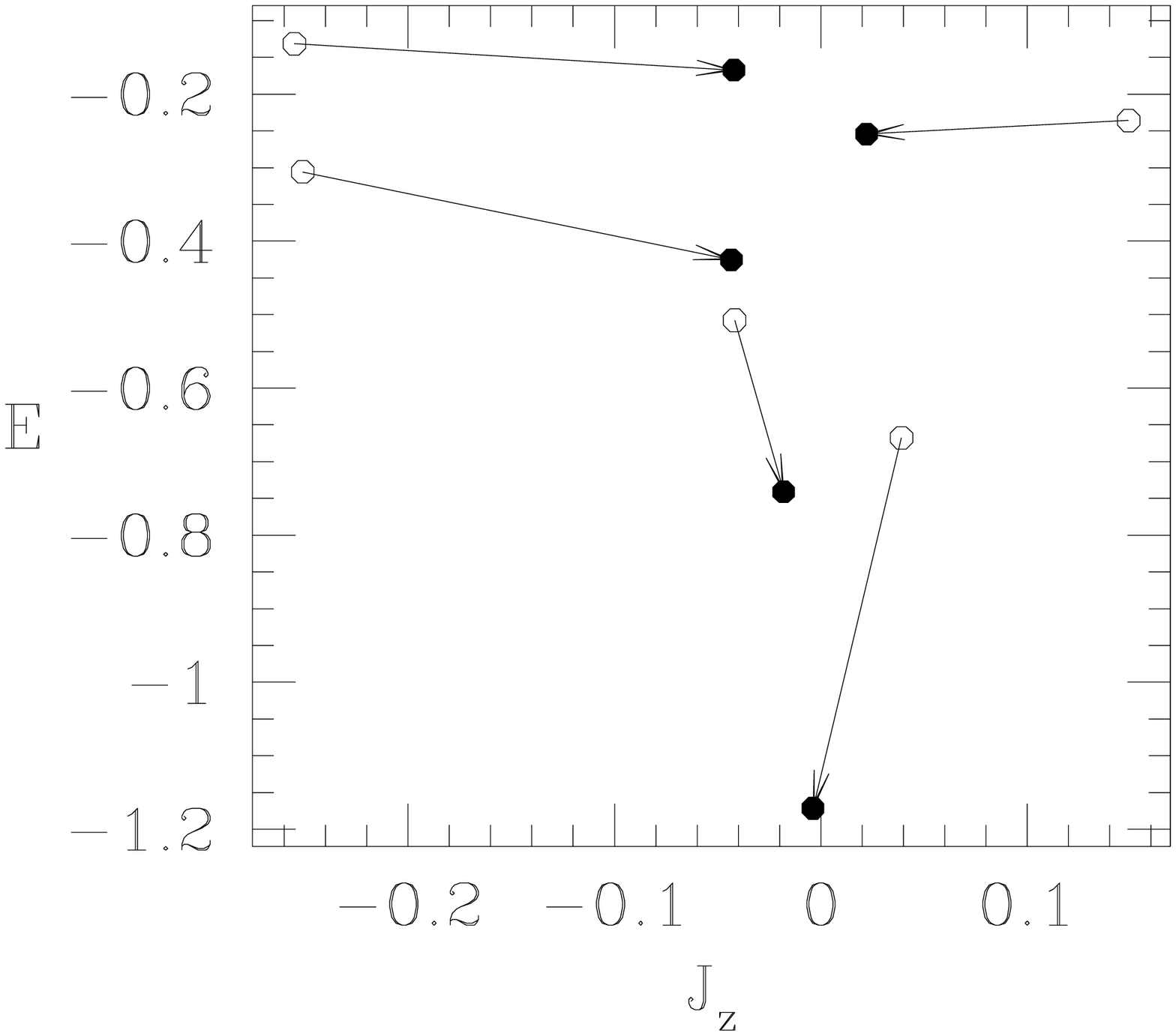,width=0.45\textwidth,angle=0} 
\psfig{file=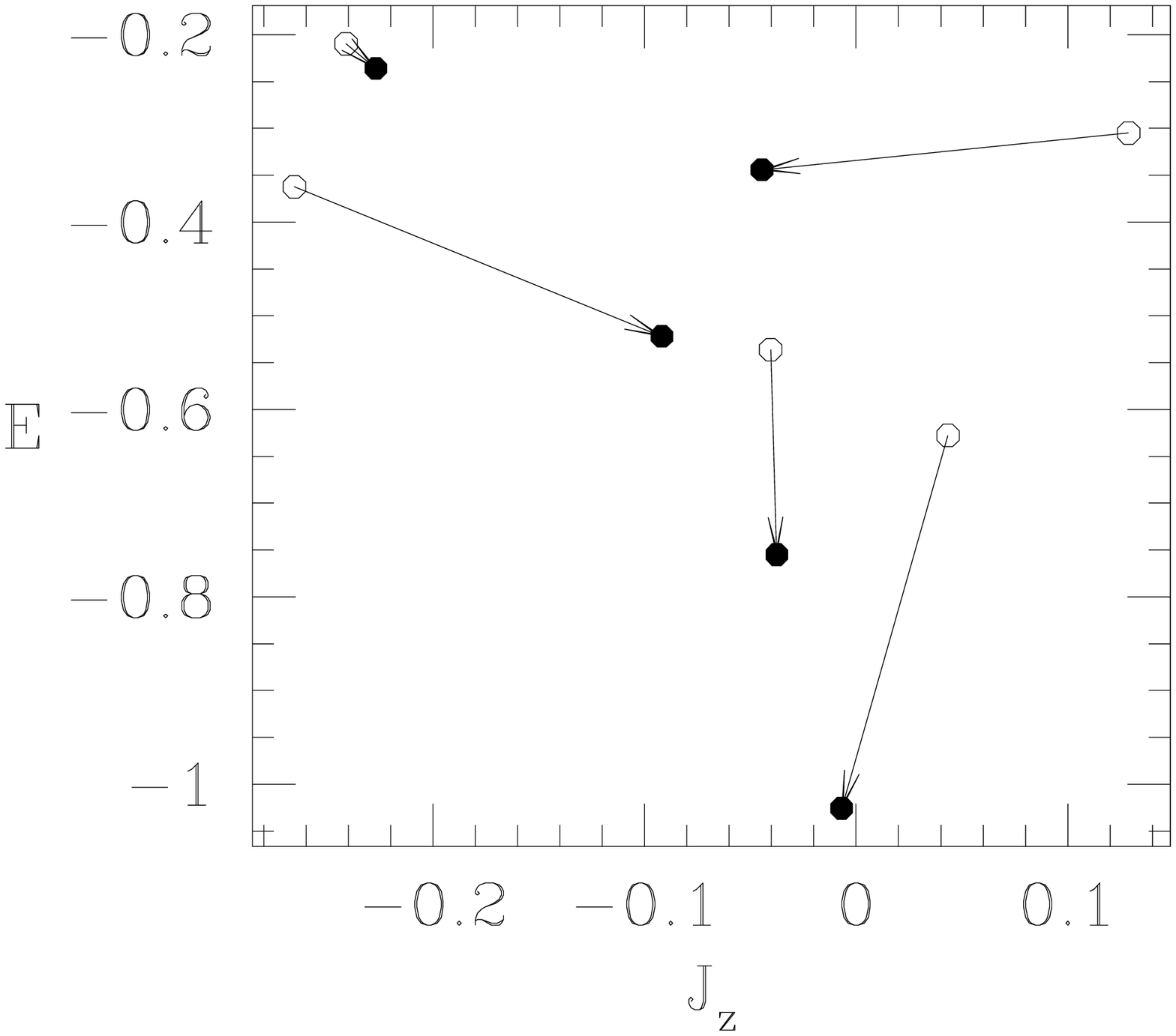,width=0.45\textwidth,angle=0}}
\vspace{0.5cm}
\centerline{\psfig{file=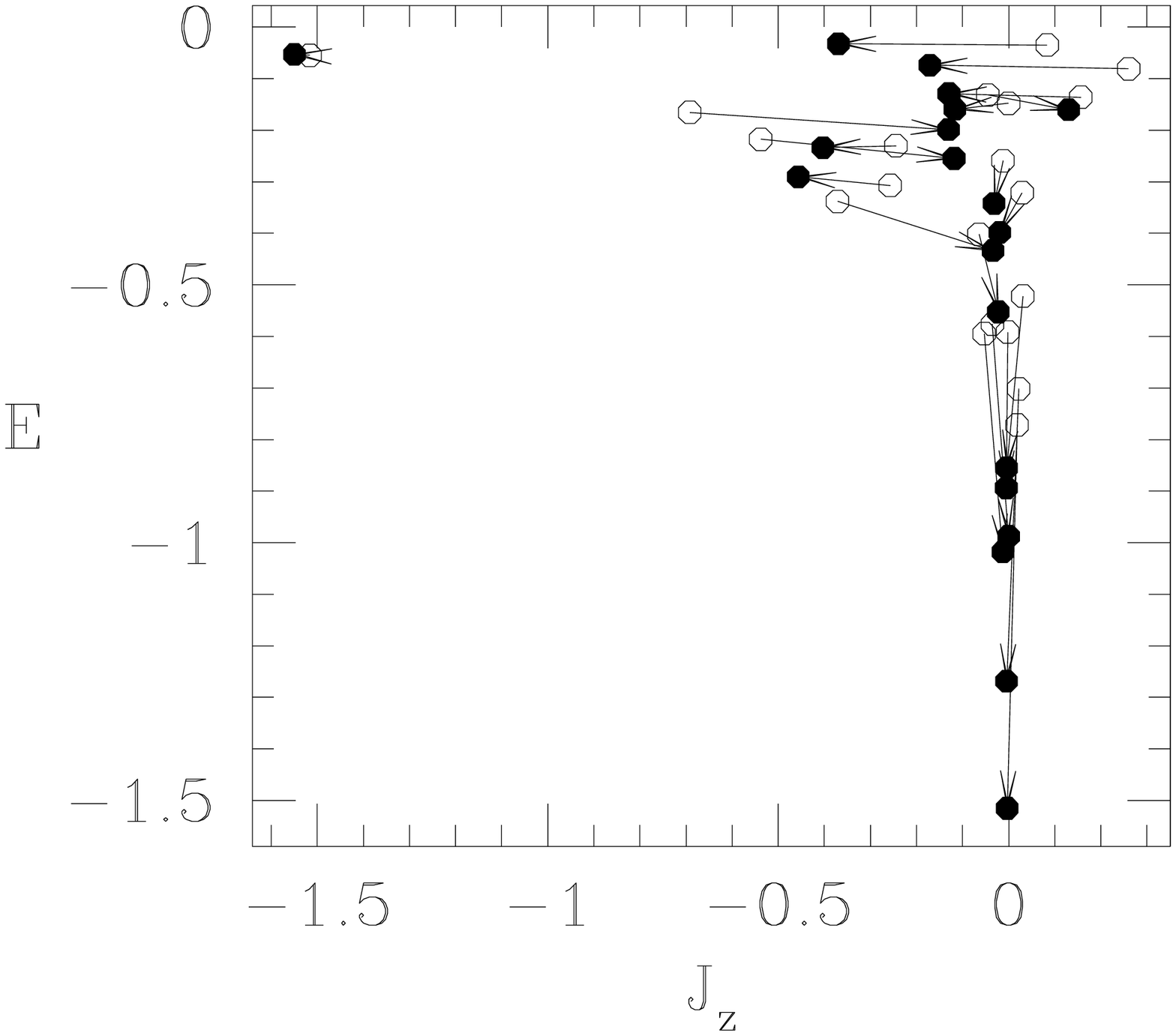,width=0.45\textwidth,angle=0}
\psfig{file=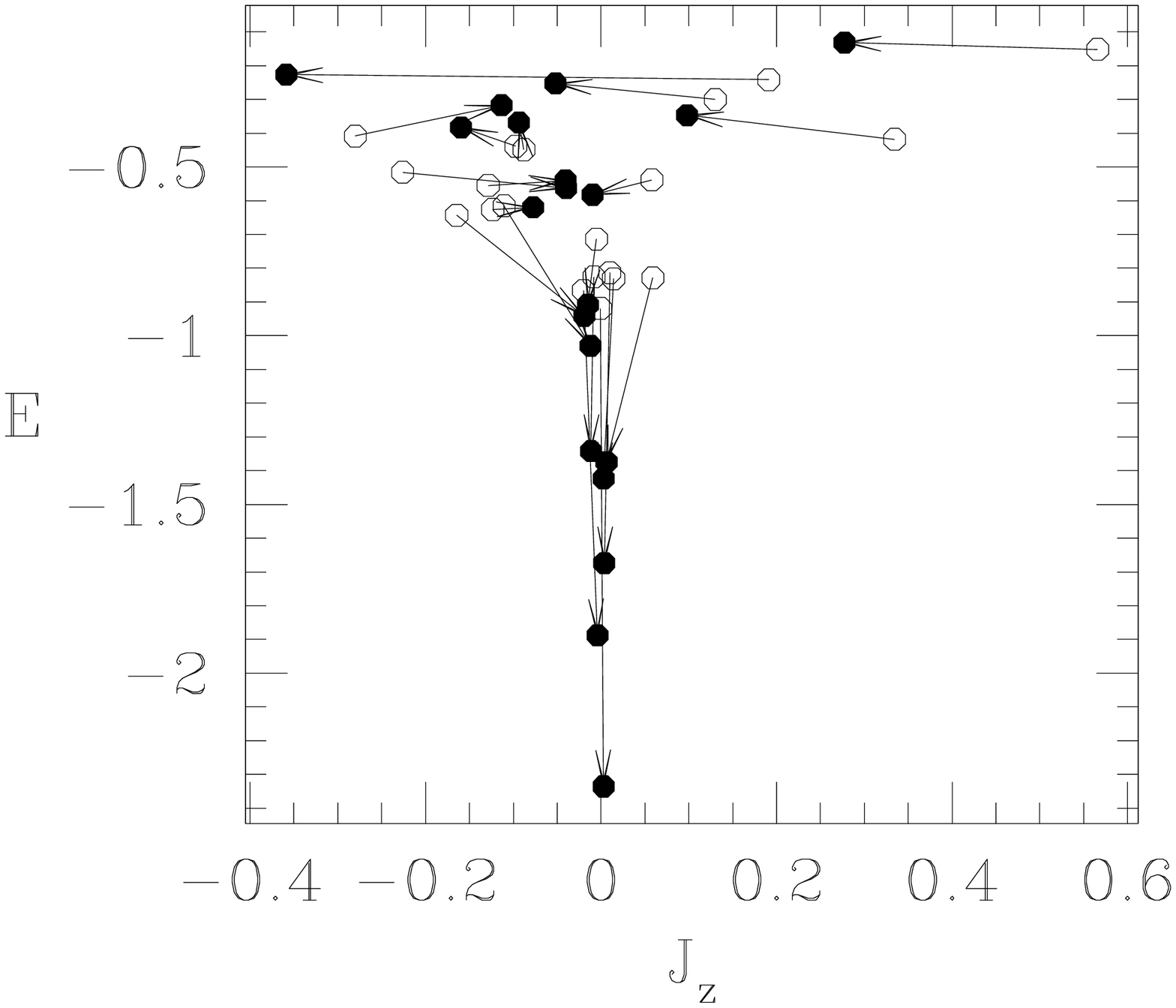,width=0.45\textwidth,angle=0}}
\vspace{0.5cm}
\caption[Clump E-J diagram for collapse cases]
{The binding energy and angular momentum for each clump for the collapse cases.
The upper left panel is for model Y; the 
upper right panel is for model Z; the 
lower left panel is for model J, while the 
lower right panel is for model P. The open circles represent initial values and the 
filled circles represent final values.}
\label{fig:clumpejco}
\end{figure}
\clearpage

\begin{figure}[htbp]
\centerline{\psfig{file=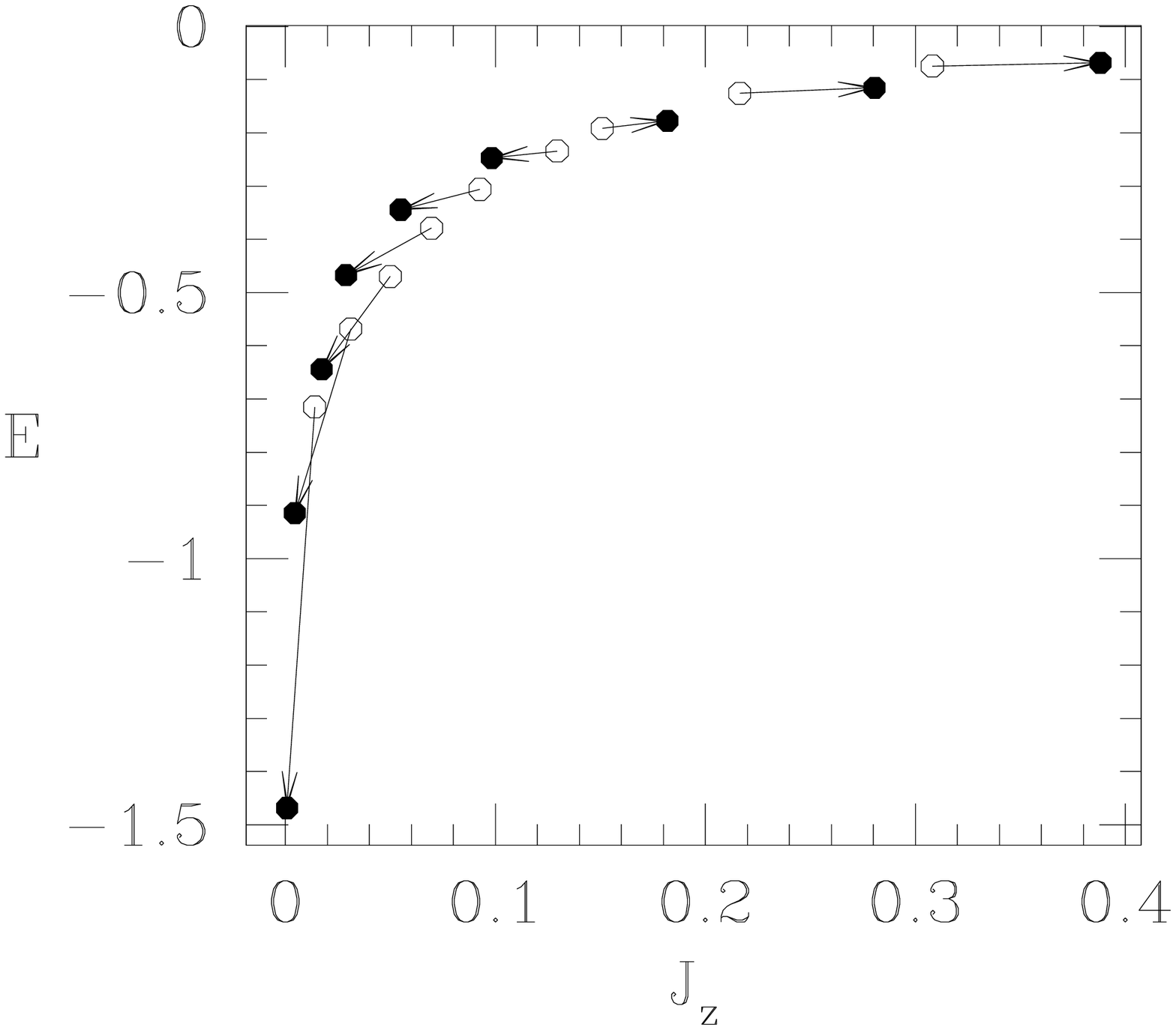,width=0.45\textwidth,angle=0} 
\psfig{file=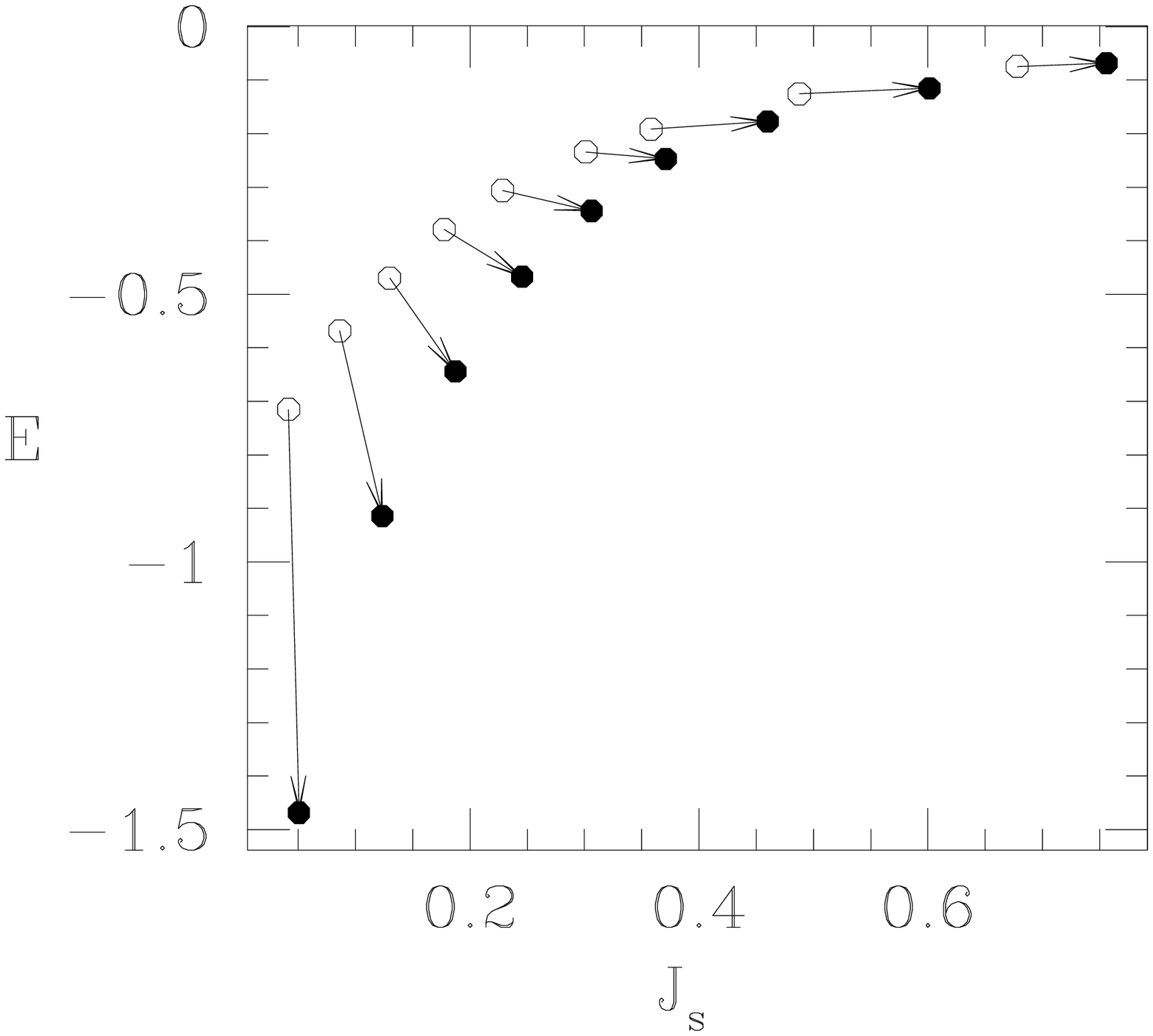,width=0.45\textwidth,angle=0}}
\vspace{0.5cm}
\centerline{\psfig{file=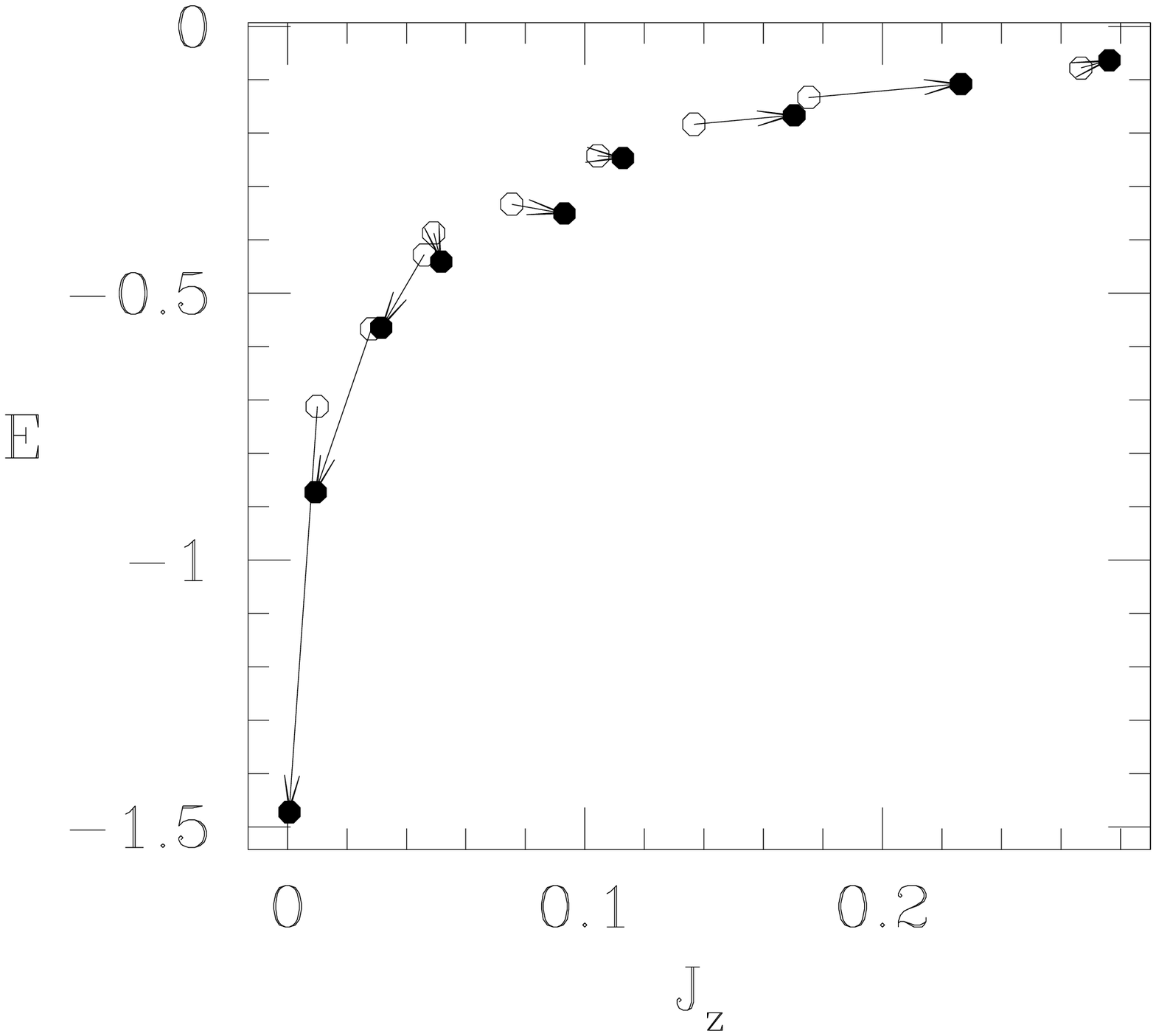,width=0.45\textwidth,angle=0}
\psfig{file=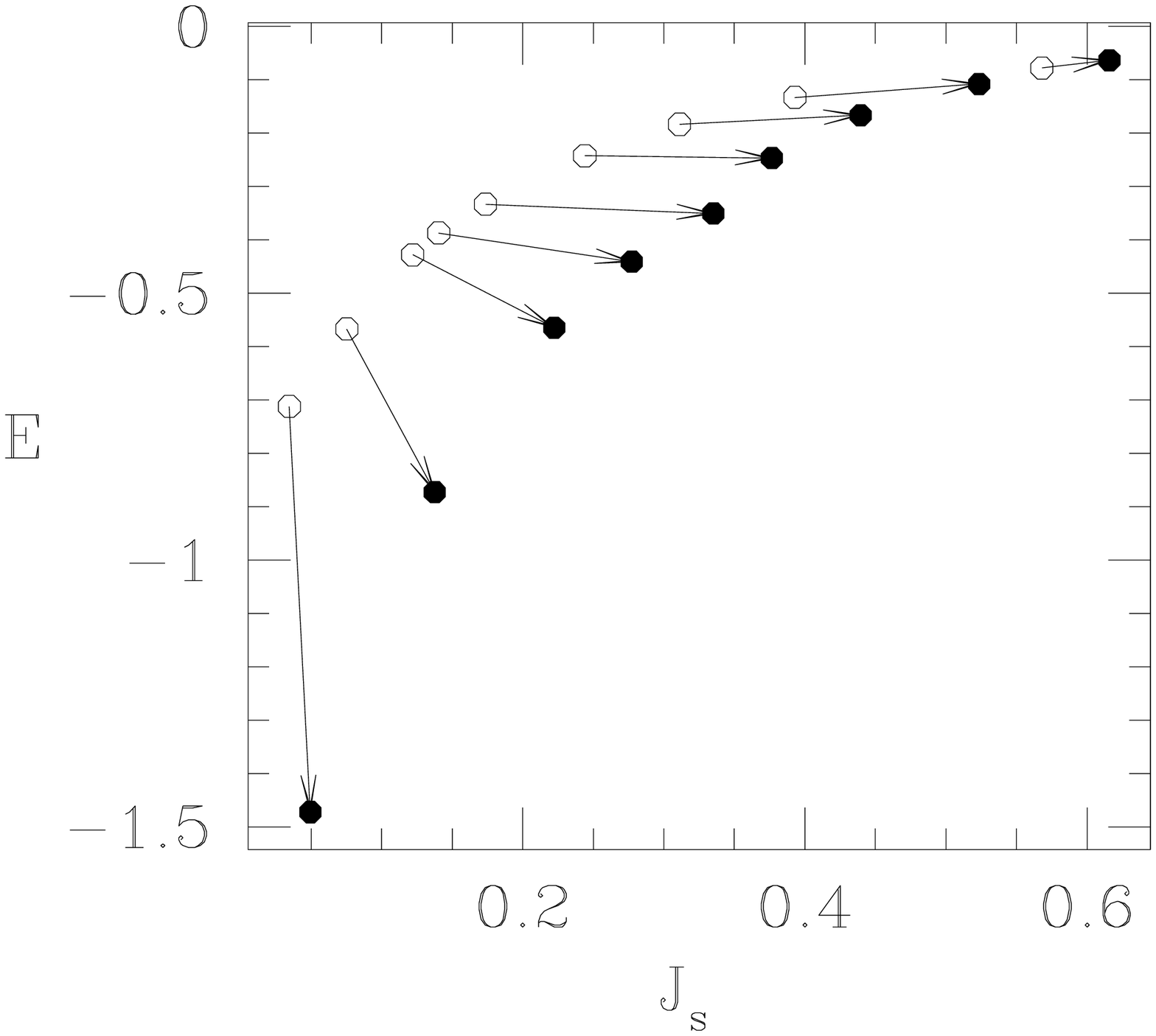,width=0.45\textwidth,angle=0}}
\vspace{0.5cm}
\caption[lindblad]
{The Lindblad diagrams for the dark matter particles.  The upper left
panel shows the evolution of the vector angular momentum $J_z$ for
model D; the upper right panel shows  
the scalar angular momentum $J_s$ for model D; the 
lower left panel shows  the vector angular momentum $J_z$ for model K, while the lower
right panel shows the scalar angular momentum $J_s$ for model K.  The most
bound $90\%$ of the DM particles are divided into nine equal size bins, 
sorted by the binding energy of each particle at the end of the 
simulation. The open circles represent initial values and the filled
circles represent final values.}
\label{fig:lindblad}
\end{figure}
\clearpage

\begin{figure}[htbp]
\centerline{\psfig{file=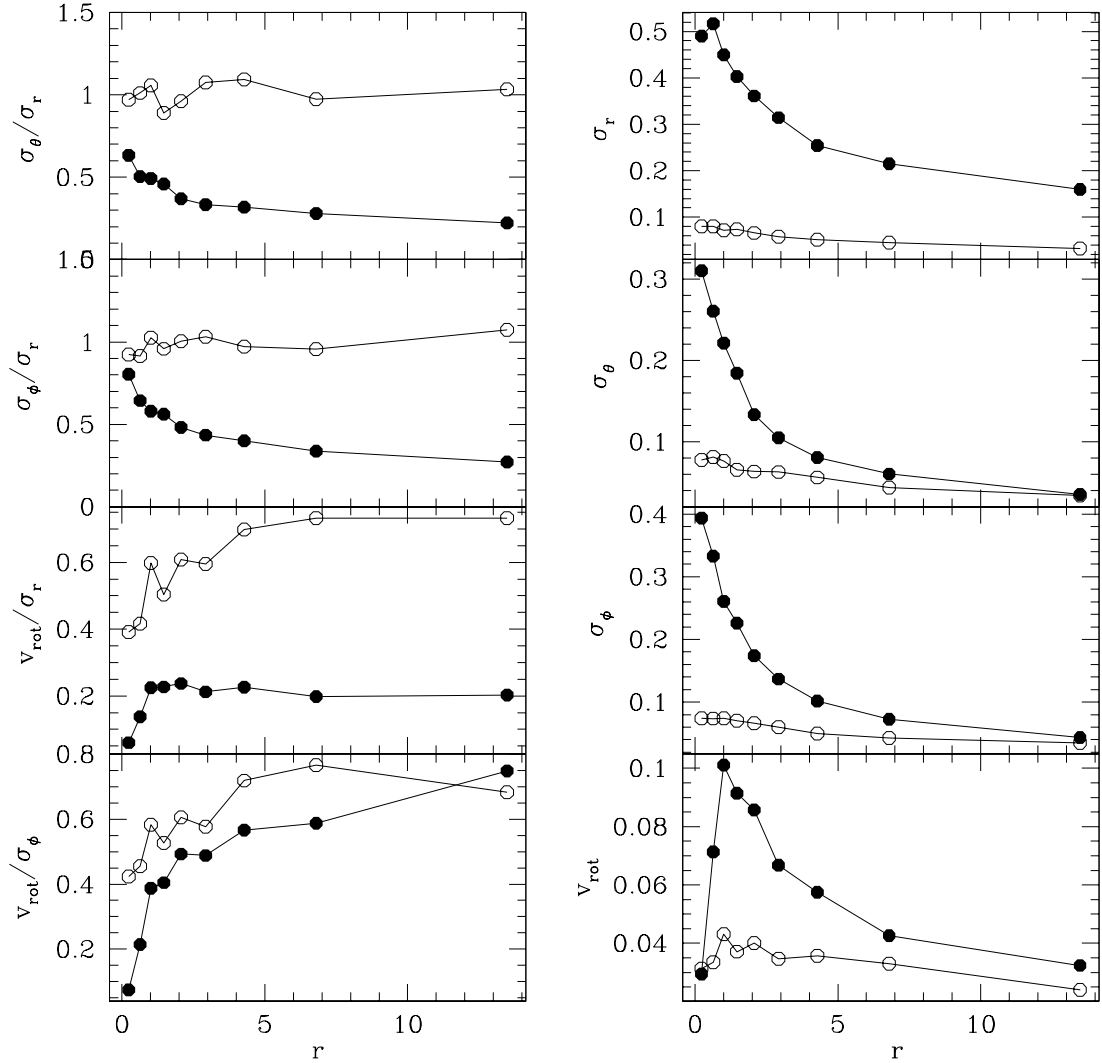,width=0.95\textwidth,angle=0}}
\vspace{1.0cm}
\caption[Kinematics of DM for model V]
{The kinematic properties of the dark halo component of model V.
The quantities shown are 
$\sigma_r$, $\sigma_\theta$, $\sigma_\phi$, the three components of 
the velocity dispersion, together with  $v_{rot}$,  the 
rotation velocity. 
The line connecting the open circles represents the initial values and 
the line connecting the filled circles represents the final values. 
Again the unit for the x-axis is the initial core radius $a$ of the halo mass distribution. }
\label{fig:dmvre10coc20d3f04}
\end{figure}

\begin{figure}[htbp]
\centerline{\psfig{file=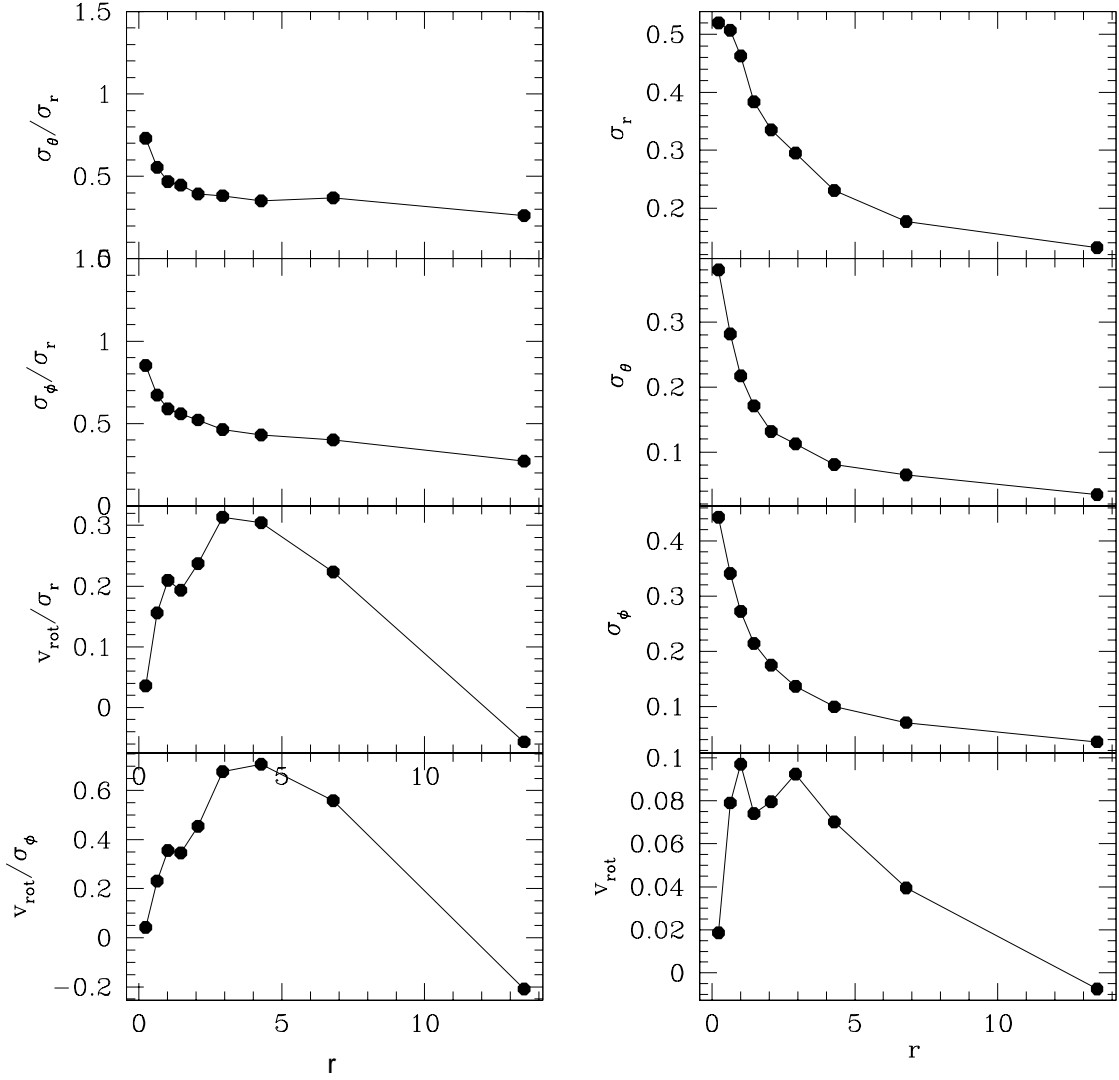,width=0.95\textwidth,angle=0}}
\vspace{1.0cm}
\caption[Kinematics of Debris for model V]
{The kinematic properties of  the debris formed in  model V. The quantities shown are 
$\sigma_r$, $\sigma_\theta$, $\sigma_\phi$, the three components of 
the velocity dispersion, together with  $v_{rot}$,  the 
rotation velocity. Again the unit for the x-axis is the initial core radius $a$ of the halo mass distribution. }

\label{fig:dedmvre10coc20d3f04}
\end{figure}
\clearpage

\begin{figure}[htbp]
\centerline{\psfig{file=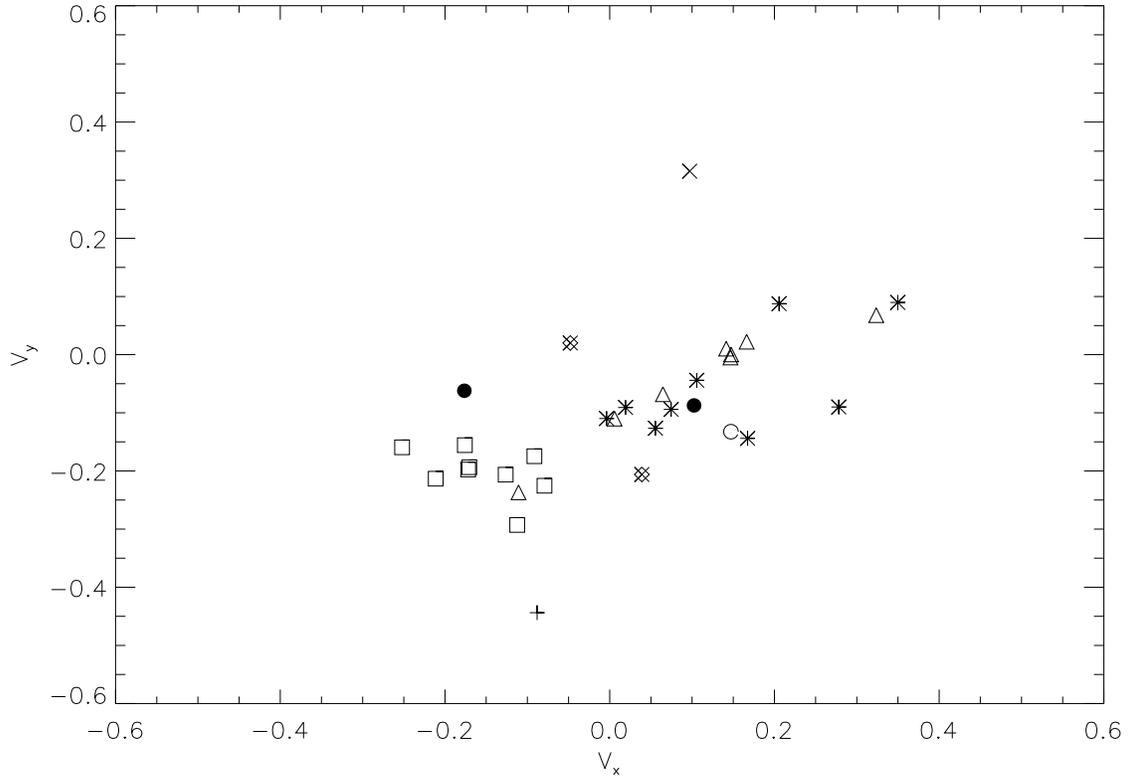,width=0.95\textwidth,angle=0}}
\vspace{1.0cm}
\caption[phase]
{The velocity components $v_x$ and $v_y$ for the clump particles within a box of  size $\sim 3$ units on a side, located at the coordinate 
(4,4,0) for the model U, at the end of simulation. The different symbols 
distinguish  particles from different initial clumps. Again the unit of length  is the initial core radius $a$ of the halo mass distribution. }
\label{fig:phase}
\end{figure}
\clearpage

\begin{figure}[htbp]
\centerline{\psfig{file=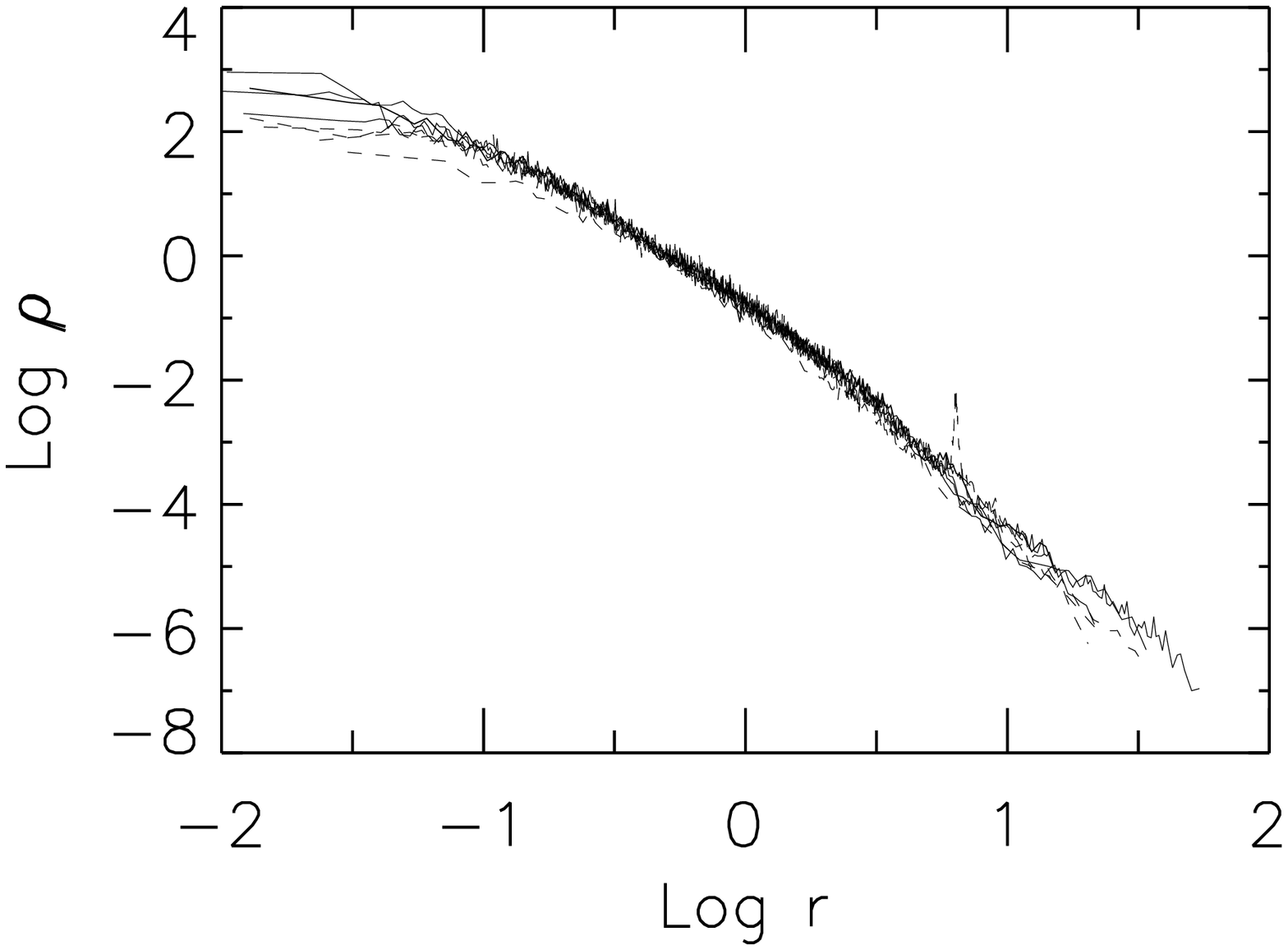,width=0.45\textwidth,angle=0} \hspace{1.0cm}
\psfig{file=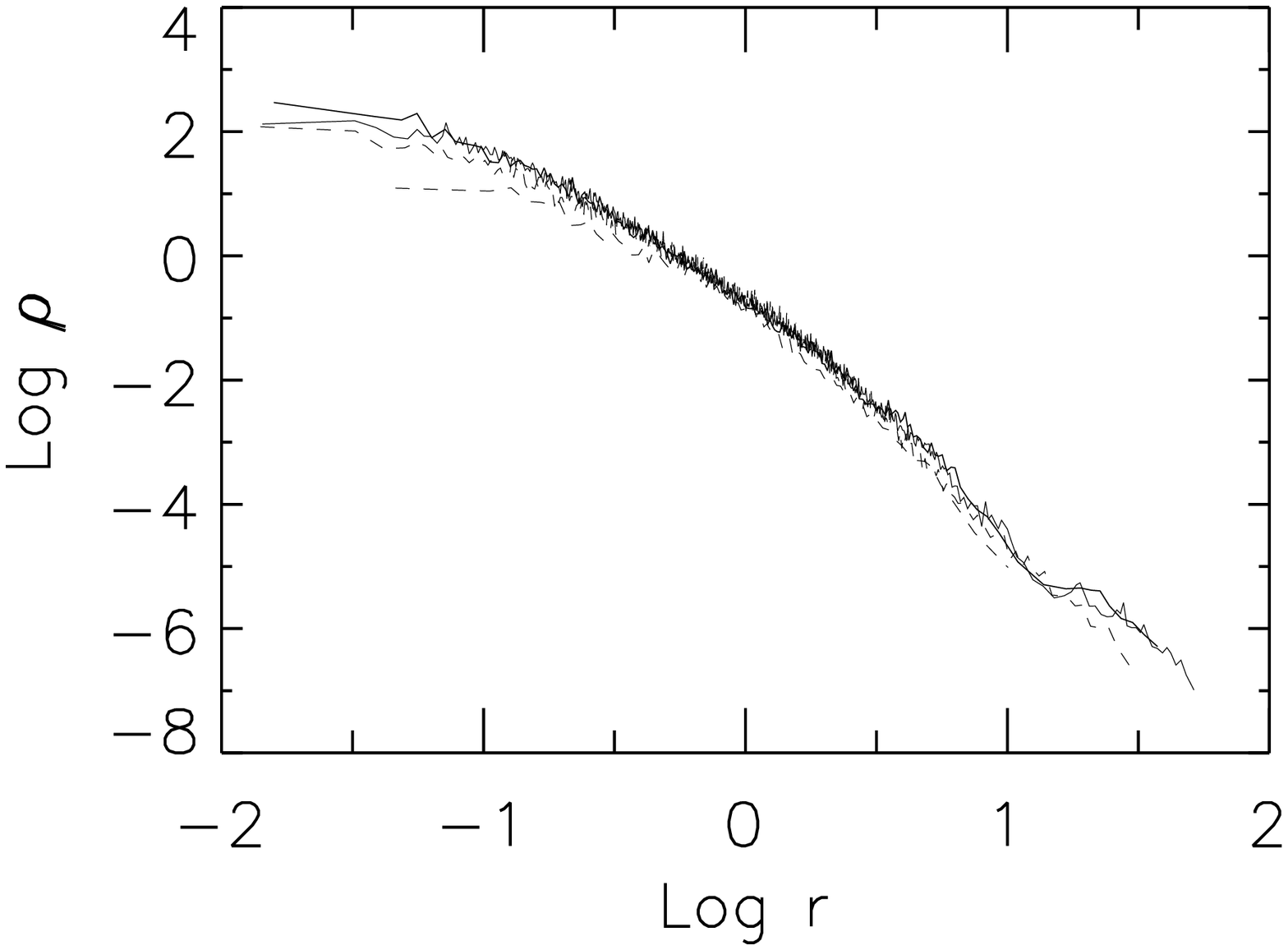,width=0.45\textwidth,angle=0}}
\vspace{1.0cm}
\centerline{
\psfig{file=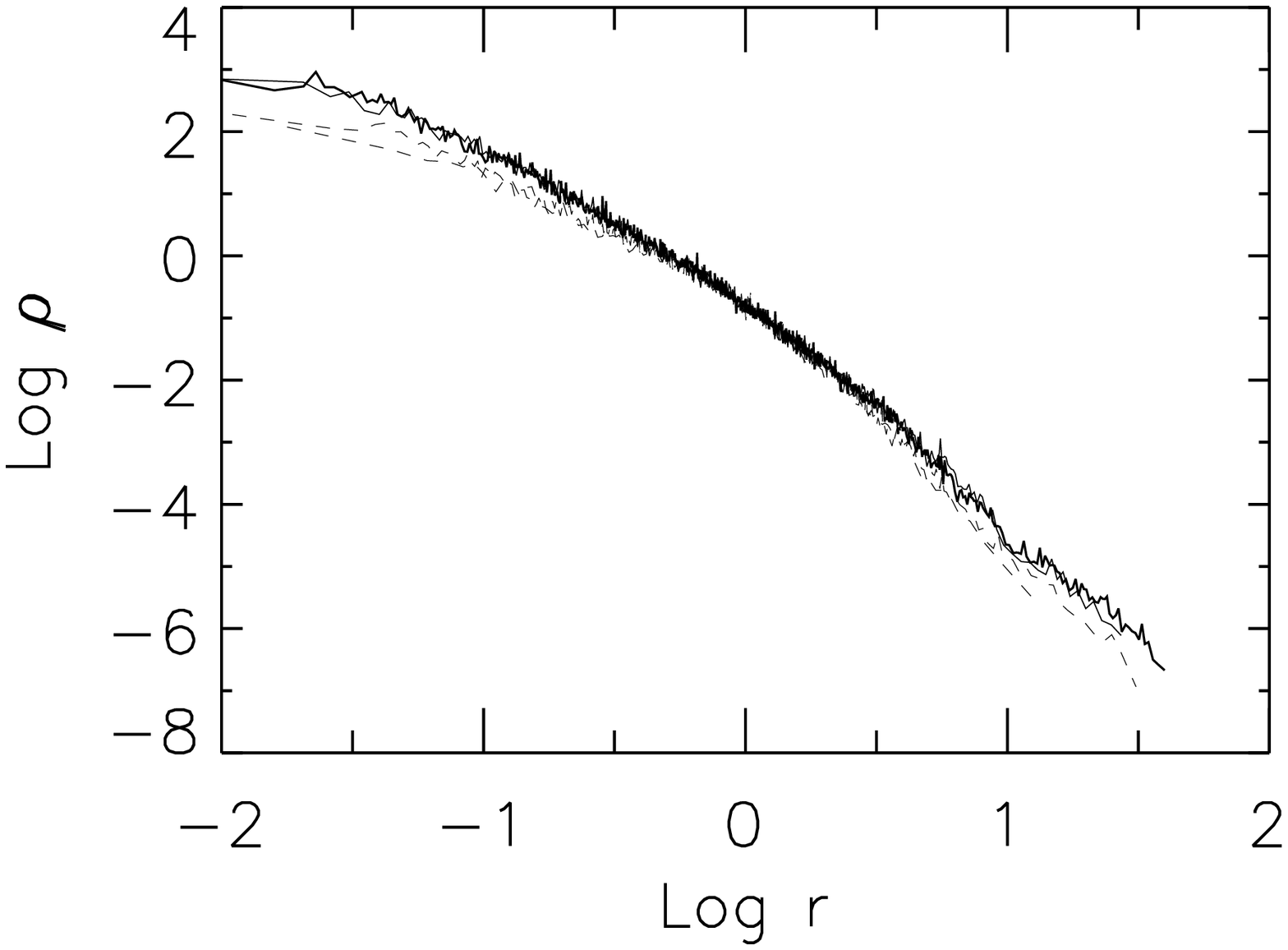,width=0.45\textwidth,angle=0}  \hspace{1.0cm}
\psfig{file=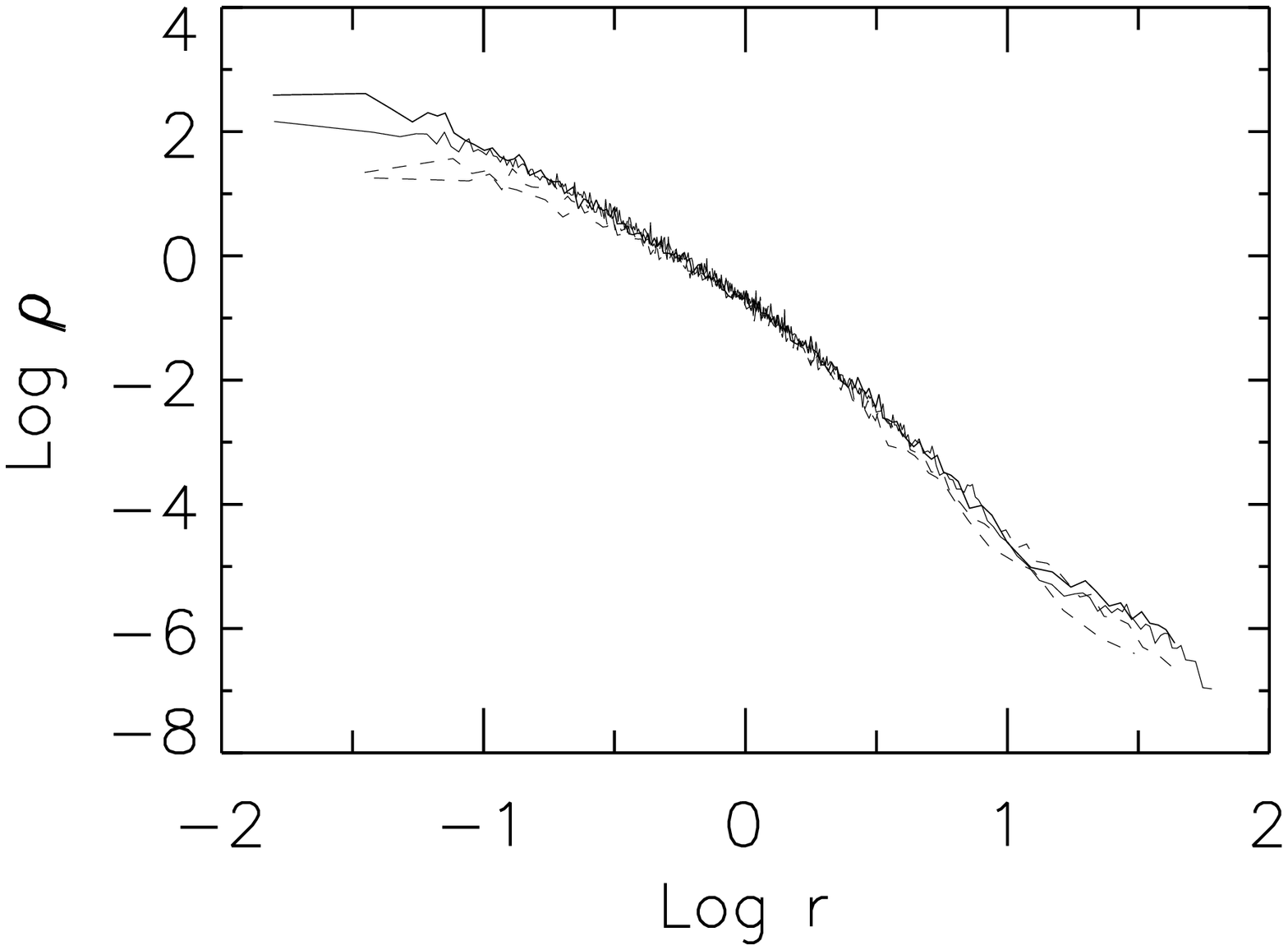,width=0.45\textwidth,angle=0}}
\vspace{1.0cm}
\caption[Density profile]
{The normalised density profiles of the debris for the
models  with 20 clumps in a Hernquist profile (upper left panel), 
20 clumps in a Plummer profile (upper right), 80 clumps in a Hernquist profile (lower left) 
and 5 clumps in a Hernquist profile (lower right). 
The solid lines correspond to the collapse cases and the dashed lines correspond to the 
virialized cases. Again the unit of length for the x-axis is the initial core radius $a$ of the halo mass distribution. }
\label{fig:rhodensity}
\end{figure}
\clearpage

\begin{figure}[htbp]
\centerline{\psfig{file=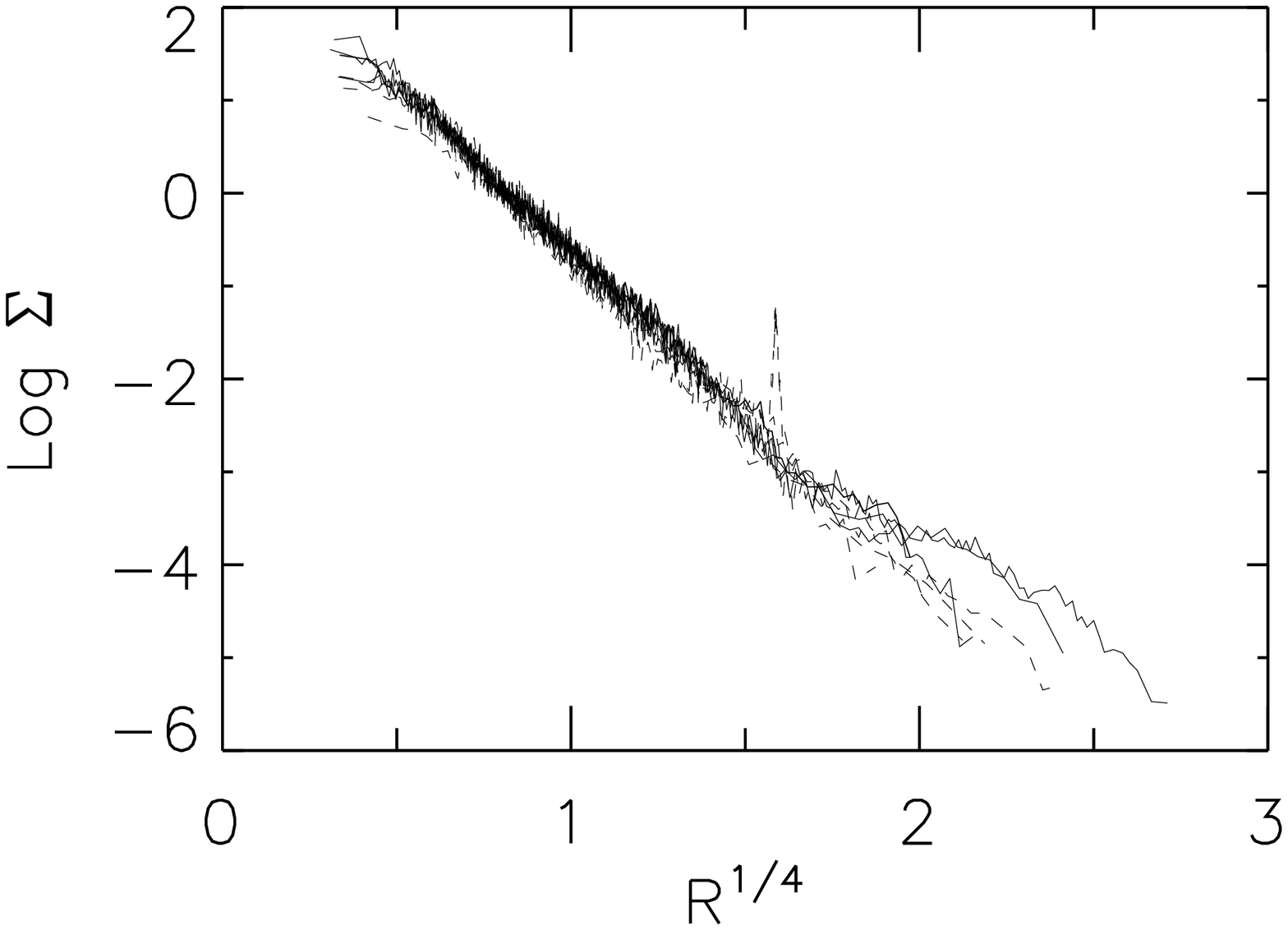,width=0.45\textwidth,angle=0}  \hspace{1.0cm}
\psfig{file=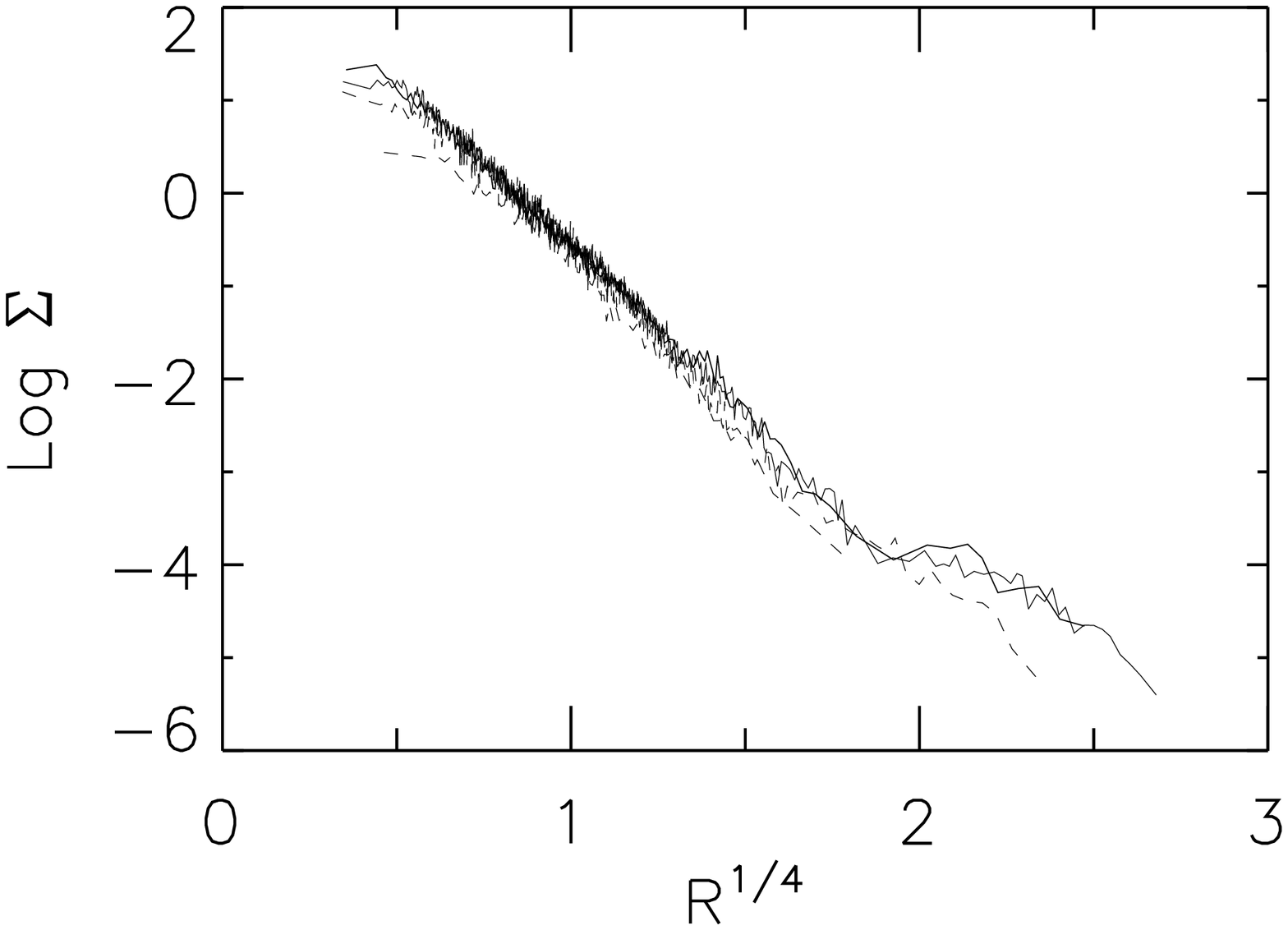,width=0.45\textwidth,angle=0}}
\vspace{1.0cm}
\centerline{
\psfig{file=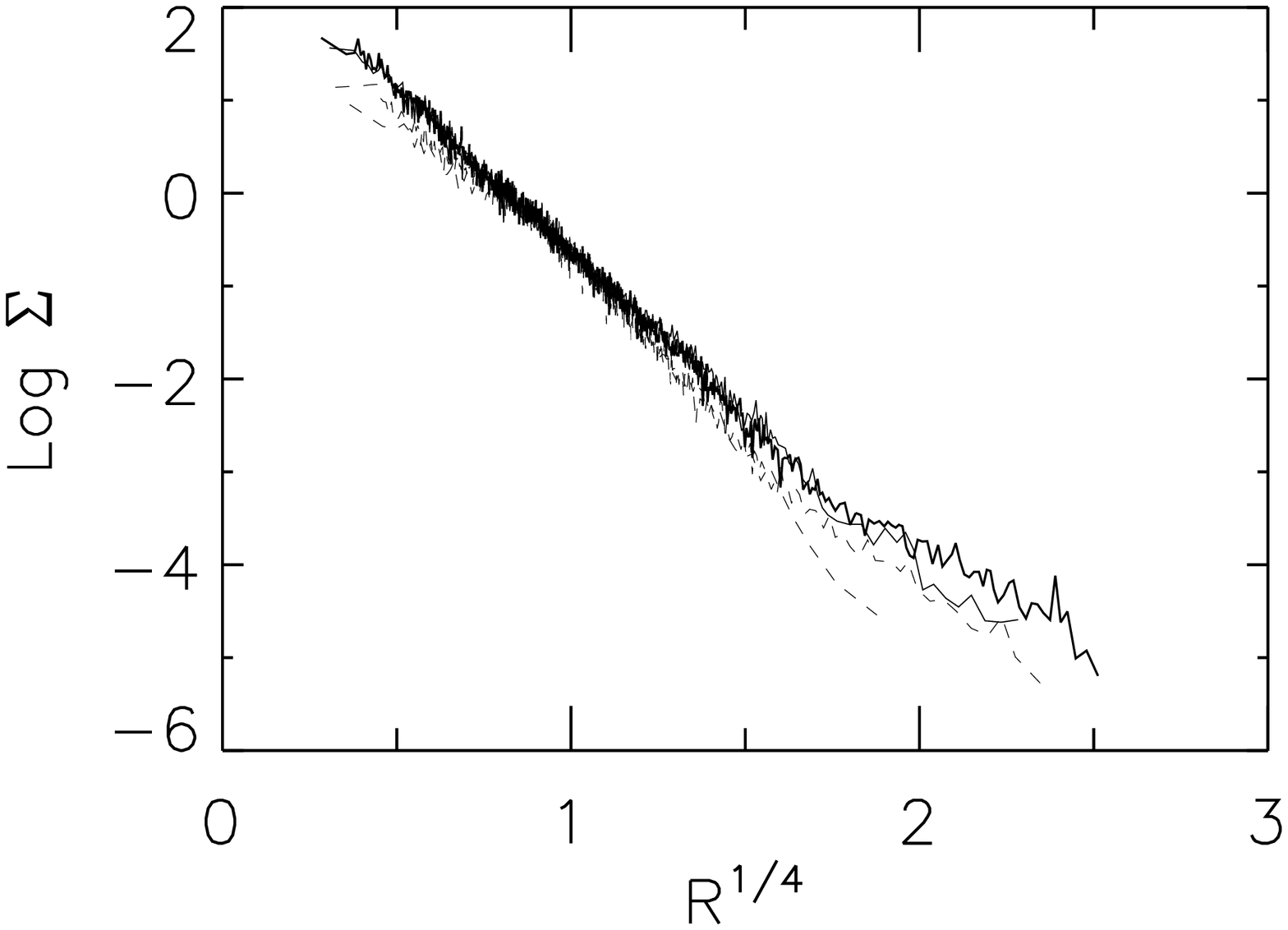,width=0.45\textwidth,angle=0}  \hspace{1.0cm}
\psfig{file=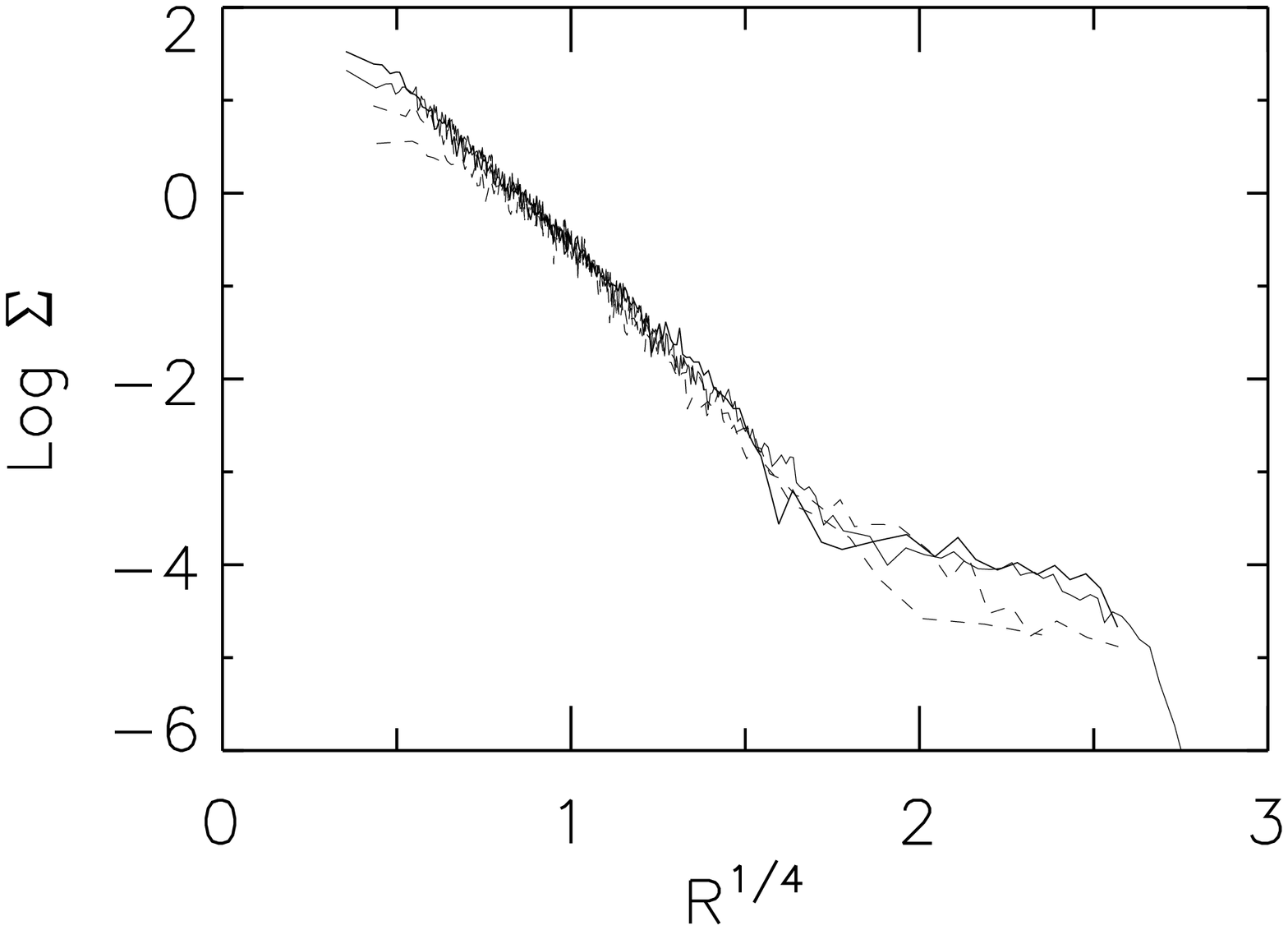,width=0.45\textwidth,angle=0}}
\vspace{1.0cm}
\caption[Surface density profile]
{The normalized surface density profiles of the debris for the
models  with 20 clumps in a Hernquist profile (upper left panel), 
20 clumps in a Plummer profile (upper right), 80 clumps in a Hernquist profile (lower left) 
and 5 clumps in a Hernquist profile (lower right). 
The solid lines correspond to the collapse cases and the dashed lines correspond to the 
virialized cases. Again the unit of length 
is the initial core radius $a$ of the halo mass distribution, 
and here we have plotted the projected distance in half-mass radii. }
\label{fig:sigdensity}
\end{figure}

\begin{figure}[htbp]
\centerline{\psfig{file=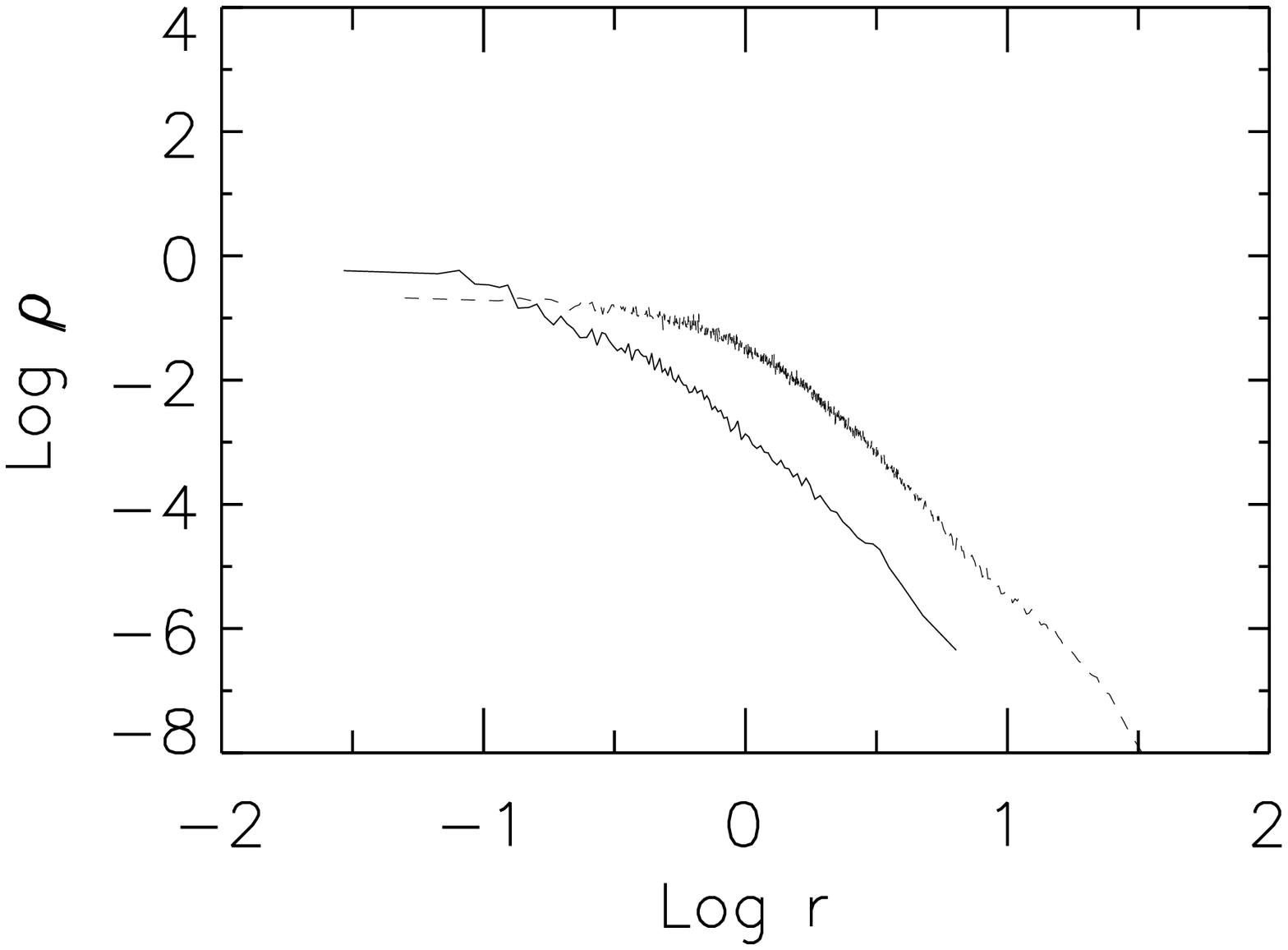,width=0.45\textwidth,angle=0} \hspace{1.0cm}
\psfig{file=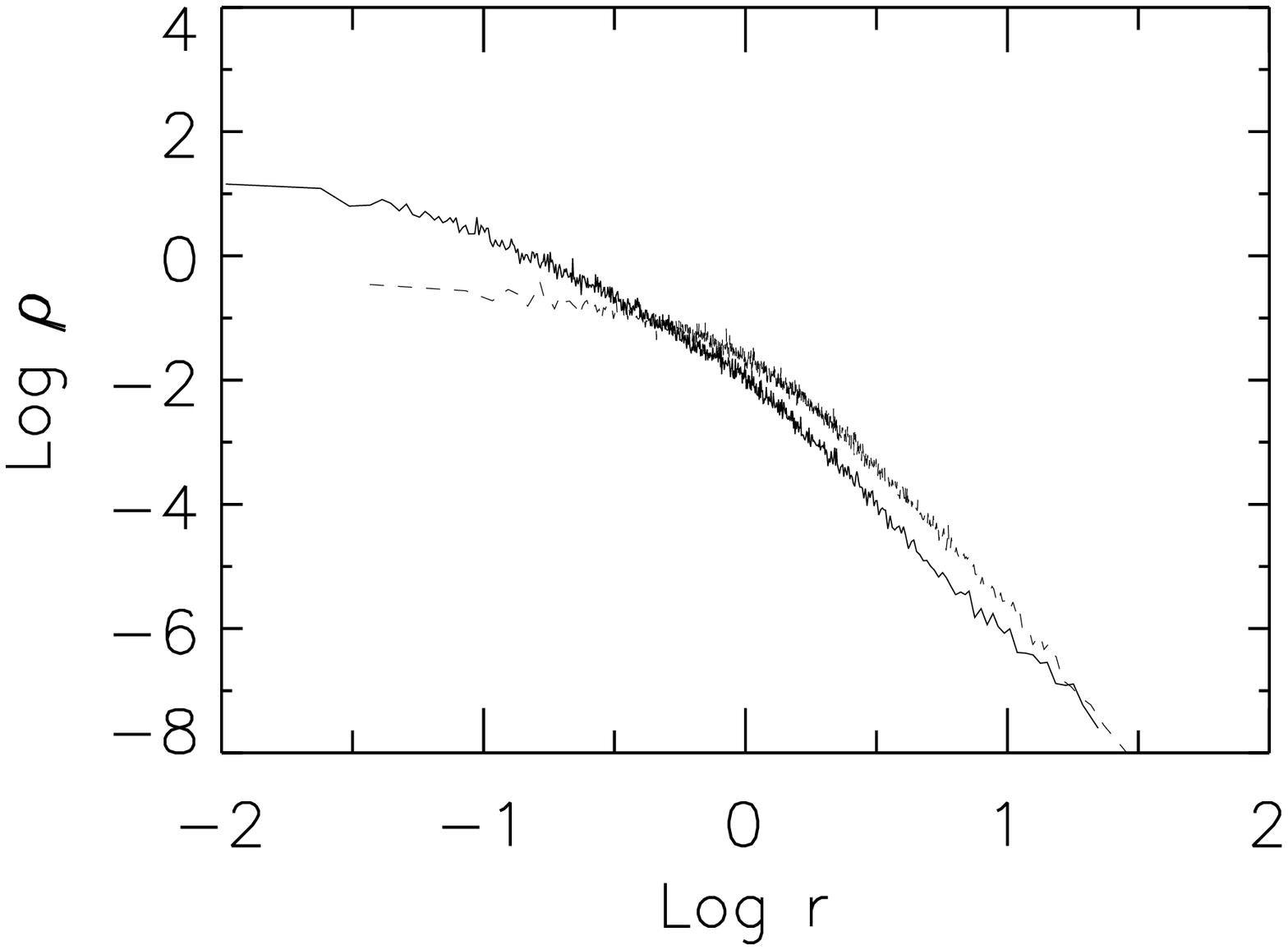,width=0.45\textwidth,angle=0}}
\vspace{1.0cm}
\caption[Density of Debris and Dark matter]
{The final density profile of each of the debris (solid lines) 
and the dark halo (dashed lines) for the 
models M (left panel) and N (right panel). Again the unit of length is the initial core radius $a$ of the halo mass distribution. }
\label{fig:plumdendedm}
\end{figure}

\begin{figure}[htbp]
\centerline{
\psfig{file=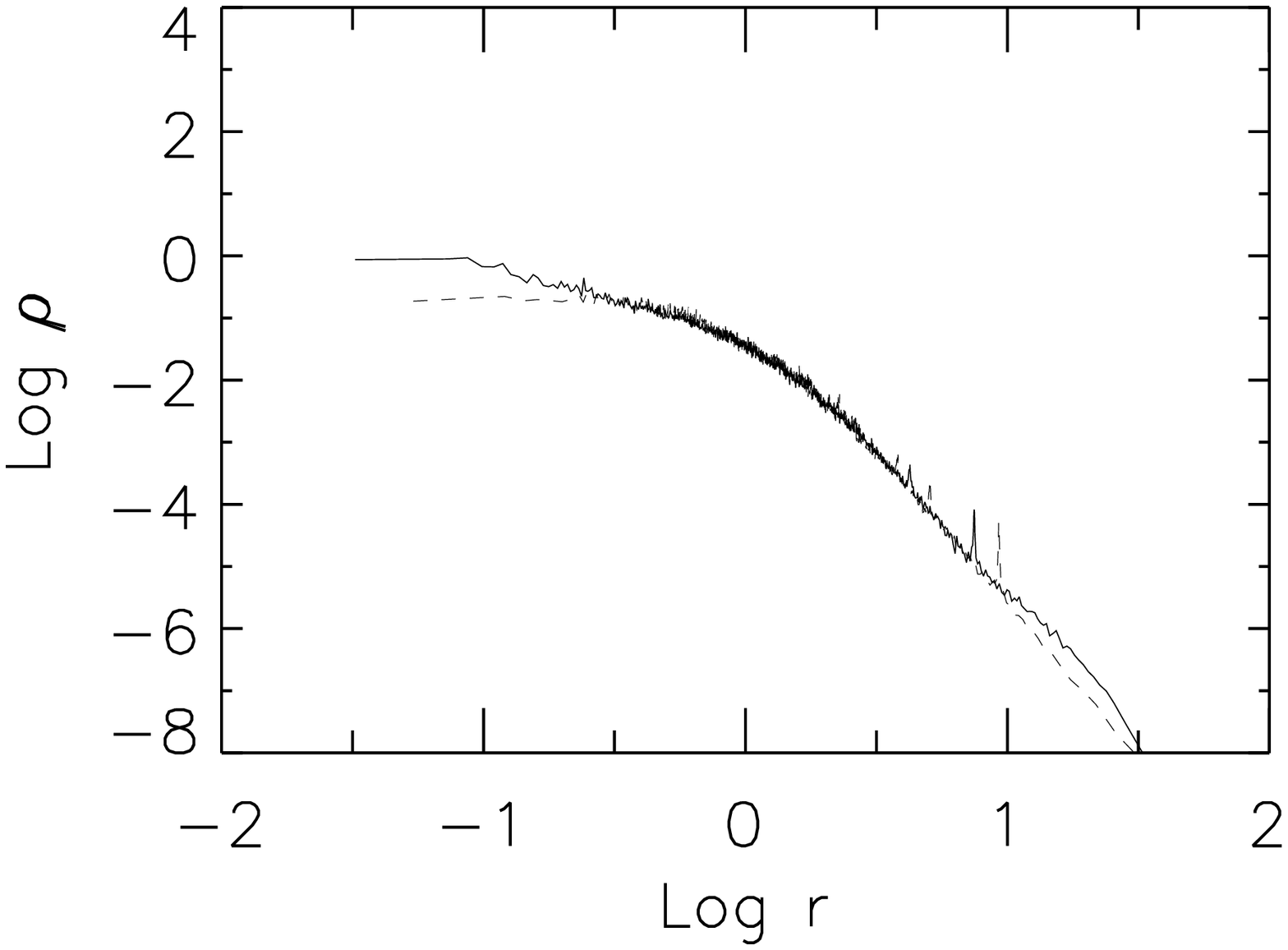,width=0.45\textwidth,angle=0} \hspace{1.0cm}
\psfig{file=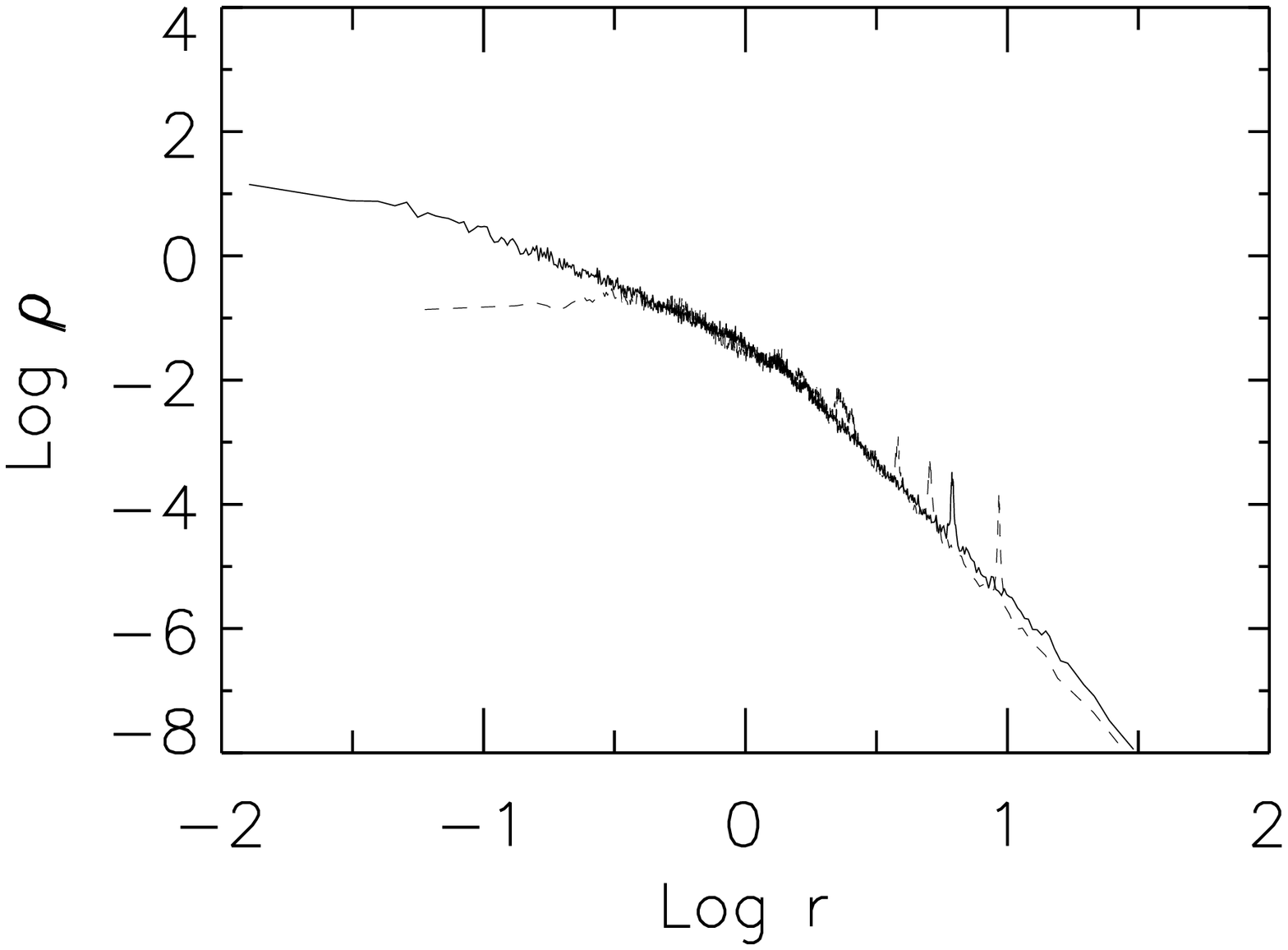,width=0.45\textwidth,angle=0}}
\vspace{1.0cm}
\caption[Density profile with all components included]
{The initial (solid lines) and final (dashed lines) density profiles,  with all components included, for 
model M (left panel) and N (right panel). Again the unit of length is the initial core radius $a$ of the halo mass distribution. }
\label{fig:plumdenall}
\end{figure}
\clearpage

\begin{figure}[htbp]
\centerline{
\psfig{file=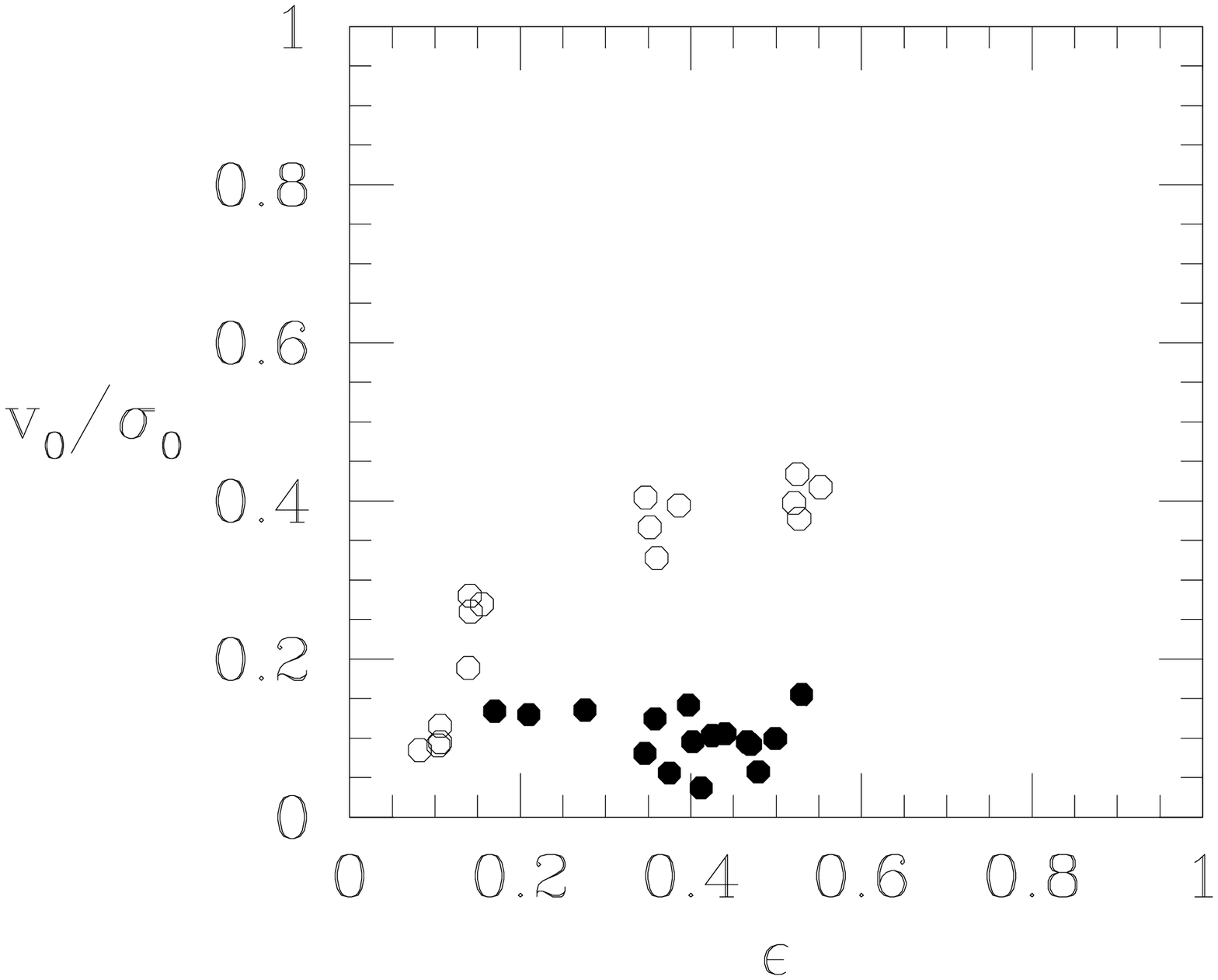,width=0.45\textwidth,angle=0} 
\hspace{0.5cm}
\psfig{file=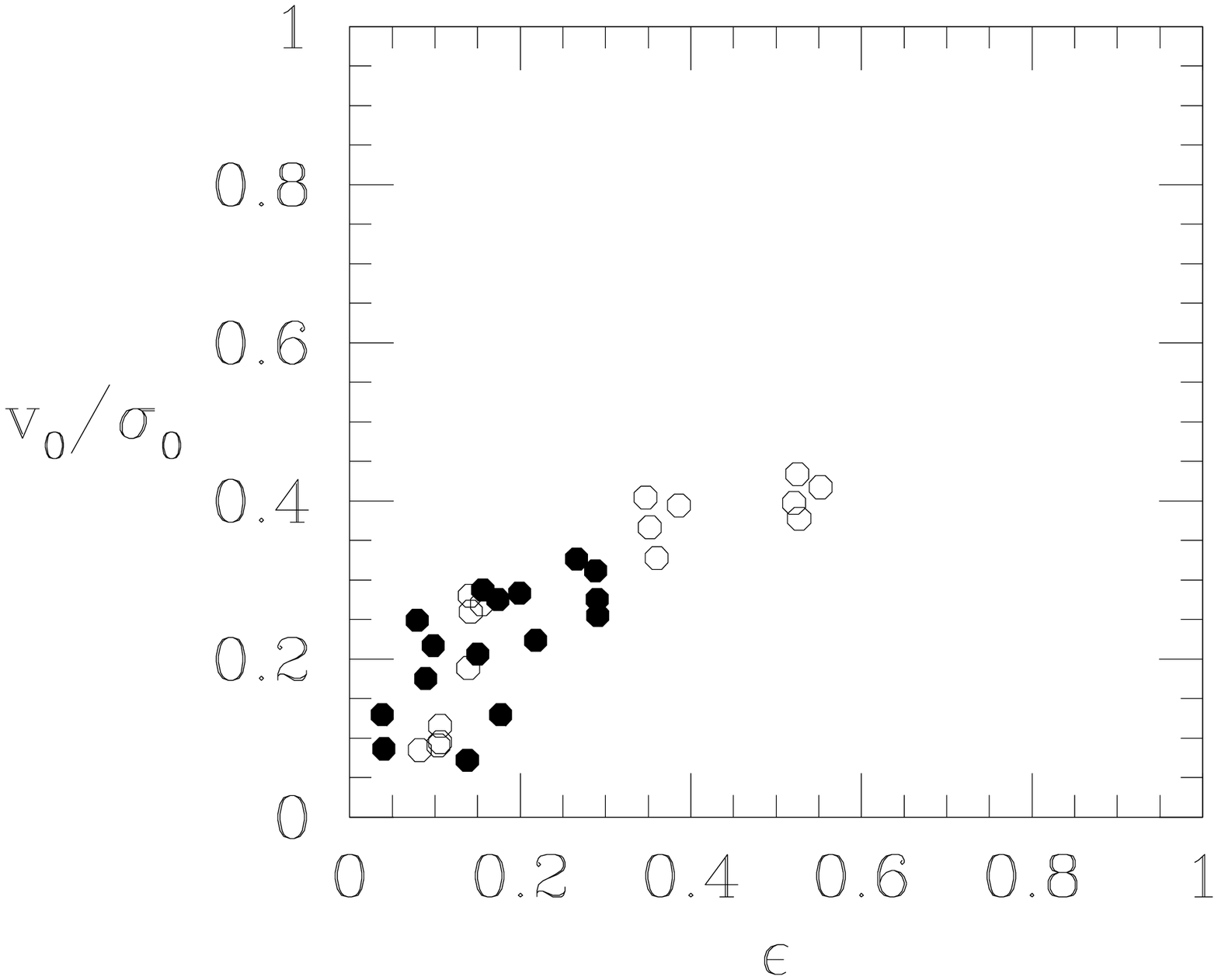,width=0.45\textwidth,angle=0}}
\vspace{0.5cm}
\centerline{
\psfig{file=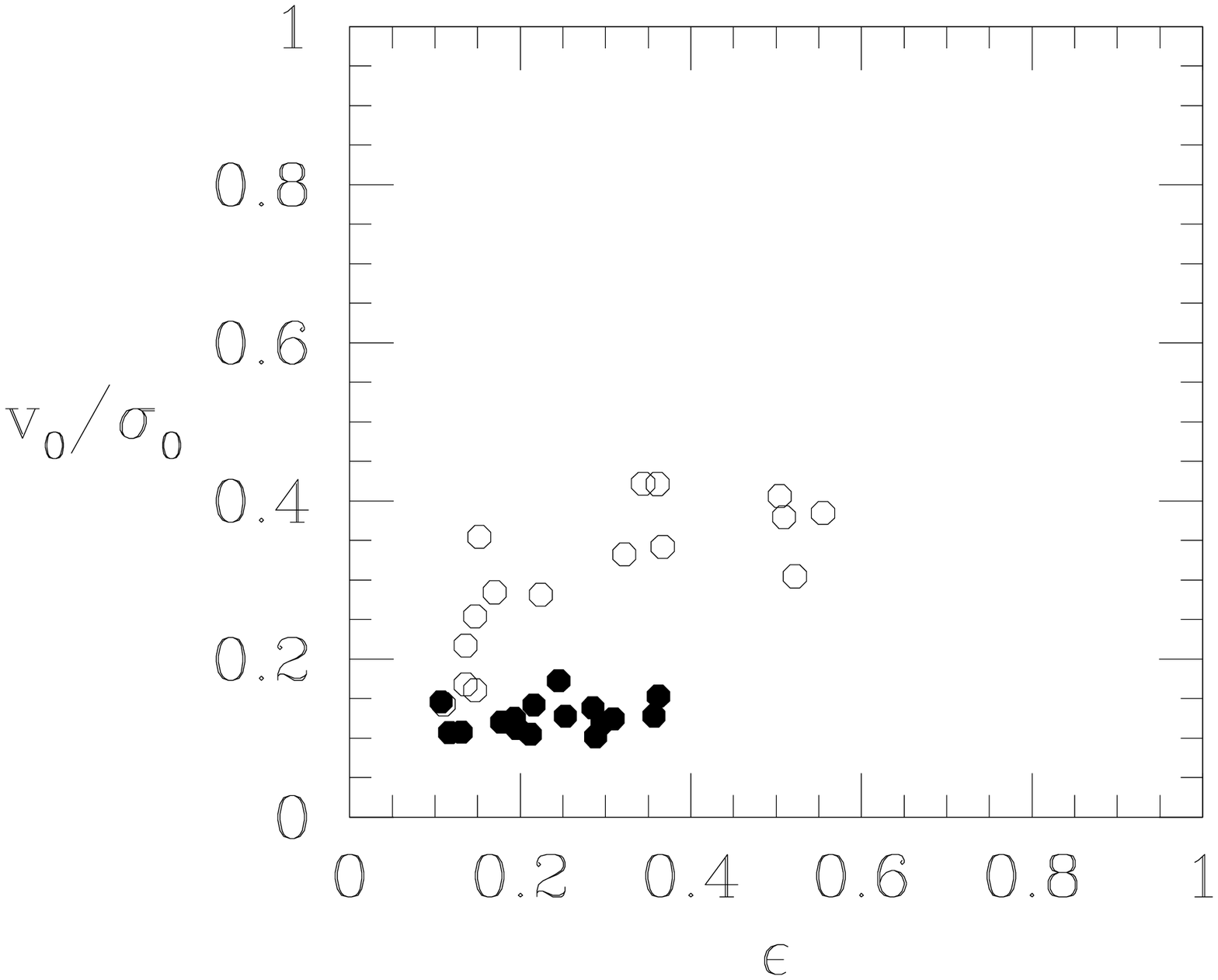,width=0.45\textwidth,angle=0} 
\hspace{0.5cm}
\psfig{file=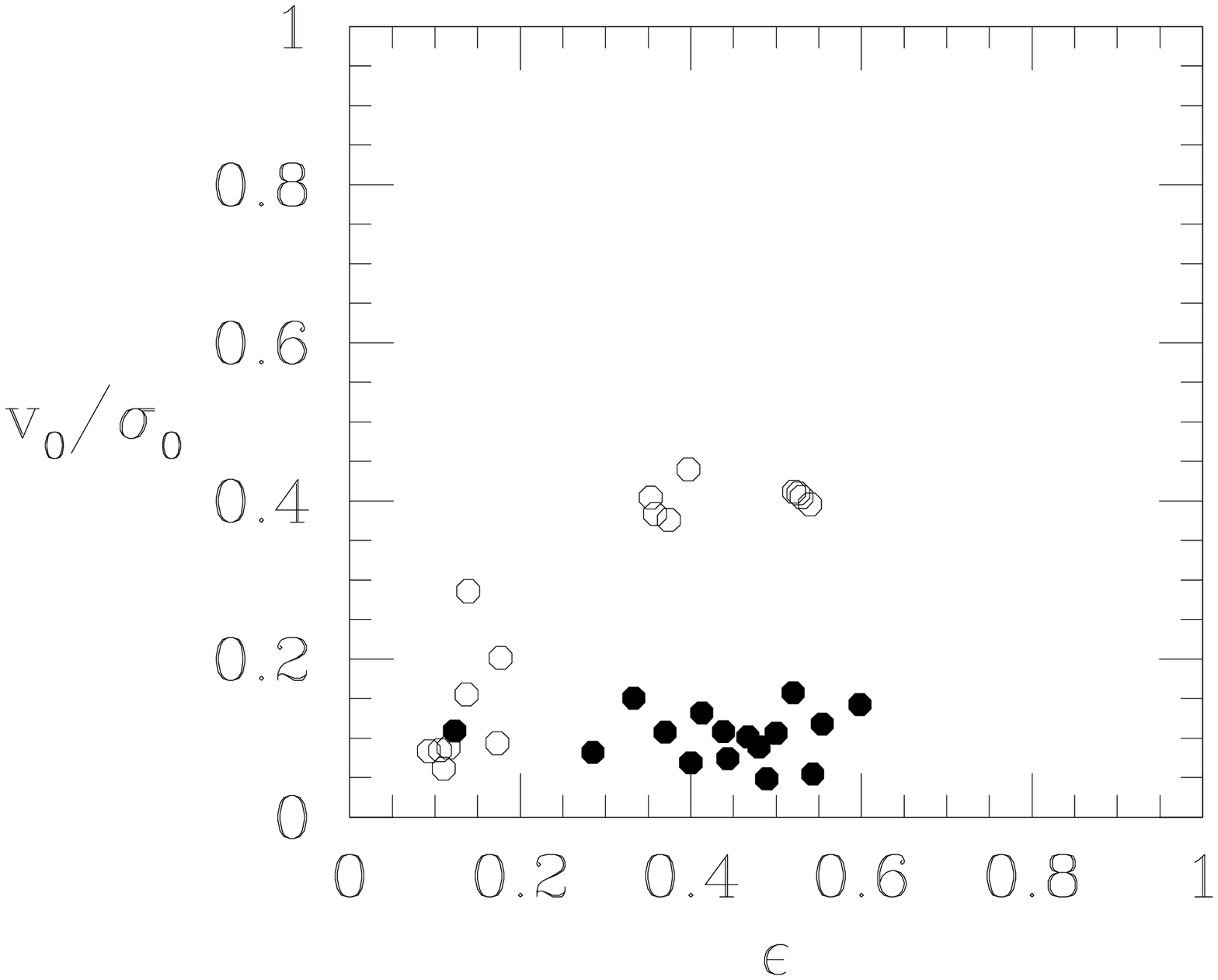,width=0.45\textwidth,angle=0}}
\vspace{0.5cm}
\caption[The relationship between the rotation parameter $v_0/\sigma_0$ and 
the ellipticity $\epsilon$ for the dark halo]
{The locations of the dark halo on the ($v_0/\sigma$, $\epsilon$) plane for the models
J (upper left panel), G (upper right), 
K (lower left), and U (lower right), 
from sixteen different viewing angles.
The open circles and filled circles represent initial and final locations respectively.}
\label{fig:contourvrvde}
\end{figure}
\clearpage 

\begin{figure}[htbp]
\psfig{file=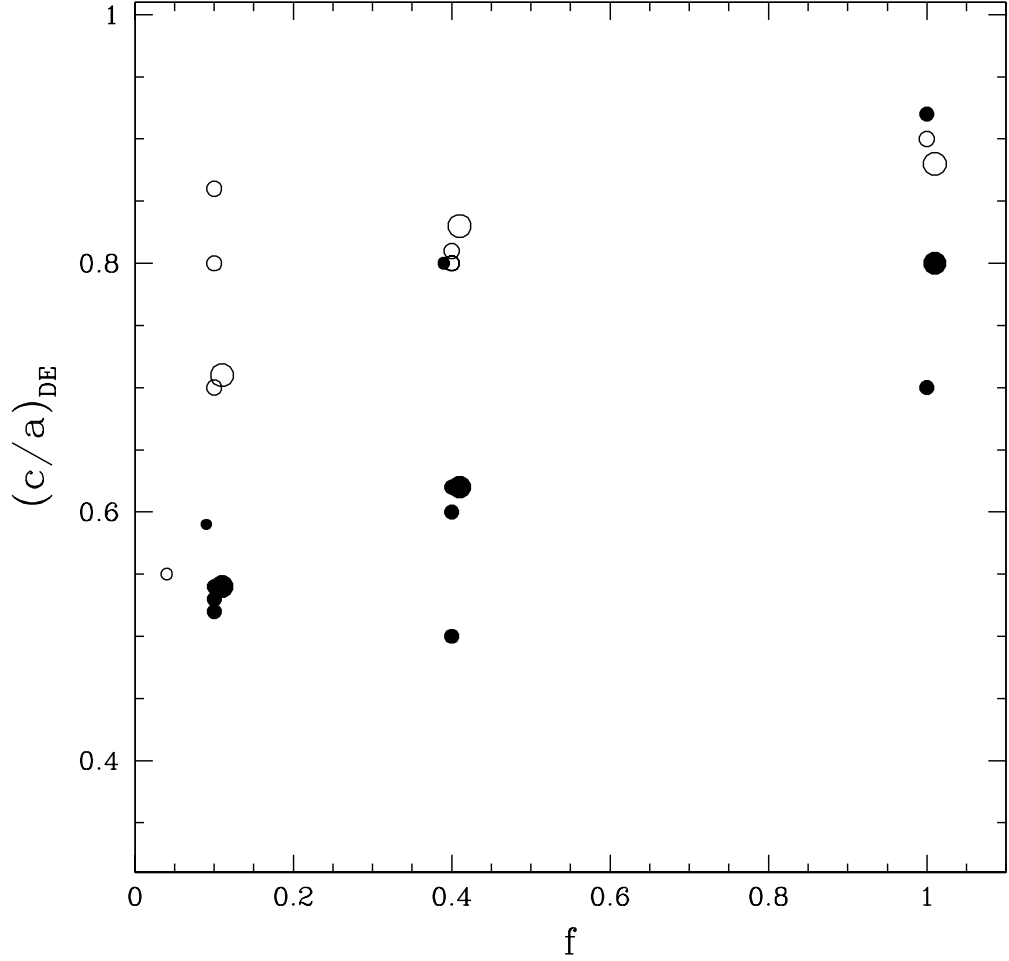,width=\textwidth,angle=0}
\caption{Axial ratio $c/a$ of the debris, plotted against clump mass
fraction $f$.  The open symbols denote models initially close to 
equilibrium, and the filled symbols are for the `collapse'
simulations. The size of the symbol is larger for simulations with
larger numbers of clumps.}
\label{shapes}
\end{figure}
\clearpage
\begin{figure}[htbp]

\caption[The scaled snapshots of the models K and A
for t=14, t=56 and t=126]
{The scaled snapshots of the models K (top panels) and A 
(bottom). 
Each image is the projection onto the XY plane of the particles in the clumpy component 
within a box of size 6 units on a side at time $t=14$ (left panels), 16 units on a 
side at $t=56$ (middle) and 30 units on a side  
at $t=126$ (right).  Again the unit of length is the initial core radius $a$ of the halo mass distribution, and the unit of time is the initial halo crossing time. }
\label{fig:cparameter}
\end{figure}

\begin{figure}[htbp]

\caption[The streaming structure in the outer halo]
{In the top panel are the snapshots of the models Z, 
K, V and E  (from left to right);
in the bottom panel are the snapshots of the models X, 
H, T and B  (from left to right);
Each image is the projection onto the XY plane of the particles in the 
clumpy component 
within a box of  size  100 units on a side, at time $t=126$. Again the unit of length is the initial core radius $a$ of the halo mass distribution. }
\label{fig:substructure}
\end{figure}

\end{document}